\documentclass[mnsc,nonblindrev]{informs1}
\OneAndAHalfSpacedXI
\usepackage{rotating}
\usepackage{bbding}
\usepackage[left=1in,top=1in,right=1in,bottom=1in]{geometry}
\usepackage{amssymb,amsmath,amsxtra,amstext,xfrac,mathtools}
\usepackage{graphicx,booktabs,enumerate,paralist,mdwlist}
\usepackage{dsfont,verbatim,latexsym}
\usepackage{mathrsfs}
\usepackage{bbm}
\usepackage{multirow}
\usepackage[table]{xcolor}
\newcolumntype{C}[1]{>{\centering\let\newline\\\arraybackslash\hspace{0pt}}m{#1}}
\usepackage{longtable,multirow,threeparttable}
\usepackage{float}
\usepackage{tikz,qtree}
\usetikzlibrary{arrows,positioning,decorations.pathreplacing}
\usepackage[normalem]{ulem}
\usepackage{mwe}
\usepackage{makecell}
\usepackage{eurosym}
\usepackage[english]{babel}
\usepackage[utf8]{inputenc}
\usepackage{fancyhdr}
\usepackage{color}
\usepackage{array}
\usepackage{subcaption}
\usepackage{url}
\usepackage[colorlinks=TRUE,citecolor=MyDarkBlue,urlcolor=MyDarkBlue,linkcolor=MyDarkBlue]{hyperref}
\definecolor{MyDarkBlue}{RGB}{158,0,0}

\usepackage{natbib}
\bibpunct[, ]{(}{)}{,}{a}{}{,}
\def\bibfont{\small}

\TheoremsNumberedThrough
\EquationsNumberedThrough

\begin{document}
\RUNAUTHOR{Xu et al.}
\RUNTITLE{Economics of Foundation Models}
\TITLE{The Economics of AI Foundation Models: Openness, Competition, and Governance}
\ARTICLEAUTHORS{
\AUTHOR{Fasheng Xu$^1$ ~~~ Xiaoyu Wang$^2$ ~~~ Wei Chen$^1$ ~~~ Karen Xie$^1$\\[-2mm]}
\AFF{$^1$School of Business, University of Connecticut}
\AFF{$^2$Faculty of Business, The Hong Kong Polytechnic University}
}

\ABSTRACT{Foundation model developers choose how open to make their models. Greater openness provides information and resources that lower the cost of adapting models for AI applications, but can also make competing models easier to use. We develop a two-period model in which an incumbent developer sets its license fee and openness, a deployer adapts the model for users, and an entrant offers a competing model in the second period. As users engage with the application, the deployer learns to adapt the incumbent's model at lower cost. This model-specific learning creates a \textit{data flywheel} that favors continued use of the incumbent's model. We obtain three main findings. First, we identify conditions under which optimal openness falls and then rises as learning strengthens. The incumbent moves from charging a high fee with full openness to restricting openness to keep the deployer from switching. When learning is sufficiently strong, the incumbent lowers its fee to encourage adaptation and usage. This builds enough experience to retain the deployer despite greater openness. Second, a full-openness mandate can make switching so attractive that the incumbent abandons this low-fee strategy. The higher fee can reduce initial adaptation despite greater openness. An \textit{openness trap} arises when the resulting first-period welfare loss exceeds the second-period gain from easier adaptation of the entrant's model. When the deployer can reuse its accumulated experience after switching models, the incumbent finds it harder to retain the deployer, and the openness trap can disappear. Third, we examine two interventions. Vertically integrating the incumbent and the deployer eliminates the fee between them but commits deployment to the incumbent's model, potentially forgoing a rival that is cheaper to adapt. Integration raises their joint profit, consumer surplus, and welfare when learning is sufficiently strong, but can lower these outcomes when learning is weak. A small subsidy for the deployer's model use can benefit both developers, the deployer, and consumers. It can also prompt the incumbent to raise its fee by more than the subsidy and reduce openness. The deployer then pays more per use even after the subsidy and faces higher adaptation costs. The resulting reduction in adaptation lowers welfare even before government spending is deducted. Together, these findings have implications for model choice and AI policy: deployers should assess how much of their adaptation expertise would carry over to another model, and policymakers should anticipate developers' pricing and openness responses.

}
\HISTORY{Current version: November, 2026}

\maketitle
\vspace{-10mm}

\section{Introduction}

Foundation models (FMs) allow applications for many different tasks to be built on a common, general-purpose AI input. In an important configuration of the AI value chain, upstream \textit{developers}, such as OpenAI and Anthropic, supply models that downstream \textit{deployers} adapt for end users \citep{cma2024fmupdate}. For example, Cursor combines models from several developers with application-specific instructions and tools. %
The deployer must evaluate model behavior, integrate data and tools, and adjust the application to its intended tasks. We use \textit{fine-tuning} broadly to refer to this costly adaptation.\footnote{This includes supervised fine-tuning, which modifies model parameters, as well as prompt engineering \citep{wei2022chain}, retrieval-augmented generation \citep{gao2023retrieval}, and agentic systems that combine reasoning, tool use, and memory \citep{acharya2025agentic}.} The developer's commercial terms and the resources it makes available therefore affect how much the deployer invests in turning model capability into a useful product.

A central choice for the developer is \textit{model openness}: the information, tools, and rights it makes available for downstream adaptation.
Openness is a matter of degree, ranging from documentation and adaptation interfaces for proprietary APIs to downloadable weights that deployers can modify directly \citep{bommasani2024considerations}.
OpenAI and Anthropic offer the former; for example, Notion builds its External Agents feature using Anthropic's Claude APIs.\footnote{See Anthropic's \href{https://claude.com/customers/notion-qa}{interview with Notion's product manager} on building the feature with Claude Managed Agents.}
Alibaba's Qwen models illustrate the latter. Releasing weights does not preclude charging for services: Alibaba also charges deployers for hosting and running these models through Alibaba Cloud Model Studio.

Greater openness can lower the deployer's cost of adapting the developer's model, encouraging current investment and usage. Downstream adaptation of open-weight models is now extensive: Hugging Face's August 2026 report counts more than 150,000 published derivatives of Qwen models alone.\footnote{See Hugging Face, \href{https://huggingface.co/blog/state-of-open-models-summer-2026}{``State of Open Models: Summer 2026 Observations.''} Its evidence concerns models published on the Hugging Face Hub, rather than application-market shares.} But publicly reusable resources can also reduce the work required to adopt a competing model later. For example, published configurations and adaptation tools can let the deployer reuse parts of its data-processing and evaluation pipeline when switching models. Supporters of open weights stress these benefits. On July 24, 2026, a coalition of model developers, cloud firms, and open-source organizations argued that open weights improve adaptability, strengthen competition, and help customers avoid lock-in and retain control over the knowledge and capabilities they build \citep{openweights2026letter}. Public resources, however, are not the only knowledge that shapes a deployer's choice of model.

Deployment generates a second source of useful knowledge, which we term the ``\textit{data flywheel effect}.'' When a deployer fine-tunes and operates an incumbent's model, usage and application-layer feedback, such as user responses, prompt adjustments, and contextual data, can help the deployer refine evaluation routines, integration choices, and application-specific adaptation of the same model family. Over time, the deployer's teams accumulate tacit expertise about the model's behavior, learning its strengths, failure modes, and most effective ways to adapt it for their product. This accumulated, model-specific know-how reduces the deployer's future cost of adapting the same model family. To the extent that such know-how is not fully portable across models, switching to an unfamiliar model requires additional recalibration, creating a form of learning-based lock-in.\footnote{Code assistants provide an illustrative setting: acceptance and rejection telemetry can help deployers refine model-specific evaluation and adaptation routines, making subsequent fine-tuning of the same model family easier while increasing the effort required to recalibrate those routines for an unfamiliar model.}

These two forms of knowledge create the central strategic tension. Publicly reusable resources help the deployer adapt a rival's model, whereas experience specific to the incumbent's model favors continuing with that model family. Greater openness can therefore support both future retention, by encouraging usage and learning, and future switching, by expanding what can be reused. The developer can also encourage current deployment through a lower license fee, making its commercial terms relevant to how much experience the deployer accumulates. To study these incentives, we develop a two-period model with an incumbent FM developer, an independent deployer, and an entrant developer whose model becomes available in period 2. The incumbent chooses its license fee and openness, anticipating the deployer's fine-tuning effort and subsequent model choice. This framework allows us to examine three research questions.

\textbf{First}, \textit{what determines an incumbent's choice of openness, and how does that choice shape downstream investment and welfare?} Developers' observed choices make this question consequential: openness varies across products and over time. Meta made Llama weights widely available through 2025, but introduced Muse Spark in 2026 through its own applications and an API without releasing downloadable weights; OpenAI retained proprietary flagship models while adding Apache-2.0 gpt-oss open-weight models to its portfolio.\footnote{\href{https://about.fb.com/news/2026/04/introducing-muse-spark-meta-superintelligence-labs/}{Meta, ``Introducing Muse Spark,'' April 8, 2026}; \href{https://openai.com/index/introducing-gpt-oss/}{OpenAI, ``Introducing gpt-oss,'' August 5, 2025}.} These product choices motivate treating openness as a strategic decision whose return depends on the developer's competitive position and prospects for future business.

\textbf{Second}, \textit{when does requiring greater public disclosure improve downstream adaptation and welfare once the developer can adjust its commercial terms?} The EU AI Act's general-purpose-model obligations and California's Transparency in Frontier Artificial Intelligence Act (SB~53) require specified documentation and disclosures, with different recipients and purposes.\footnote{The European Commission's \href{https://digital-strategy.ec.europa.eu/en/faqs/guidelines-obligations-general-purpose-ai-providers}{guidelines and questions and answers} distinguish documentation supplied to authorities and downstream firms from public training-content summaries. The \href{https://oag.ca.gov/sb53}{California Department of Justice's SB~53 guidance} describes publication of safety frameworks and other reporting obligations.} %
Our policy analysis focuses on publicly reusable information and resources that affect downstream adaptation and competition. We use a full-openness requirement as a limiting benchmark to examine how the incumbent's pricing response shapes the effects of public disclosure on adaptation and welfare.

\textbf{Third}, \textit{how do vertical integration and government subsidies reshape adaptation incentives, competition, and welfare in the AI value chain?} Both interventions can change the relationship between model development and application deployment. \textit{Vertical integration}, illustrated by Databricks' acquisition of MosaicML, brings model development and downstream application capabilities under common ownership.\footnote{See Databricks, \href{https://www.prnewswire.com/news-releases/databricks-completes-acquisition-of-mosaicml-301881059.html}{``Databricks Completes Acquisition of MosaicML,'' July 19, 2023}.} It can coordinate adaptation incentives, but committing deployment to the integrated developer can limit access to competing models. \textit{Government subsidies} instead seek to encourage AI development and use by reducing resource costs, as illustrated by the EU's AI Factories initiative.\footnote{See the \href{https://www.consilium.europa.eu/en/press/press-releases/2024/06/17/council-adopts-regulation-on-use-of-supercomputing-in-ai-development/}{Council of the European Union's June 17, 2024 announcement}. The initiative supports access to computing infrastructure; our per-unit model-use subsidy isolates the effect of lower deployment costs rather than reproducing that program's design.} Their effect on downstream investment depends on whether developers adjust their license fees and openness in response.

Our analysis yields three main findings. \textbf{First}, we identify conditions under which optimal openness falls and then rises as model-specific learning strengthens. When learning is weak, retaining the deployer would cost too much current revenue. The incumbent therefore chooses \textit{Harvest}: it offers full openness to expand current usage and monetizes that usage at a high license fee before the deployer switches to the entrant. As learning makes retention worthwhile, the incumbent switches to \textit{Defend}, keeping the high fee but restricting openness to limit how much its public disclosures lower the deployer's cost of adapting the entrant's model. Within Defend, stronger learning allows the incumbent to relax that restriction while retaining the deployer. With still stronger learning, the incumbent chooses \textit{Dominate}, pairing a low fee with greater, but still partial, openness. The lower fee encourages initial fine-tuning and usage, and the resulting experience supports retention despite greater public reuse. License revenue from the retained deployer in the second period justifies the lower current fee. Openness thus falls at the switch from Harvest to Defend, rises within Defend, and rises again when Dominate becomes optimal.

\textbf{Second}, a full-openness requirement can create an \textit{openness trap}, reducing welfare by changing the incumbent's pricing incentives. Under Dominate, the incumbent charges a low fee because current deployment builds experience that helps it retain the deployer. In the competitive setting we study, requiring full openness makes the entrant so attractive that retention is no longer feasible. The incumbent therefore switches to Harvest and raises its current fee. The higher fee can outweigh the direct benefit of greater openness and reduce first-period adaptation. The trap arises when the resulting first-period welfare loss exceeds the second-period gain from easier adaptation of the entrant's model. We characterize when this welfare-loss region exists; outside it, the requirement weakly improves welfare. An extension shows that when the deployer can reuse its accumulated experience after switching models, the incumbent finds it harder to retain the deployer, and the trap can disappear.

\textbf{Third}, we examine two interventions that change these incentives in different ways. Integration affects both the incentive to invest in adaptation and the deployer's choice of model. In our full-integration benchmark, the integrated firm eliminates the internal license fee and chooses full openness, raising initial adaptation and the experience it produces. The benchmark also commits deployment to the incumbent, potentially forgoing a rival that is cheaper to adapt. When model-specific learning is sufficiently strong, the gains from adaptation and accumulated experience outweigh this loss of choice, and integration raises the incumbent and deployer's joint profit, consumer surplus, and welfare relative to decentralization. With weaker learning, it can lower these outcomes.

A subsidy lowers the deployer's payment for model use at a given license fee, but the incumbent still receives the full posted fee and can adjust both its fee and its openness. The additional usage generated by the subsidy can therefore make high-fee strategies more attractive to the incumbent. A sufficiently small subsidy can move the incumbent from Defend to Harvest, raising both developers' profits, the deployer's profit, and consumer surplus. When the incumbent would choose Dominate without the subsidy, the subsidy can instead induce a switch to Defend. The incumbent then raises its fee by more than the subsidy and reduces openness, so the deployer pays more per use even after the subsidy and faces higher adaptation costs. Adaptation and welfare fall, even before deducting subsidy expenditure. Adoption support should therefore be assessed together with developers' pricing and openness responses.

These findings connect research on AI learning and strategic openness through the incumbent's incentive to support downstream adaptation. In the AI flywheel literature, usage data can improve algorithm performance and thereby shape contracting and pricing \citep{gurkan2022contracting}. Research on platform openness and FM regulation examines how access and disclosure affect downstream innovation \citep{parker2018innovation,qiu2025formal}. We study how experience that lowers the deployer's own adaptation cost for a model family changes the developer's joint choice of license fee and openness under future competition. The incumbent may charge a low initial license fee to encourage adaptation because the resulting experience helps it retain the deployer and earn future license revenue.
Disclosure mandates and adoption subsidies can each make this low-fee strategy less attractive, although they induce different subsequent model choices. The analysis links policy effects to the developer's joint pricing and openness decisions; the transferable-learning extension further identifies how the disclosure result depends on the portability of deployment experience.

The remainder of this paper is structured as follows. Section \ref{Literature Review} reviews the relevant literature. Section \ref{Model Setup} outlines the model setup. Section \ref{benchmark} presents the equilibrium analysis of the baseline model. Sections \ref{vi} and \ref{subsidy} analyze the effects of vertical integration and government subsidies, respectively. Section \ref{conclusion} discusses the theoretical and managerial implications and concludes. Appendix \ref{sec:extensions} examines five targeted generalizations involving transferable deployment experience, benefits of earlier openness when continuing with the incumbent, partial vertical integration, endogenous entry, and the strength of the public spillover.

\section{Literature Review}\label{Literature Review}

Our paper contributes to the growing literature on the economic implications of generative AI and foundation models, particularly large language models \citep[e.g.,][]{castro2023human,acemoglu2024simple}. Studies address content creation and knowledge contribution \citep{burtch2024consequences,borwankar2023unraveling,shan2025examining,chen2024large,zou2023welfare}, labor and productivity \citep{eloundou2023gpts,noy2023experimental,brynjolfsson2025generative}, organizational design \citep{ide2025artificial,xu2025generative}, marketing, preference measurement, and agentic product search \citep{brand2023using,goli2024frontiers,dong2026product}, automated bargaining \citep{liang2026negotiate}, platform monetization \citep{ji2026monetizing}, financial forecasting \citep{lopez2023can}, healthcare \citep{moor2023foundation}, and copyright \citep{yang2024generative,gans2024copyright}. Within this literature, we study the developer--deployer relationship, focusing on how openness and pricing shape downstream application investment and how deployment experience affects subsequent competition.

The closest mechanism comparison is with research on data-driven learning and ``AI flywheels.'' \citet{gurkan2022contracting} model a two-period relationship between a firm and an external machine-learning specialist in which usage data improve algorithm performance and affect pricing and contracting. Our model instead holds base-model quality fixed and lets deployment generate experience that lowers the deployer's future adaptation cost for a model family. Because this experience affects subsequent model choice, it changes the incumbent's incentive to support current adaptation through pricing and openness. Public disclosure also makes the competing model easier to adapt, so the incumbent weighs the retention benefit of deployment experience against the competitive effect of public reuse.

Formal models of FM markets also study openness and downstream adaptation. \citet{qiu2025formal} examine how a regulator's open-source definition and penalties affect a generalist's model release and a specialist's fine-tuning effort. \citet{liu2025generative} study application-layer product development with open-source LLMs, including fine-tuning uncertainty and limited collaboration. We examine how the prospect of retaining the deployer shapes the incumbent's pricing and openness decisions when current deployment builds experience and disclosure benefits a future competitor.

Research on technology adoption and AI use examines customer-service interactions, procurement outcomes, and healthcare technology adoption \citep{schanke2021estimating,cui2022ai,adida2025provider}. Related models study the elicitation of human judgment, learning about machine performance, and the allocation of cognitive effort \citep{ibrahim2021eliciting,de2023your,boyaci2024human}. Our focus is on application investment across organizational boundaries: a developer sets the terms for using and adapting a model, while a separate deployer invests in the user-facing application. This setting allows us to examine how openness, integration, and subsidies affect adaptation investment and the distribution of surplus among developers, the deployer, and end-users.

More broadly, research on technology adoption and IT investment examines how competition, adoption timing, and implementation risk shape investment \citep[e.g.,][]{erat2006introduction,milliou2011timing,tang2022sooner,choudhary2023sequential}. Foundation models are general-purpose inputs whose application-level value often depends on substantial post-adoption adaptation. We capture this feature by making the deployer's fine-tuning effort endogenous and allowing both model openness and accumulated experience to affect its cost. This investment links the incumbent's pricing and disclosure choices to downstream product quality, user engagement, and future switching frictions.

Our paper also relates to research on competition between proprietary and open-source software, including the roles of licensing, service provision, and network effects \citep[e.g.,][]{sen2007strategic,cheng2011impact,casadesus2011mixed,august2018generating,august2021competition}. %
\citet{zhu2012research} examine how switching costs affect pricing and profits when proprietary software competes with open-source software. In our model, the switching friction comes from the adaptation-cost savings a deployer obtains through operating experience with a model family. These savings are model-family-specific in the baseline and partly transferable in an extension, allowing us to examine how the portability of experience changes the incumbent's retention incentive.

Research on strategic platform openness examines how access and control affect innovation and competition \citep[e.g.,][]{parker2018innovation,chen2022new}. In our model, openness is public technical disclosure that lowers adaptation costs for the incumbent's model and can also benefit a competing model. We study how these benefits interact with model-family-specific experience to shape the developer's joint pricing and disclosure decisions and the welfare consequences of requiring greater openness.

\section{Model Setup}\label{Model Setup}
We study a two-period AI value chain ($t\in\{1,2\}$) in which an incumbent foundation model (FM) developer chooses its license fee and model openness before facing an entrant. Greater openness lowers the downstream deployer's adaptation cost and encourages usage, which builds experience that can favor continued use of the incumbent's model. The same disclosure also creates public knowledge that can lower the cost of adopting the entrant's model. The incumbent therefore chooses price and openness jointly, balancing current revenue against these competing effects on its future business.

\subsection{Players and Timing}
We study a vertically separated AI value chain in which a developer supplies a general-purpose model and an independent deployer adapts it for end-users, a structure documented by \citet{cma2024fmupdate}. For example, a deployer building an enterprise assistant can purchase model access while investing in domain-specific evaluation, workflow integration, and adaptation. The developer controls the license fee and the resources made available for adaptation, while the deployer decides how much to invest in the product users experience. This separation makes openness economically consequential: it changes the cost of creating downstream value and can also affect the deployer's subsequent dependence on the original developer. Appendix \ref{appen-openness-mapping} develops the developer-operated managed-access configuration that motivates this relationship.

We represent the value chain with two FM developers, one downstream \textit{deployer}, and a unit mass of end-users. We refer to a developer in its role of selling model access as the \textit{supplier}. The \textit{incumbent} (developer 1) is the sole model supplier in period 1 and chooses a per-unit license fee $w_1$ and openness level $\eta_1$. The deployer licenses the model and invests in fine-tuning its application, whose use generates revenue. An \textit{entrant} (developer 2) offers an alternative model at the beginning of period 2.

We take the availability of the entrant's model in period 2 as given, so the baseline asks how the incumbent competes when the deployer has another model to choose from. First-period price and openness still affect the deployer's subsequent model choice. Appendix \ref{appendix-Developer 2 Entry Cost} endogenizes participation by introducing an entry cost and a binding incumbent tariff posted before the entrant decides whether to enter.

Figure~\ref{f-b-sequence} summarizes the sequence of decisions. In period 1, the incumbent sets $(w_1,\eta_1)$, the deployer chooses fine-tuning effort $Q_1$, and users choose their engagement. In period 2, the developers offer license fees $(w_2,\tilde{w}_2)$ and openness levels $(\eta_2,\tilde{\eta}_2)$, with the entrant's fee fixed at the low level specified below. After observing these terms, the deployer selects a model and chooses its fine-tuning effort, after which users engage with the application.

\usetikzlibrary{fit,positioning}
\begin{figure}[ht]
\centering
\footnotesize
\begin{tikzpicture}[scale=0.9]
\linespread{1}
\draw [thick,->] (-2.5,0) -- (15,0);
\foreach \x in {-2.5,5.8, 15}{
    \draw (\x cm,5pt) -- (\x cm,-5pt);
}
\node[align=left,below] at (2,-0.1) {Period 1 - One FM Developer};
\node[align=left,below] at (11,-0.1) {Period 2 - Two FM Developers};

\node (p1dev1) [align=left,below] at (-0.5,2) {\textbf{Developer 1}:\\ (1) license fee $w_1$\\ (2) openness level $\eta_1$};
\node (p1dep)  [align=left,below] at (3.5,2)  {\textbf{Deployer} \& \textbf{User}:\\ (1) fine-tuning effort $Q_1$\\ (2) usage level $\alpha_1$};

\node (p2devs) [align=left,below] at (8.1,2.3)  {\textbf{Developer 1} \& \textbf{2}:\\ (1) license fee $w_2$, $\Tilde{w}_2$\\ (2) openness level $\eta_2$, $\Tilde{\eta}_2$};
\node (p2dep)  [align=left,below] at (12.5,2.3) {\textbf{Deployer} \& \textbf{User}:\\ (1) select developer\\(2) fine-tuning effort $Q_2$\\ (3) usage level $\alpha_2$};

\node[draw, dashed, rounded corners=5pt, inner sep=4pt, fit=(p1dev1) (p1dep)] {};
\node[draw, dashed, rounded corners=5pt, inner sep=4pt, fit=(p2devs) (p2dep)] {};

\end{tikzpicture}

\caption{Timeline of the Two-Period AI Value Chain Model}
\label{f-b-sequence}
\end{figure}

\subsection{Deployer Effort and User Engagement}
The deployer adapts a general-purpose FM into an application for end-users. We normalize the relationship between fine-tuning effort and application quality so that $Q_t$ denotes both the effort chosen in period $t$ and the resulting quality. An end-user chooses engagement $\alpha_t$ to maximize $U_t=Q_t\alpha_t-\alpha_t^2/2$ \citep{fainmesser2023digital}, yielding $\alpha_t^*(Q_t)=Q_t$. With a unit mass of users, the deployer earns revenue $\theta Q_t$, where $\theta$ is revenue per unit of engagement. This revenue can come from advertising, premium subscriptions, or enterprise licensing. After paying the per-unit license fee $w_t$, the deployer receives net revenue of $\theta-w_t$ per unit of engagement, before adaptation costs.

The model distinguishes upstream developer strategy from application-level operations. The incumbent chooses pricing and openness to attract deployment and sustain demand across deployment cycles. The deployer chooses fine-tuning effort for the application and model configuration used in the current cycle. We represent this adaptation decision as a project evaluated on its operating return within that cycle. Deployment also generates model-family-specific experience that lowers subsequent adaptation costs; this learning is a by-product of current operations, and its continuation value lies outside the current project's effort objective. Accordingly, the incumbent maximizes two-period profit, while the deployer chooses effort to maximize current-period operating profit.

Model openness $\eta$ summarizes how easily a professional deployer can inspect, adapt, and integrate a model. In period 1, the deployer's fine-tuning cost is
\begin{equation}
\label{eq:cost_period1}
C_1(Q_1) = \frac{cQ_1^2}{1+\eta_1}.
\end{equation}
Here, $c$ scales adaptation costs, and greater openness $\eta_1$ lowers the cost of achieving a given application quality. The openness index combines access to weights and architecture, documentation and tooling, modularity and interface stability, and licensing terms for adaptation. These features can reduce adaptation work by making the model easier to inspect and by supporting shared implementation resources, including those distributed through GitHub and Hugging Face \citep{kapoor2024societal, white2024model}.

Openness determines how easily the model can be adapted, while the fee $w$ is the price of using the developer's API or managed deployment service. We study a supplier that controls both model disclosure and the price of that service. Appendix \ref{appen-openness-mapping} maps the openness dimensions to observable artifacts and develops the managed-access reference case.

\subsection{Incumbent Pricing and Openness}
For tractability, in each period the incumbent chooses a per-unit license fee $w_t$ from a fixed menu containing a high fee $w_H$ and a low fee $w_L$, where $0\leq w_L\leq w_H\leq\theta/2$.\footnote{The exogenous price menu satisfies $w_H\leq\theta/2$. For a fixed openness level, current-period access revenue is proportional to $w(\theta-w)$, which increases up to $\theta/2$. We study the incumbent's intertemporal trade-off within this increasing-revenue range.} Its first-period openness satisfies $\eta_1\in[0,\overline{\eta}]$.

A lower fee increases the revenue the deployer retains per unit of engagement, while greater openness reduces its adaptation cost. Both can encourage fine-tuning and usage, thereby strengthening the model-specific learning that favors continued use of the incumbent. Openness also creates a public spillover that lowers the cost of adapting the entrant's model, as specified in Section \ref{section-new-entrant-competition}. The incumbent chooses $w_1$ and $\eta_1$ jointly because the resulting usage affects current access revenue and future learning, while disclosure also changes the deployer's alternative in period 2.

\subsection{Period-2 Competition and Adaptation Costs}\label{section-new-entrant-competition}
When the entrant becomes available in period 2, the deployer compares the returns from adapting each supplier's model. We model the entrant as a price-competitive follower whose license fee is fixed at $\tilde{w}_2=w_L$ and whose openness $\tilde{\eta}_2$ is chosen to maximize its period-2 profit. If the deployer adopts the entrant's model, its fine-tuning cost is
\begin{equation}
\label{eq:cost_entrant}
C_{2E}(Q_2) = \frac{cQ_2^2}{(1+\eta_1)(1+\tilde{\eta}_2)}.
\end{equation}
The factor $1+\eta_1$ captures public knowledge inherited from the incumbent's first-period disclosure, while $1+\tilde{\eta}_2$ captures the adaptability of the entrant's own model. Public architecture information, configurations, and adaptation tools can support compatibility with a later model, allowing the deployer to reuse parts of its training, data-processing, and evaluation pipeline when it switches. The cost saving accrues to the deployer because it performs this adaptation work.\footnote{Appendix \ref{appen-three k channels} develops the reduced-form interpretation and institutional pathway for this public spillover. The channel is distinct from the data flywheel $k$, which captures model-specific deployment know-how that does not transfer to the entrant in the baseline. Appendix \ref{appen-knowledge spillover on incumbent} allows first-period openness $\eta_1$ also to lower the deployer's continuation cost with the incumbent.}

Not every feature that eases adaptation of the incumbent's model creates the same opportunity for reuse with an entrant. The baseline links these effects through a single reduced-form index, with entrant-facing adaptation efficiency multiplied by $1+\eta_1$. Appendix \ref{appen-spillover-strength} varies the strength of this link.

Continuing with the incumbent offers a different potential cost advantage: experience accumulated through first-period use can make the deployer more efficient at adapting the same model family. The corresponding period-2 cost is
\begin{equation}
    C_{2I}(Q_2) = \frac{cQ_2^2}{(1+k\alpha_1)(1+\eta_2)}.
    \label{eq:cost_incumbent}
\end{equation}
The parameter $k\geq0$ measures the strength of the \textit{data flywheel}: first-period engagement $\alpha_1$ generates model-specific deployment know-how, represented by $1+k\alpha_1$. For $k>0$, more engagement lowers the cost of continuing with the incumbent. Holding openness fixed, this saving improves the incumbent's position relative to the entrant because the accumulated know-how does not transfer across model families in the baseline. We refer to this experience-based switching friction as learning-based lock-in; it arises without an explicit switching charge.\footnote{At a given engagement level, a larger $k$ means a greater model-specific cost saving. A code assistant whose evaluation routines and workflow integrations are closely tailored to one model family illustrates the kind of deployment in which this channel can matter; a standardized, model-agnostic workflow provides a contrasting case. The baseline isolates the model-family-specific component. Appendix \ref{appen-data flywheel on entrant} allows a weaker transferable component, $1+\hat{k}Q_1$, with $k\ge\hat{k}$, to lower the deployer's adaptation cost with the entrant.}

In period 2, greater openness reduces the cost of adapting a developer's model without creating competition in a subsequent period. Both developers therefore choose maximal openness, $\eta_2=\tilde{\eta}_2=\overline{\eta}$, in the equilibrium analyzed below. The cost functions thus distinguish the adaptability of the model currently in use, the public knowledge created by first-period disclosure, and the model-specific experience accumulated through first-period engagement. Appendix \ref{appen-three k channels} provides a detailed interpretation and institutional mapping of these channels.

\subsection{Interpretation and Scope}
The following assumptions define the setting in which we study the interaction of openness, pricing, and deployment experience.

\textbf{Deployer-Side Learning.} The data flywheel reduces the deployer's adaptation cost through application-specific know-how accumulated during use. It does not improve the upstream base model's intrinsic capability. Appendix \ref{appen-three k channels} provides the institutional and technical motivation for this distinction.

\textbf{Intrinsic Model Capability.} We abstract from endogenous R\&D and normalize intrinsic model capability in the baseline. The comparison therefore concerns settings in which capability differences are small enough for pricing, openness, public spillovers, and deployment experience to determine the deployer's model choice. Application quality remains endogenous through fine-tuning effort $Q_t$. Appendix \ref{appen-capacity normalization} explains the normalization and distinguishes intrinsic capability from the performance of the adapted application.

\textbf{A Concrete Reference Case: Developer-Operated Managed Access}: Alibaba's Qwen through Alibaba Cloud Model Studio illustrates the supplier configuration in our model. Alibaba releases adaptable Qwen models and sells managed deployment of its own models, placing disclosure and service pricing within the same supplier organization. IBM Granite through IBM watsonx.ai provides a second example of this configuration. Appendix \ref{appen-openness-mapping} develops the institutional mapping and specifies the pricing and incentive assumptions used to represent it.

\textbf{Irreversibility of Disclosure.} The non-recallable component of openness is represented by a reduced-form intertemporal constraint: after choosing $\eta_1\in[0,\bar{\eta}]$, the incumbent chooses $\eta_2\in[\eta_1,\bar{\eta}]$, while the entrant chooses $\tilde{\eta}_2\in[0,\bar{\eta}]$. Disclosed weights, architectural information, and public tooling cannot be fully withdrawn, even though contractual access terms may later change. As shown in Section \ref{benchmark}, the constraint is nonbinding in the baseline equilibrium because $\eta_2=\tilde{\eta}_2=\bar{\eta}$. It nevertheless makes first-period disclosure a credible commitment to maintain at least that level of openness.

\section{Equilibrium Analysis of Strategic Openness}\label{benchmark}
The incumbent chooses its first-period price and openness to earn current revenue and influence the deployer's subsequent model choice. We solve by backward induction: period-2 competition determines when the incumbent can retain the deployer, and this retention condition reduces its first-period problem to three candidate strategies. We then interpret these strategies and examine how a full-openness mandate changes the incumbent's choice and social welfare.

\subsection{Period 2: Competition for the Deployer}
Given the developers' period-2 offers and the inherited learning state, the deployer chooses the model that yields the higher current-period operating profit. If it continues with the incumbent, its objective is
$$\Pi_{2I}(Q_2)=(\theta-w_2)Q_2-\frac{cQ_2^2}{(1+k\alpha_1)(1+\eta_2)},$$
where user engagement equals application quality, $\alpha_2^*(Q_2)=Q_2$. The optimal effort and resulting profit are
\begin{align*}
Q_2^*&=\frac{(1+k\alpha_1)(1+\eta_2)(\theta-w_2)}{2c},&
\Pi_{2I}^*&=\frac{(1+k\alpha_1)(1+\eta_2)(\theta-w_2)^2}{4c}.
\end{align*}
If it switches to the entrant, its objective is
$$\Pi_{2E}(\tilde Q_2)=(\theta-\tilde w_2)\tilde Q_2-\frac{c\tilde Q_2^2}{(1+\eta_1)(1+\tilde\eta_2)},$$
with optimal effort and profit
\begin{align*}
\tilde Q_2^*&=\frac{(1+\eta_1)(1+\tilde\eta_2)(\theta-\tilde w_2)}{2c},&
\Pi_{2E}^*&=\frac{(1+\eta_1)(1+\tilde\eta_2)(\theta-\tilde w_2)^2}{4c}.
\end{align*}
The comparison depends on license fees and adaptation costs: prior deployment experience favors continued use of the incumbent, whereas public spillovers from its first-period openness favor adoption of the entrant.

In the terminal period, greater openness lowers the deployer's adaptation cost and expands usage if that model is selected, without creating a subsequent competitive disadvantage. The equilibrium considered below therefore has $\eta_2=\tilde\eta_2=\bar\eta$. The entrant's fee is fixed at $\tilde w_2=w_L$. We study competition within the maintained range $0\leq k<K_{\max}$, where
\[
K_{\max}:=\min\left\{\frac{2c\bar\eta}{(1+\bar\eta)(\theta-w_L)},\frac{2c(2\theta-w_H-w_L)(w_H-w_L)}{(\theta-w_H)^2(\theta-w_L)}\right\}.
\]
The first bound ensures that full first-period openness prevents retention even when the incumbent matches the entrant's low fee; the second ensures that the incumbent cannot retain the deployer at the high second-period fee $w_H$. We consequently consider the equilibrium with $w_2=w_L$. At these common period-2 terms, the deployer stays with the incumbent when $\Pi_{2I}^*\geq\Pi_{2E}^*$, with ties resolved in the incumbent's favor. Substituting $\alpha_1=Q_1=(1+\eta_1)(\theta-w_1)/(2c)$ gives
\begin{equation}
\frac{2c(1+\eta_1)}{2c+k(1+\eta_1)(\theta-w_1)}\leq1.
\end{equation}

Equivalently, retention requires $kQ_1\geq\eta_1$: the incumbent-specific learning benefit must at least match the public spillover that lowers the cost of adopting the entrant. A lower first-period fee increases $Q_1$ and strengthens learning without directly expanding that spillover. Greater first-period openness raises both $Q_1$ and the spillover. %

\subsection{Period 1: Price and Openness}
The incumbent anticipates this model-choice response when maximizing its two-period profit. For each first-period fee, retention places an upper bound on openness. The next lemma shows why this bound is the relevant candidate whenever the incumbent chooses to retain the deployer.

\begin{lemma}\label{b-l-etabar}
\begin{enumerate}[$(a)$]
\item \textbf{Retention thresholds.} Define $\bar\eta_H=\frac{k(\theta-w_H)}{2c-k(\theta-w_H)}$ and $\bar\eta_L=\frac{k(\theta-w_L)}{2c-k(\theta-w_L)}$.
If the first-period fee is $w_H$ (respectively, $w_L$), the incumbent retains the deployer in period 2 if and only if $\eta_1\leq\bar\eta_H$ (respectively, $\eta_1\leq\bar\eta_L$).
The thresholds satisfy $\bar\eta_H\leq\bar\eta_L$, with strict inequality when $k>0$ and $w_H>w_L$.

\item \textbf{Profit within an outcome.} For a fixed first-period fee and a fixed period-2 model choice, the incumbent's two-period profit is weakly increasing in $\eta_1$.
\end{enumerate}
\end{lemma}

At a fixed fee, openness increases current usage and raises current access revenue. If the incumbent retains the deployer, its profit is $\pi_{win}(w_1,\eta_1)=w_1\alpha_1+w_L\alpha_2$; the additional first-period usage can also build a learning advantage that supports second-period usage and access revenue. If the deployer switches, the incumbent earns only $\pi_{lose}(w_1,\eta_1)=w_1\alpha_1$. Thus the cost of greater openness comes from changing the period-2 model choice, rather than from a negative effect on revenue within either outcome.

This distinction reduces the incumbent's choice to the highest openness compatible with its intended outcome. To retain the deployer at a first-period fee $w_H$ or $w_L$, it can choose $\bar\eta_H$ or $\bar\eta_L$, respectively. To concede the deployer, it can choose full openness $\bar\eta$. In the latter case, $w_H$ maximizes current revenue over the price menu because $w(\theta-w)$ is increasing for $w\leq\theta/2$. The incumbent therefore needs to compare only three candidate price--openness pairs.

\begin{figure}[ht]
	\centering
    \begin{subfigure}[b]{0.55\textwidth}
    \centering
        \includegraphics[width=\textwidth]{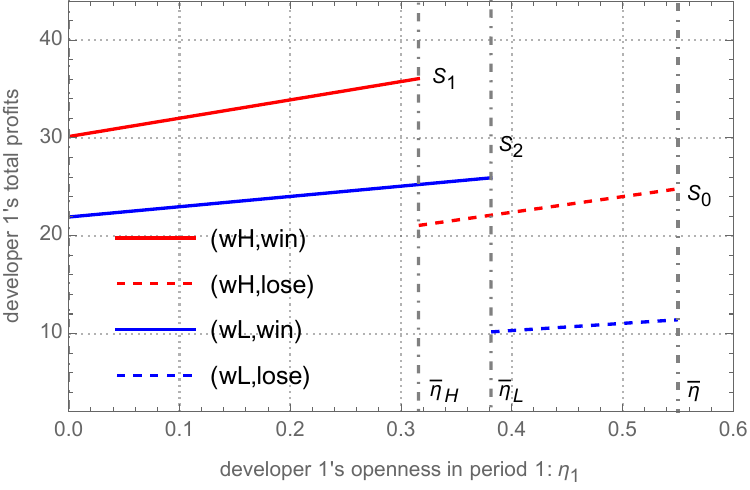}
    \end{subfigure}~~~
   	\caption{Deriving the Incumbent's Optimal Strategy: Profit Trade-Offs with Respect to Openness\\ ($\theta=10$, $c=0.5$, $k=0.03$, $w_H=2$, $w_L=0.8$, $\bar{\eta}=0.55$)}\label{b-f-etahbar and etalbar}
\end{figure}

Figure \ref{b-f-etahbar and etalbar} plots profit along the retention and switching branches for each fee. The solid retention branches are feasible up to and including $\bar\eta_H$ or $\bar\eta_L$; the dashed branches apply when the incumbent loses the deployer. Their relevant endpoints are $S_1=(w_H,\bar\eta_H)$, $S_2=(w_L,\bar\eta_L)$, and $S_0=(w_H,\bar\eta)$. The comparison is between the profits at these three points, rather than between openness levels alone. We refer to these candidates as follows:
\begin{enumerate}[(1)]

\item \textbf{Harvest:} The incumbent chooses $(w_1,\eta_1)=(w_H,\bar\eta)$ to maximize current revenue and concedes the period-2 deployer ($S_0$). Full openness supports current adaptation, while the incumbent forgoes the future revenue that retaining the deployer would generate.
\item \textbf{Defend:} The incumbent chooses $(w_1,\eta_1)=(w_H,\bar\eta_H)$ to retain the deployer while preserving its high first-period fee ($S_1$). It limits public spillovers by restricting openness, accepting lower current usage as the cost of retention.
\item \textbf{Dominate:} The incumbent chooses $(w_1,\eta_1)=(w_L,\bar\eta_L)$ to retain the deployer with a lower first-period fee and a higher retention-compatible openness level ($S_2$). The lower fee encourages usage and model-specific learning, permitting more openness without losing the deployer.

\end{enumerate}

As $k$ rises, a given amount of prior deployment creates a larger continuation cost advantage, allowing the incumbent to offer more openness while retaining the deployer. The next proposition determines which of these strategies maximizes the incumbent's profit. %

\begin{proposition}
\label{b-p-equilibrium-k}
For $0\leq k<K_{\max}$, there exist thresholds $\bar{k}_1$ and $\bar{k}_2$ such that the equilibrium outcomes are as follows.
\begin{enumerate}[$(a)$]
    \item If $k \leq \bar{k}_1$, the incumbent adopts a \textbf{Harvest} strategy. The deployer selects the entrant in period 2, and the equilibrium outcomes are:
    \begin{equation*}
    (w_1^*,\eta_1^*,Q_1^*,Q_2^*) = \left(w_H,\bar{\eta},\frac{(1+\bar{\eta})(\theta-w_H)}{2c},\frac{(1+\bar{\eta})^2(\theta-w_L)}{2c}\right).
    \end{equation*}

    \item If $\bar{k}_1 < k \leq \bar{k}_2$, the incumbent adopts a \textbf{Defend} strategy. The deployer selects the incumbent in period 2, and the equilibrium outcomes are:
    \begin{equation*}
    (w_1^*,\eta_1^*,Q_1^*,Q_2^*) = \left(w_H,\bar{\eta}_H,\frac{\theta-w_H}{2c-k(\theta-w_H)},\frac{(1+\bar{\eta})(\theta-w_L)}{2c-k(\theta-w_H)}\right).
    \end{equation*}

    \item If $k>\bar{k}_2$, the incumbent adopts a \textbf{Dominate} strategy. The deployer selects the incumbent in period 2, and the equilibrium outcomes are:
    \begin{equation*}
    (w_1^*,\eta_1^*,Q_1^*,Q_2^*) = \left(w_L,\bar{\eta}_L,\frac{\theta-w_L}{2c-k(\theta-w_L)},\frac{(1+\bar{\eta})(\theta-w_L)}{2c-k(\theta-w_L)}\right).
    \end{equation*}
\end{enumerate}
Some regions may be empty within the maintained range; the threshold formulas are given in the proof. All three regions have positive length if and only if $0<\bar{k}_1<\bar{k}_2<K_{\max}$.
\end{proposition}

\begin{figure}[ht]
	\centering \footnotesize
    \begin{subfigure}[b]{0.464\textwidth}
    \centering
        \includegraphics[width=\textwidth]{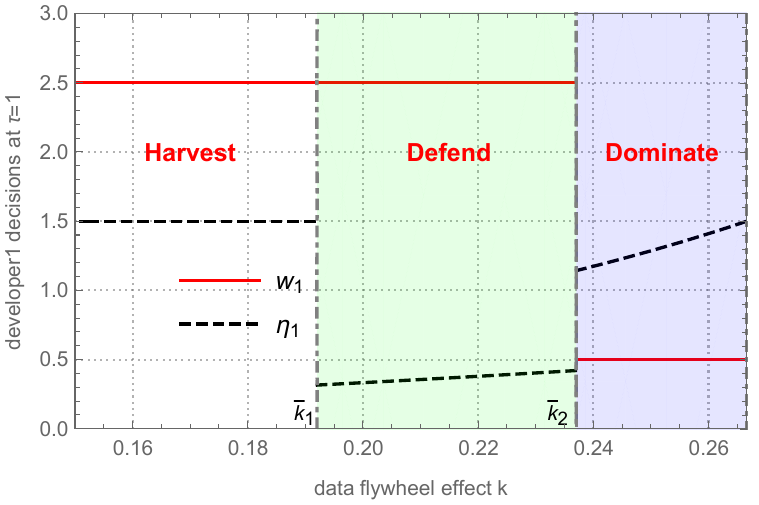}
       \caption{developer 1: $w_1$ and $\eta_1$}\label{b-f-w1 and eta1}
    \end{subfigure}
    ~~~
    \begin{subfigure}[b]{0.458\textwidth}
    \centering
        \includegraphics[width=\textwidth]{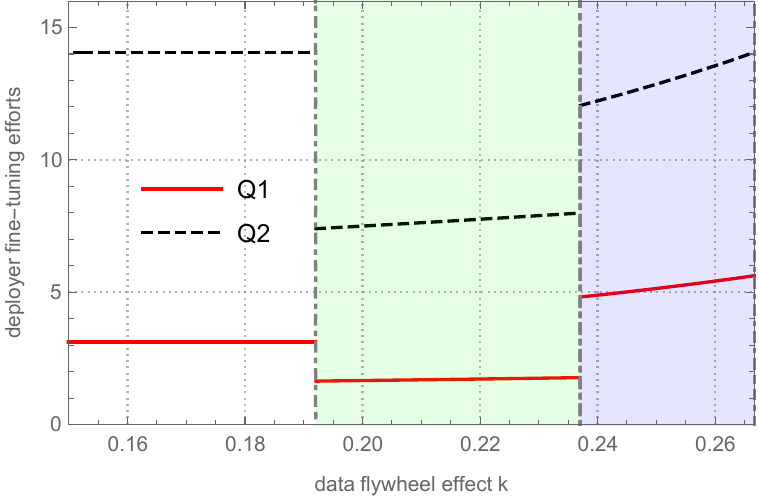}
       \caption{deployer: $Q_1$ and $Q_2$}\label{b-f-q1 and q2}
    \end{subfigure}
    	\caption{Equilibrium Strategy and Outcomes as a Function of the Data Flywheel Effect $k$\\ ($\theta=5$, $c=1$, $w_H=2.5$, $w_L=0.5$, $\bar{\eta}=1.5$)}\label{b-f-equilibrium}
\end{figure}

Proposition \ref{b-p-equilibrium-k} explains why stronger learning need not lead to greater openness. Figure \ref{b-f-equilibrium} illustrates a case in which stronger learning leads the incumbent from Harvest to Defend and then to Dominate. At low $k$, retention would require giving up too much current adaptation and revenue, and Harvest is optimal. Just above $\bar k_1$, future access revenue makes retention worthwhile: the incumbent switches to Defend and reduces openness from $\bar\eta$ to $\bar\eta_H$ while keeping the high fee. This discrete restriction of openness lowers first-period fine-tuning, as shown in Figure \ref{b-f-q1 and q2}.

Within Defend, a higher $k$ relaxes the retention constraint, so the incumbent gradually increases $\bar\eta_H$. Once $k$ exceeds $\bar k_2$, the additional usage and learning induced by a low fee justify giving up revenue per unit of first-period use. The incumbent switches to Dominate, lowering its fee to $w_L$ and raising openness to $\bar\eta_L$. Both changes encourage initial fine-tuning; the resulting experience also supports greater period-2 effort. Thus, in the illustrated three-region configuration, openness falls at the Harvest-to-Defend switch, rises within Defend, and jumps upward at the switch to Dominate. The managerial implication is that price and openness must be evaluated jointly: stronger model-specific learning can first make restricting openness worthwhile and later support a lower-price, more open strategy. Figure \ref{b-f-each party profit} in Appendix \ref{appen-baseline-welfare} reports the corresponding payoffs and welfare.

\subsection{Interpreting the Equilibrium Strategies}\label{regimes-practice}
Harvest, Defend, and Dominate describe joint price--openness choices and their consequences for one deployer relationship.
Table \ref{tab:regime-proxies} summarizes the fee, openness level, and second-period model choice under each strategy. Harvest concedes the deployer to an available alternative; it does not imply that the incumbent exits the market. Defend and Dominate both retain the deployer, but use different combinations of price and openness; neither label establishes market-wide dominance. Because the baseline takes the entrant's availability as given, these outcomes describe competition for deployment conditional on that availability. Appendix \ref{appendix-Developer 2 Entry Cost} shows how an endogenous participation decision can change the strategy map.

\begin{table}[ht]
\centering \small
\caption{Price--Openness Strategies and Deployer Continuation}\label{tab:regime-proxies}
\setlength{\arrayrulewidth}{0.2mm}
\renewcommand{\arraystretch}{1.4}
\begin{tabular}{ C{3.3cm} | C{3.1cm} | C{5cm} | C{3.7cm} }
\hline \hline
\cellcolor{black!4} Strategy & \cellcolor{black!4} First-period fee & \cellcolor{black!4} First-period openness & \cellcolor{black!4} Period-2 model choice \\
\hline \hline
\textit{Harvest} ($k\leq\bar k_1$) & $w_H$ & Full: $\bar\eta$ & Entrant \\
\hline
\textit{Defend} ($\bar k_1<k\leq\bar k_2$) & $w_H$ & Retention limit: $\bar\eta_H$ & Incumbent \\
\hline
\textit{Dominate} ($k>\bar k_2$) & $w_L$ & Retention limit: $\bar\eta_L$; higher than $\bar\eta_H$ for $k>0$ and $w_H>w_L$ & Incumbent \\
\hline \hline
\end{tabular}
\end{table}

These predictions direct attention to price, adaptation resources, and continuation of the developer--deployer relationship. License fees and disclosure records help characterize the first two choices. The Foundation Model Transparency Index provides systematic evidence on disclosed information \citep{bommasani20242024}, but its score does not directly measure the adaptation-cost reduction represented by $\eta$. That mapping requires evidence on which disclosed resources reduce the work of adapting the model and which also facilitate adoption of an alternative.

Application counts, fine-tuned derivatives, dependencies on model families, and usage-share trajectories can provide complementary evidence about ecosystem breadth and customer persistence. For example, a lower license fee accompanied by more useful adaptation resources and persistent deployment relationships would motivate examining the Dominate mechanism. Such observations do not identify $k$ or establish an equilibrium regime: capability differences, complementary products, and marketing can produce similar patterns. The model does not predict a growing or shrinking market-wide ecosystem in any of its three regimes.

The learning mechanism suggests a more direct measurement strategy. Application teams assigned different amounts of prior incumbent use can be compared on fresh adaptation tasks with a common quality target. The cost reduction on incumbent tasks measures the benefit of continued use; the corresponding reduction on entrant tasks measures how much experience transfers. Appendix \ref{appen-three k channels} develops the implied cost ratios, identifying conditions, and additional measurements needed to compare these estimates with the policy thresholds.

\subsection{Openness Requirements and the Openness Trap}\label{Openness Trap}
Policy debates weigh the potential benefits of open foundation models for competition and innovation against their societal risks \citep{bommasani2024considerations}. We examine one economic channel in this debate: how an openness requirement changes the incumbent's incentives when setting the license fee. The effect depends on what must be disclosed and to whom. For example, the EU AI Act distinguishes technical documentation supplied to authorities or downstream firms from a publicly available summary of training content.\footnote{See the \href{https://digital-strategy.ec.europa.eu/en/faqs/general-purpose-ai-models-ai-act-questions-answers}{European Commission's questions and answers on general-purpose AI models}.} These duties need not make model resources publicly reusable to the same extent. We study an idealized full-openness mandate that sets $\eta_1=\bar\eta$ and expands the publicly reusable adaptation resources represented by the model's openness index. The experiment asks whether the incumbent's price response can offset the adaptation benefit, identifying a competitive effect to consider alongside safety, accountability, and compliance objectives.

Under the baseline assumptions, full openness prevents the incumbent from retaining the deployer in period 2. The incumbent therefore charges the high first-period fee $w_H$ to maximize current revenue, implementing Harvest. The mandate leaves an unregulated Harvest outcome unchanged and relaxes the openness restriction under Defend without changing the fee sequence. Under Dominate, it both increases openness and raises the first-period fee from $w_L$ to $w_H$, creating opposing effects on initial fine-tuning.

The welfare comparison asks whether the adaptation gains outweigh the consequences of that price response.\footnote{Social welfare consists of developer 1's profit, developer 2's profit, the deployer's profit, and consumer surplus.} Let $W_H$ denote Harvest welfare and $W_D(k)$ the welfare of the Dominate candidate. Define $k_W$ by $W_D(k_W)=W_H$, with the formula given in the proof. This welfare crossing is distinct from the threshold at which the incumbent selects Dominate, and it need not lie in the admissible Dominate region.

\begin{proposition}\label{b-p-welfare}
Within $0\leq k<K_{\max}$, mandating full openness leaves total social welfare unchanged in Harvest and strictly increases it in Defend. In Dominate, it lowers welfare if and only if $k>k_W$. Equivalently, under the strategy-selection convention used in Proposition~\ref{b-p-equilibrium-k}, the mandate lowers welfare if and only if $\bar{k}<k<K_{\max}$, where $\bar{k}:=\max\{\bar{k}_2,k_W\}$. This welfare-loss region is nonempty if and only if $\max\{0,\bar{k}_2,k_W\}<K_{\max}$; otherwise, the mandate weakly increases welfare throughout the admissible range.
\end{proposition}

\begin{figure}[ht]
	\centering \footnotesize
    \begin{subfigure}[b]{0.46\textwidth}
    \centering
        \includegraphics[width=\textwidth]{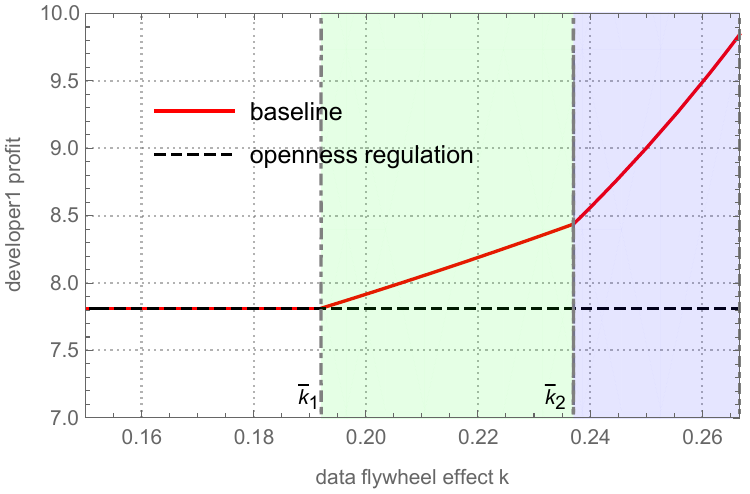}
       \caption{developer 1 profit}\label{b-f-regulate-developer1}
    \end{subfigure}
    ~~~
    \begin{subfigure}[b]{0.45\textwidth}
    \centering
        \includegraphics[width=\textwidth]{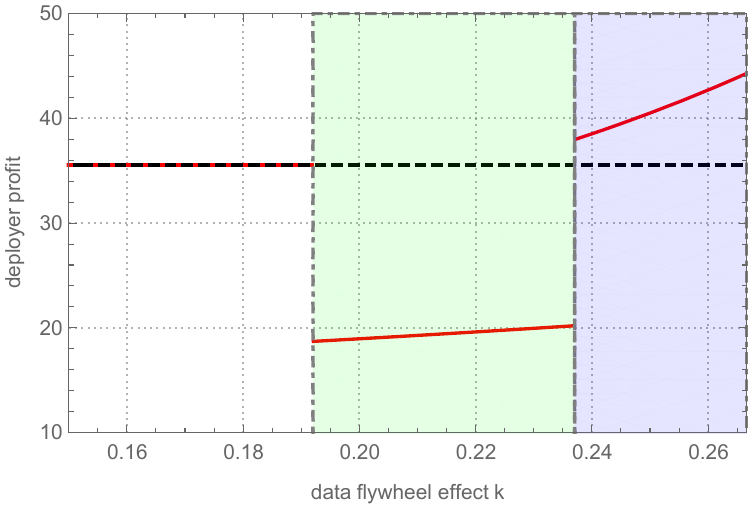}
       \caption{deployer profit}\label{b-f-regulate-deployer}
    \end{subfigure}
    \begin{subfigure}[b]{0.465\textwidth}
    \centering
        \includegraphics[width=\textwidth]{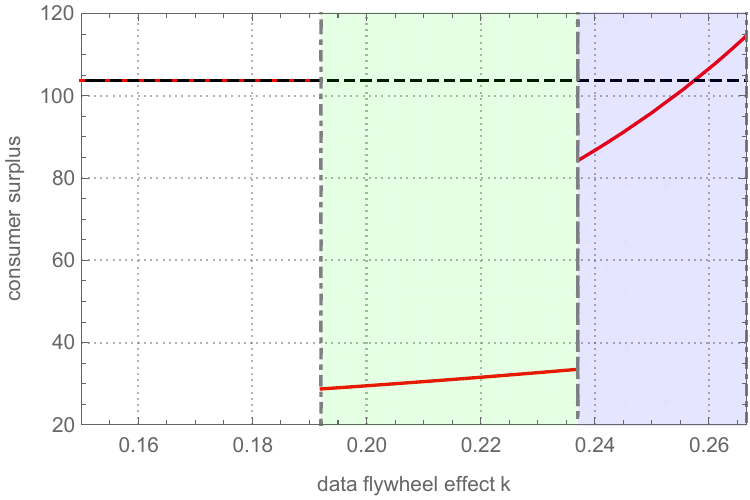}
       \caption{consumer surplus}
    \end{subfigure}
    ~~~
    \begin{subfigure}[b]{0.465\textwidth}
    \centering
        \includegraphics[width=\textwidth]{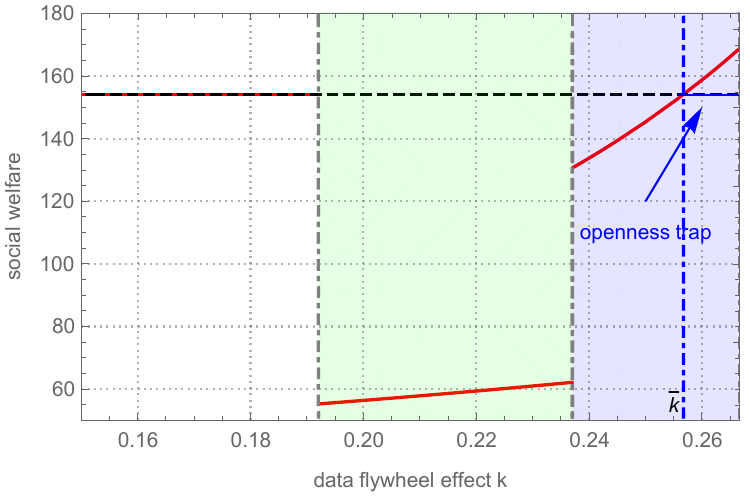}
       \caption{social welfare}
    \end{subfigure}
    	\caption{Impacts of Openness Regulation on Each Party's Profit and Social Welfare\\ ($\theta=5$, $c=1$, $w_H=2.5$, $w_L=0.5$, $\bar{\eta}=1.5$)}\label{b-f-open trap}
\end{figure}

The source of the welfare difference is clearest when the two periods are considered separately. Starting from Defend, the mandate raises first-period effort through greater openness at the same fee. At a retention boundary, $kQ_1=\eta_1<\bar\eta$, so full public disclosure also makes the entrant less costly to adapt in period 2 than the unregulated incumbent and raises period-2 effort. Both changes raise welfare at the common fee sequence. Starting from Dominate, the same period-2 adaptation gain is present, but the higher first-period fee can reduce initial effort despite greater openness. An openness trap occurs only when the first-period welfare loss exceeds the period-2 gain. Proposition \ref{b-p-welfare} identifies this comparison precisely through $k>k_W$, together with the requirement that Dominate is the unregulated choice. Figure \ref{b-f-open trap} illustrates a nonempty loss region and the corresponding effects on developer profit, deployer profit, and consumer surplus.

\textit{Other possible firm responses.} The baseline identifies a price response to an openness requirement when model availability is fixed. Innovation, differentiation, or complementary revenues could change both that response and its welfare consequences; the analysis does not order welfare effects across richer choice sets. When entry is endogenous, the fee response can still lower welfare at low entry costs. At the calibration in Figure \ref{b-f-equilibrium}, however, a mandate that ends exclusion at intermediate entry costs raises welfare net of the entry cost (Appendix \ref{appendix-Developer 2 Entry Cost}).

\textit{Recipients of disclosure.} Confidential access by a regulator need not generate the same entrant-facing spillover as public release. Evaluating it as an alternative requires comparing oversight needs, leakage, compliance costs, and safety benefits, as well as the competitive effect studied here \citep{guha2023ai}. The model identifies why the audience for disclosure matters, but does not establish that confidential disclosure prevents the openness trap or yields higher total welfare.

The practical implication is to assess an openness requirement together with the license-fee response it induces and the adaptation resources it makes reusable. Appendix \ref{appen-spillover-strength} varies the entrant-facing benefit of openness separately from its benefit for adapting the incumbent's model. At the illustrative parameters and within the competitive region examined there, weakening public spillovers eliminates the welfare-loss region before it eliminates openness restriction. The policy concern therefore depends on how strongly the particular disclosure requirement facilitates adoption of an alternative relative to the incumbent-specific learning advantage. Appendix~\ref{appen-data flywheel on entrant} adds a related qualification: when part of the deployer's experience also lowers its cost of adapting the entrant's model, the incumbent finds the deployer harder to retain, and the welfare-loss region can disappear.

An initial assessment would examine whether experience primarily lowers adaptation costs within one model family, whether the released resources reduce the work of adopting another model, and whether low license fees help the incumbent build future demand. The price--openness combinations in Table \ref{tab:regime-proxies} and the accompanying discussion of ecosystem evidence organize this inquiry. The focus on model choice also connects to the Competition and Markets Authority's emphasis on switching frictions and interoperability \citep{cma2024fmupdate}. Adaptation costs matter through their effects on model choice and deployment, even when they are not the largest expenditure category.

A quantitative assessment would compare deployment and welfare with and without the mandate, allowing fees and effort to adjust. It would require estimates of adaptation costs, revenues, and the benefits of the proposed disclosure on a common scale, with uncertainty in those estimates taken into account. A full policy evaluation would also include compliance costs and safety benefits.

\section{The Welfare Effects of Vertical Integration}\label{vi}
AI firms combine model development with downstream products and deployment services, making the organization of model access an economically relevant choice. For example, Databricks completed its acquisition of MosaicML in July 2023, bringing model-training and customization capabilities into its data and AI platform with the stated aim of supporting enterprise AI applications.\footnote{See Databricks' acquisition announcement, \url{https://www.databricks.com/blog/databricks-mosaicml}. The example illustrates coordination across stages of the AI value chain; it does not establish that the acquisition foreclosed competing models.} Such combinations raise a concrete question: when do the gains from coordinating model access outweigh the costs of limiting the deployer's alternatives?

In our integration benchmark, removing the internal license fee increases the return to adaptation, while requiring deployment to use the incumbent's model prevents switching to an entrant that may be less costly to adapt. The first change raises current effort and the model-specific experience carried into the next period; the second can sacrifice the adaptation gains available from switching models. We study this trade-off by integrating the incumbent developer and the deployer into a single firm, denoted by subscript $v$, and comparing its outcomes with the decentralized equilibrium. The main benchmark ties all deployment to the incumbent; Appendix \ref{appendix-Partial Vertical Integration} allows a share of operations to use the entrant.

\subsection{Equilibrium under Vertical Integration}
The integrated benchmark removes the internal license fee and requires the deployer to use the incumbent's model in both periods. Developer 2 is therefore excluded from this deployment relationship in period 2. We retain this project-level operating mandate under integration, allowing ownership to change internal license fees and access to competing models while holding the organization of adaptation decisions fixed. The operating unit chooses $Q_{tv}$ to maximize current-period profit, taking openness and inherited learning as given. The owner chooses openness to maximize two-period profit, anticipating the operating unit's decisions. Integration eliminates the internal fee, raising net revenue per unit of engagement from $\theta-w_t$ to $\theta$ while preserving the operating unit's decision horizon. Because the operating unit cannot switch to the entrant in this benchmark, the owner benefits from the adaptation-cost reduction due to openness without losing deployment to a competing supplier. It therefore chooses $\eta_{1v}=\eta_{2v}=\bar{\eta}$ under the maintained operating mandate.

Proposition \ref{v-p-equilibrium} in Appendix \ref{appen-vertical integration} gives the resulting efforts:
\[
Q_{1v}^*=\frac{(1+\bar{\eta})\theta}{2c},\qquad
Q_{2v}^*=\frac{(1+\bar{\eta})(1+kQ_{1v}^*)\theta}{2c}.
\]
The first expression shows how removing the internal license fee and choosing full openness raise initial effort relative to decentralization. The second shows how that effort lowers subsequent adaptation costs through accumulated experience. If the decentralized deployer would remain with the incumbent, integration removes the license fee and increases the experience accumulated on the same model family. If it would switch to the entrant, the comparison also depends on the entrant's adaptation advantage from public spillovers. Integration can then reduce period-2 effort despite raising effort in period 1.

\subsection{Welfare Analysis}
We compare three outcomes with their decentralized counterparts. The profit comparison is between the integrated firm's profit and the combined profit of the incumbent and deployer before integration; it excludes the entrant's profit. Consumer surplus measures the effect on users, while social welfare includes the profits of both developers and the deployer as well as consumer surplus. Thus a gain in the incumbent and deployer's joint profit need not be a gain for every firm or for society. The next proposition states the comparisons as the strength of model-specific learning varies.

\begin{proposition}\label{v-p-profit comparison}
There exist thresholds $\bar{k}_{dv}$, $\bar{k}_{cv}$, and $\bar{k}_{sv}$ in $[0,K_{\max}]$ such that:
\begin{enumerate}[$(a)$]
    \item \textbf{Outcome comparisons.} Integration increases joint profit if $k>\bar{k}_{dv}$ and reduces it if $k<\bar{k}_{dv}$. The same holds for consumer surplus with $\bar{k}_{cv}$ and for social welfare with $\bar{k}_{sv}$.
    \item \textbf{Joint gains.} If $k>\max\{\bar{k}_{dv},\bar{k}_{cv}\}$, integration yields a win-win outcome for joint profit and consumer surplus.
    \item \textbf{Joint losses.} If $k<\min\{\bar{k}_{dv},\bar{k}_{cv}\}$, integration yields a lose-lose outcome for joint profit and consumer surplus.
\end{enumerate}
If the Harvest region is nonempty, no threshold exceeds its upper endpoint; otherwise, all three thresholds equal zero.
Equality holds at a crossing within Harvest, but a cutoff reached through a strategy change need not be an indifference point for the compared payoff measures.
\end{proposition}

The welfare trade-off originates in period 2. Integration raises first-period effort through the removal of the internal license fee and full openness. When the decentralized outcome is Harvest, however, the deployer would use the entrant's model in period 2 and benefit from the incumbent's public disclosures. Tying deployment to the incumbent can forgo enough of this adaptation advantage to offset the first-period gain. Within Harvest, decentralized outcomes do not vary with $k$, whereas integration benefits from stronger learning on the incumbent's model. When the decentralized outcome instead retains the incumbent, integration eliminates the license fee and increases accumulated deployment experience without sacrificing a rival model that would otherwise have been used.

\begin{figure}[ht]
	\centering \footnotesize
    \begin{subfigure}[b]{0.45\textwidth}
    \centering
        \includegraphics[width=\textwidth]{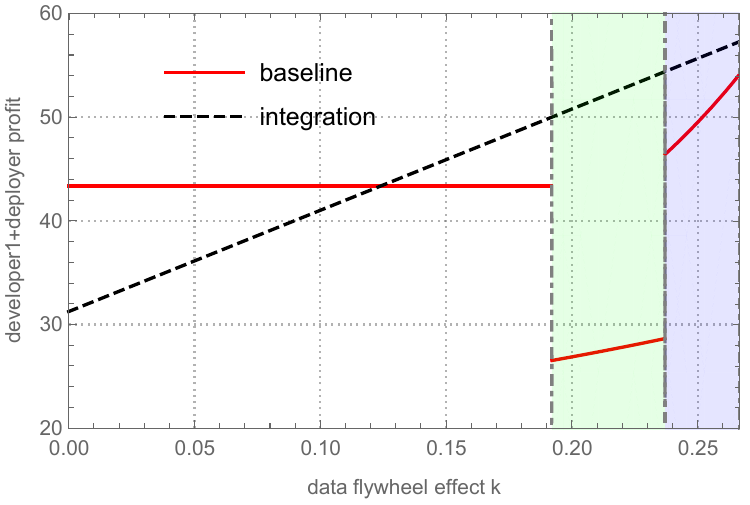}
       \caption{Incumbent and deployer joint profit}\label{v-f-developer1 and deployer}
    \end{subfigure}
    ~~~
    \begin{subfigure}[b]{0.45\textwidth}
    \centering
        \includegraphics[width=\textwidth]{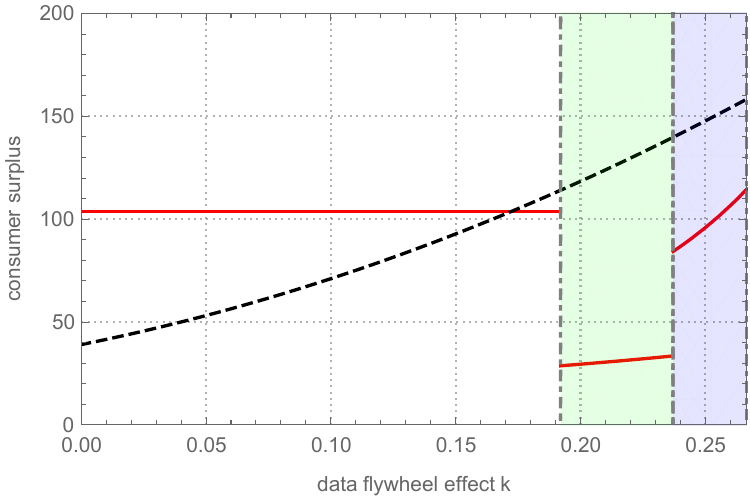}
       \caption{consumer surplus}\label{v-f-consumer surplus}
    \end{subfigure}
\caption{Joint Profit and Consumer Surplus under Decentralization and Integration\\ ($\theta=5$, $c=1$, $w_H=2.5$, $w_L=0.5$, $\bar{\eta}=1.5$)}\label{v-f-profit comparison}
\end{figure}

Figure \ref{v-f-profit comparison} illustrates the resulting joint-profit and consumer-surplus comparisons at the stated calibration. In each panel, the integrated outcome rises with $k$ and crosses the Harvest outcome before the decentralized strategy changes. The two crossings need not coincide because joint profit and consumer surplus value changes in effort differently. Neither panel by itself establishes the social-welfare comparison, which must also include the entrant's lost profit when the decentralized deployer switches models.

\subsection{Implications for Evaluating Integration}
The comparison directs attention to the license fees removed by integration, the rival models that remain available, and the model-specific experience that the deployer would lose by switching. These are distinct features of an arrangement: a commercial partnership may alter payments or favor one model without tying every operation to it. The FTC's 2025 study, which examines Microsoft's partnership with OpenAI alongside other AI partnerships, documents variation in equity, revenue-sharing, control, and exclusivity rights.\footnote{See the FTC's summary of its study, \url{https://www.ftc.gov/news-events/news/press-releases/2025/01/ftc-issues-staff-report-ai-partnerships-investments-study}. These contractual features motivate distinguishing arrangements; they do not establish that any particular partnership implements the benchmark studied here.} Applying our comparison therefore requires evidence on effective license payments and on adaptation costs at a common application-quality target, including the cost of using a rival after experience with the incumbent.

Appendix \ref{appendix-Partial Vertical Integration} shows why the extent of rival access matters. When an integrated firm can use the entrant for some operations, it does so when the entrant offers a higher operating return. Switching also raises effort and consumer surplus on those operations, increasing welfare relative to full foreclosure. Relative to decentralization, this improvement reduces any welfare loss from integration and enlarges any gain. Stronger model-specific learning can nevertheless make the incumbent preferable for every operation. Welfare can fall at this transition because the firm does not internalize the resulting losses to consumers and the entrant. Partial integration therefore need not preserve the single welfare cutoff of the full-integration benchmark. Evaluating an arrangement requires distinguishing access to rival models from their actual use. Our comparison holds the deployer's operating objective fixed, so these results alone do not determine the welfare effect of a particular merger or access remedy.

\section{The Impact of Government Subsidies for AI Adoption}\label{subsidy}
Public support for AI takes several forms. The European Union's AI Factories initiative provides access to supercomputing infrastructure and related services, while Japan's 2024 support program subsidized projects expanding the computing resources available for AI development.\footnote{See the EU Council's description of AI Factories, \url{https://www.consilium.europa.eu/en/press/press-releases/2024/06/17/council-adopts-regulation-on-use-of-supercomputing-in-ai-development/}, and Japan's Ministry of Economy, Trade and Industry announcement, \url{https://www.meti.go.jp/english/press/2024/0419_001.html}. These are computing-access and infrastructure programs, rather than the per-unit model-access subsidy analyzed below.} These initiatives motivate a broader incidence question: how much of the benefit reaches downstream deployment once model developers can adjust their commercial terms?

We study a per-unit subsidy, interpreted as a reimbursement tied to the deployer's billed model usage. Because the license fee is also charged per unit, this subsidy directly lowers the deployer's cost of using the model and raises the return to fine-tuning at fixed license fees and openness. The incumbent can respond by changing both the license fee and the model's adaptability, potentially offsetting that direct benefit. We compare the resulting equilibrium with the unsubsidized benchmark to determine who gains, who loses, and how deployment changes.

\subsection{Equilibrium Analysis with Government Subsidy}
Let $s$ be the subsidy per unit of model usage in each period $t\in\{1,2\}$, and use subscript $g$ for subsidized outcomes. At a posted license fee $w$, the deployer pays $w-s$, the government contributes $s$, and the selected developer receives $w$. At fixed terms, the deployer retains more revenue from each unit of usage, making additional fine-tuning more profitable. The developer can respond through price and openness. We compare equilibria within their common maintained competition range and restrict attention to $0<s<\bar{s}$, where the proof of Proposition \ref{f-p-welfare comparison} defines a sufficient cutoff with $\bar{s}\leq w_L$. The cutoff depends on the primitives and can equal zero; the result is conditional on a nonempty subsidy interval.\footnote{The subsidy illustrations use $w_L=0.8$, for which the sufficient cutoff is $\bar{s}\approx0.530$. The displayed subsidy $s=0.5$ therefore lies within the interval covered by Proposition~\ref{f-p-welfare comparison}.}

Within this domain, the thresholds satisfy $\bar{k}_1<\bar{k}_{1g}<\bar{k}_2<\bar{k}_{2g}$. The subsidy therefore changes Defend to Harvest for $\bar{k}_1<k<\bar{k}_{1g}$ and Dominate to Defend for $\bar{k}_2<k<\bar{k}_{2g}$, whenever these intervals lie in the common competition range. In the first interval, the incumbent chooses full openness and gives up period-2 deployment; in the second, it retains the deployer but switches to the high first-period fee. Appendix \ref{appen-free trial} derives the subsidized equilibrium, and Figure \ref{f-f-kbar comparison} summarizes the threshold ordering.

\begin{figure}[ht]
	\centering \footnotesize
\begin{tikzpicture}[scale=0.9]
\linespread{0.8}
\foreach \x in {0,3,7}{
    \draw (\x cm,5pt) -- (\x cm,-5pt);
}
\node[align=center] at (3,0.5) {$\bar{k}_1$};
\fill[green!15] (3.01,-0.16) rectangle (6.99,0.16);
\node[align=center] at (7,0.5) {$\bar{k}_2$};
\fill[blue!15] (7.01,-0.16) rectangle (10.8,0.16);
\draw [thick,->] (0,0) -- (11,0)node[below]{~~$k$};
\node at (-1.9,0) {Baseline};
\node[align=center] at (0,0.5) {0};

\node[align=center,gray] at (1.5,0.35) {\textbf{Harvest}};
\node[align=center,gray] at (5,0.35) {\textbf{Defend}};
\node[align=center,gray] at (9,0.35) {\textbf{Dominate}};

\draw [decorate,decoration={brace,amplitude=5pt,mirror,raise=2ex}] (0.05,0) -- (2.95,0) node[midway,yshift=-2em]{$(w_H,\bar{\eta})$};
\draw [decorate,decoration={brace,amplitude=5pt,mirror,raise=2ex}] (3.05,0) -- (6.95,0) node[midway,yshift=-2em]{\textcolor{black}{$(w_H,\bar{\eta}_H)$}};
\draw [decorate,decoration={brace,amplitude=5pt,mirror,raise=2ex}] (7.05,0) -- (10.8,0) node[midway,yshift=-2em]{\textcolor{black}{$(w_L,\bar{\eta}_L)$}};
\end{tikzpicture}

\begin{tikzpicture}[scale=0.9]
\linespread{0.8}
\foreach \x in {0,4,8}{
    \draw (\x cm,5pt) -- (\x cm,-5pt);
}
\node[align=center] at (4,0.5) {$\bar{k}_{1g}$};
\node[align=center] at (8,0.5) {$\bar{k}_{2g}$};
\node[align=center] at (0,0.5) {0};

\fill[green!15] (4.01,-0.16) rectangle (7.99,0.16);
\fill[blue!15] (8.01,-0.16) rectangle (10.8,0.16);
\draw [thick,->] (0,0) -- (11,0)node[below]{~~$k$};
\node at (-1.9,0) {Subsidy};

\node[align=center,gray] at (2,0.35) {\textbf{Harvest}};
\node[align=center,gray] at (6,0.35) {\textbf{Defend}};
\node[align=center,gray] at (9.5,0.35) {\textbf{Dominate}};

\draw [decorate,decoration={brace,amplitude=5pt,mirror,raise=2ex}] (0,0) -- (3.95,0)  node[midway,yshift=-2em]{$(w_H,\bar{\eta})$};
\draw [decorate,decoration={brace,amplitude=5pt,mirror,raise=2ex}] (4.05,0) -- (7.95,0) node[midway,yshift=-2em]{\textcolor{black}{$(w_H,\bar{\eta}_{Hg})$}};
\draw [decorate,decoration={brace,amplitude=5pt,mirror,raise=2ex}] (8.05,0) -- (10.8,0) node[midway,yshift=-2em]{\textcolor{black}{$(w_L,\bar{\eta}_{Lg})$}};
\end{tikzpicture}

\caption{Strategy Thresholds With and Without an Access Subsidy (Schematic)}
\label{f-f-kbar comparison}
\end{figure}

\begin{figure}[ht]
	\centering \footnotesize
    \begin{subfigure}[b]{0.45\textwidth}
    \centering
        \includegraphics[width=\textwidth]{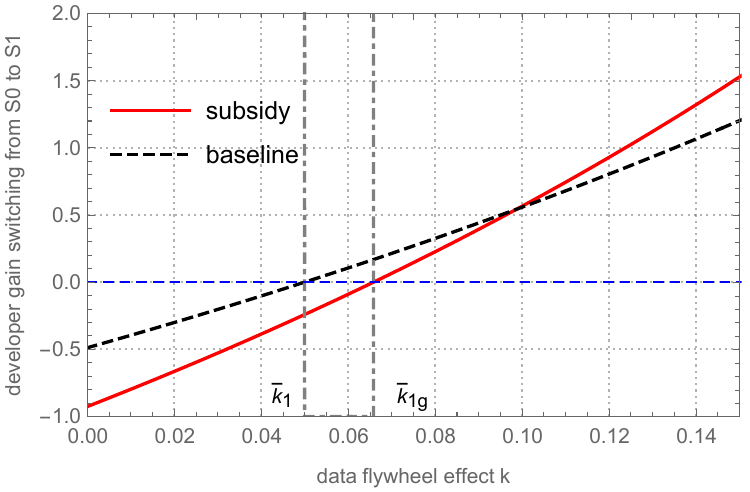}
\caption{Defend profit minus Harvest profit}\label{fig-9a}
    \end{subfigure}
    ~~~
    \begin{subfigure}[b]{0.45\textwidth}
    \centering
        \includegraphics[width=\textwidth]{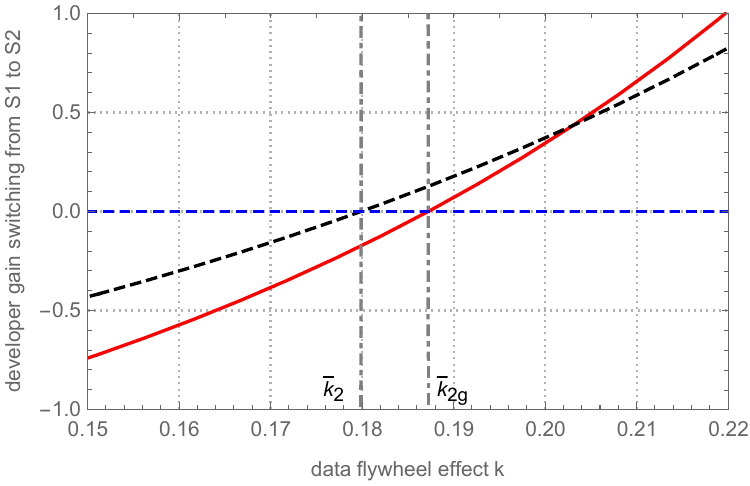}
\caption{Dominate profit minus Defend profit}\label{fig-9b}
    \end{subfigure}
\caption{Incumbent Profit Differences and Strategy Thresholds With and Without a Subsidy\\ ($\theta=5$, $c=1$, $w_H=2.5$, $w_L=0.8$, $\bar{\eta}=1.5$, $s=0.5$)}\label{fig-9}
\end{figure}

\textbf{Harvest versus Defend.} At the original threshold $\bar{k}_1$, both strategies charge the same high first-period fee. Harvest uses full openness to expand current deployment, while Defend restricts openness to retain the deployer. The subsidy raises deployment under both strategies. It also allows Defend to relax its openness restriction while retaining the deployer, strengthening learning and continuation revenue. Within the sufficient subsidy domain, however, Harvest's increase in current access revenue exceeds Defend's combined current and future gain at the original threshold. The incumbent therefore prefers Harvest there and requires stronger model-specific learning to justify restricting openness and retaining the deployer, shifting the transition to $\bar{k}_{1g}$ (Figure \ref{fig-9a}).

\textbf{Defend versus Dominate.} At the original threshold $\bar{k}_2$, both strategies retain the deployer. Dominate's lower fee and greater openness generate a larger increase in usage and continuation revenue when the subsidy is introduced. Defend, however, earns the higher first-period fee on each additional unit of usage. Within the sufficient subsidy domain, Defend's advantage in additional current revenue outweighs Dominate's advantage in additional future revenue. The subsidy therefore makes Defend preferable at the original threshold, even though learning strengthens under both strategies, and a stronger flywheel is needed to induce Dominate (Figure \ref{fig-9b}).

\subsection{Private Payoffs and Welfare}
The two strategy changes have different consequences for deployment. A Defend-to-Harvest switch raises openness to its full level and allows the entrant to serve the deployer in period 2. A Dominate-to-Defend switch retains the incumbent but raises the first-period fee and reduces openness relative to the unsubsidized allocation. Social welfare sums both developers' profits, the deployer's profit, and consumer surplus before deducting government subsidy expenditure. The next proposition compares the resulting private payoffs and welfare.

\begin{proposition}\label{f-p-welfare comparison}
Compare subsidized and unsubsidized equilibria within their common maintained competition range, and suppose $0<s<\bar{s}$.%
\begin{enumerate}[$(a)$]
\item \textbf{Defend to Harvest.} If $\bar{k}_1<k<\bar{k}_{1g}$, the subsidy induces a switch from Defend to Harvest and increases both developers' profits, the deployer's profit, and consumer surplus.
\item \textbf{Dominate to Defend.} If $\bar{k}_2<k<\bar{k}_{2g}$, the subsidy induces a switch from Dominate to Defend and reduces $Q_1$, $Q_2$, social welfare, deployer profit, and consumer surplus.
\end{enumerate}
\end{proposition}

\begin{figure}[ht]
	\centering \footnotesize
    \begin{subfigure}[b]{0.45\textwidth}
    \centering
        \includegraphics[width=\textwidth]{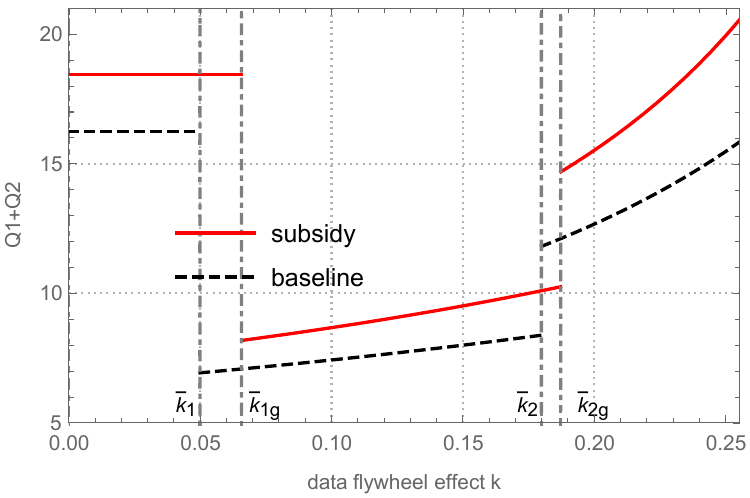}
\caption{Total fine-tuning effort $Q_1+Q_2$}\label{f-f-q1+q2 compare}
    \end{subfigure}
    ~
    \begin{subfigure}[b]{0.45\textwidth}
    \centering
        \includegraphics[width=\textwidth]{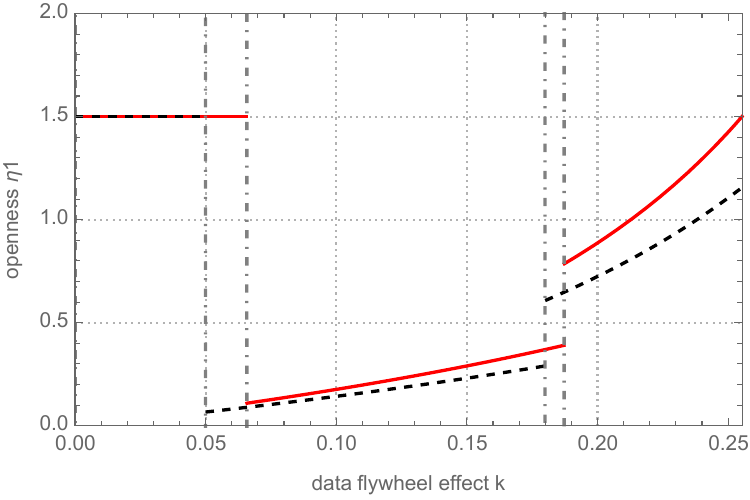}
\caption{First-period openness $\eta_1$}\label{f-f-openness compare}
    \end{subfigure}
     ~
	\caption{Impact of Government Subsidy on Fine-Tuning and Strategic Openness\\ ($\theta=5$, $c=1$, $w_H=2.5$, $w_L=0.8$, $\bar{\eta}=1.5$, $s=0.5$)}\label{f-f-equilibrium compare}
\end{figure}

In the first interval, full openness lowers the deployer's first-period adaptation cost and expands the public knowledge available for adapting the entrant's model in period 2. Together with the subsidy's reduction in the license payment per unit, these cost reductions raise effort and deployer profit in both periods. Consumers gain from higher effort, and the entrant earns access revenue from serving the deployer in period 2. The incumbent also gains: its subsidized Defend option is more profitable than baseline Defend, and it chooses Harvest because Harvest is at least as profitable. These are gains in private payoffs; they do not by themselves establish a net welfare gain after government expenditure is deducted. Figure \ref{f-f-equilibrium compare} illustrates the changes in total effort and first-period openness.

In the second interval, the subsidy induces a switch from Dominate to Defend. The incumbent raises its first-period fee from $w_L$ to $w_H$ and reduces openness relative to the unsubsidized allocation. Within the sufficient subsidy domain, $s<w_H-w_L$, so the fee increase exceeds the subsidy and the deployer's net license payment per unit rises. The higher net license payment per unit, together with the higher adaptation cost, reduces first-period effort. The resulting loss of model-specific learning is large enough to outweigh the subsidy benefit in period 2, so effort and consumer surplus fall in both periods. The incumbent's profit rises while its joint profit with the deployer falls, leaving the deployer worse off. The paper's welfare measure also falls before deducting subsidy expenditure; accounting for that expenditure would reinforce the loss. The result identifies a conditional reversal of the subsidy's direct benefit, rather than a claim that the incumbent captures all public spending.

\subsection{Implications for Subsidy Design}
The adverse switching interval shows why an assessment of adoption support should track both effective license payments and the resources required to adapt a model. An incumbent can offset financial support through a higher fee and lower openness, reducing deployment even though the subsidy lowers the deployer's payment at fixed terms. Both interventions can prompt withdrawal of the low-fee strategy supporting initial deployment. Their consequences for retention differ: the mandate moves Dominate to Harvest and induces switching, whereas the adverse subsidy moves Dominate to Defend and preserves retention. The comparison concerns a per-unit access subsidy; evaluating computing allocations, infrastructure grants, or other programs requires mapping their instruments to the incentives they change.

This incidence question connects to the broader literature on how market responses determine who benefits from a tax or subsidy \citep{weyl2013pass,fullerton2002tax}. In housing, \citet{collinson2018benefits} find that a uniform increase in voucher generosity raises rents with little improvement in neighborhood or unit quality. In agriculture, \citet{ricker2013fertilizer} find that subsidized fertilizer partly displaces commercial purchases. These findings illustrate different ways in which the intended benefit can be reduced; neither is evidence for the particular welfare reversal in our model. Our analysis adds an adjustment in model openness to the license-fee response, showing why payments alone need not capture the effect on downstream investment.

The analysis motivates examining eligibility conditions that address both price and adaptability. A restriction on the license fee alone may leave the incumbent free to reduce openness, while an openness requirement alone may leave it free to raise the fee. Whether such conditions preserve downstream benefits requires a separate analysis of supplier participation and responses through other contract or quality choices. Their effectiveness is therefore a policy-design question, rather than a result established here.

\section{Discussion and Conclusion}\label{conclusion}
The same disclosures that make a foundation model easier to adapt can also make a competing model easier to use. This paper shows how that trade-off links the incumbent's license fee and openness to the deployer's adaptation effort, subsequent model choice, and the effects of policy.

\subsection{Summary of Key Findings}
The analysis yields three main findings. First, stronger learning need not lead to greater openness. When all three equilibrium regions occur, the incumbent moves from Harvest to Defend as retaining future business becomes more valuable. It initially reduces openness to limit the entrant's adaptation advantage, then increases openness within Defend as stronger learning helps it retain the deployer. At the switch to Dominate, it lowers the license fee and increases openness further, using greater current deployment to strengthen future retention.

Second, a full-openness mandate can induce a Dominate incumbent to abandon retention and raise its current license fee. The deployer switches to the entrant in the second period, while the higher license fee can reduce first-period effort despite greater openness. The openness trap occurs only when the resulting first-period welfare loss exceeds the second-period gain from easier adaptation of the entrant's model.

Third, integration and subsidies change adaptation incentives as well as competition for deployment. Under the maintained rule that adaptation effort maximizes current-period profit, full integration removes the internal license fee and ties deployment to the incumbent. The gains from greater adaptation and learning must therefore be weighed against the adaptation advantage forgone when the entrant would otherwise have been used. For sufficiently small subsidies, the direction of the incumbent's response matters: a Defend-to-Harvest switch can improve private payoffs, whereas a Dominate-to-Defend switch reduces downstream effort and welfare. The favorable private comparison does not establish a gain net of subsidy expenditure.

\subsection{Implications for Strategy and Policy}
For model developers, the analysis links pricing and openness through their effects on current adoption and future competition. Both choices shape the deployer's incentive to adapt the model, while openness also affects public spillovers and the adoption that generates model-specific learning. Their joint value therefore depends on how much of the resulting knowledge supports continuation with the incumbent and how much benefits an alternative model. For deployers, continued access to a model and the ability to reuse accumulated experience with another model are distinct considerations. Released weights can help preserve access and control, while evaluation routines, workflow integrations, and adaptation expertise may remain tied to a particular model family. An assessment of strategic dependence should therefore identify what remains useful after switching models and what requires rebuilding or recalibration. The matched-task comparisons in Appendix \ref{appen-three k channels} outline a way to assess these differences through adaptation costs.

For policymakers, disclosure requirements and adoption support call for assessing how the developer adjusts its license fee and openness, and how those responses change adaptation. The transferable-learning result motivates studying measures that help deployers reuse experience after switching models; it does not establish the effectiveness or welfare benefit of a particular portability policy. Such an assessment would need to identify how the measure changes transferability, its implementation costs, and suppliers' responses. Confidential oversight raises a different question because information supplied to regulators need not become available to competing developers. Integration requires weighing stronger adaptation incentives against access to an alternative model. These distinctions identify the choices and evidence needed to evaluate each intervention in its institutional setting.

\subsection{Scope and Future Research}

The extensions in Appendix~\ref{sec:extensions} identify distinct limits: transferable experience can eliminate Dominate and the welfare reversal, while partial integration changes the welfare comparison by preserving access to a rival. The spillover-strength illustration also shows that weaker public reuse can remove the welfare loss while openness remains restricted. Restricted openness alone therefore does not establish that a mandate would lower welfare. Endogenous entry introduces another role for disclosure: it can determine whether a rival participates. A mandate can still reduce welfare at low entry costs, while the baseline calibration illustrates a gain when it removes exclusion at intermediate costs. These comparisons highlight the need to assess both the incumbent's license-fee response and the mandate's effect on entry.

Several generalizations remain open, including multiple entrants, a longer horizon or a continuous-time formulation, endogenous base-model quality and R\&D investment, continuous license fees, and forward-looking adaptation. Finally, the model represents openness through its effect on adaptation costs. Separating disclosures of model weights, training data, and tools would allow their effects on current adaptation and public reuse to differ, connecting the analysis more directly to the resources covered by a disclosure policy.

\makeatletter
\renewcommand{\@biblabel}[1]{}
\let\OLDthebibliography\thebibliography
\renewcommand{\thebibliography}[1]{
  \OLDthebibliography{#1}
  \setlength{\itemsep}{0pt}
  \setlength{\parskip}{0pt}
  \setlength{\baselineskip}{16pt}
}
\makeatother
\bibliographystyle{ormsv080}
\bibliography{references}

\newpage
\ECSwitch

\vspace{-20pt}

\begin{APPENDICES}
\renewcommand{\theHsection}{EC.\arabic{section}}
\renewcommand{\theHsubsection}{\theHsection.\arabic{subsection}}
\renewcommand{\theHsubsubsection}{\theHsubsection.\arabic{subsubsection}}
\renewcommand{\theHequation}{EC.\arabic{equation}}
\renewcommand{\theHproposition}{EC.\arabic{proposition}}
\renewcommand{\theHlemma}{EC.\arabic{lemma}}
\renewcommand{\theHtheorem}{EC.\arabic{theorem}}
\renewcommand{\theHfigure}{EC.\arabic{figure}}
\renewcommand{\theHtable}{EC.\arabic{table}}

\section{Supplemental Materials}
This appendix connects the model's primitives to deployment practice and provides additional payoff comparisons and equilibrium derivations supporting the main analysis.

\subsection{Economic Interpretation and Real-World Mapping of Model Openness}\label{appen-openness-mapping}

\textbf{Openness as an Economic Index.} Openness $\eta$ summarizes how the resources and rights a developer makes available reduce the cost of downstream adaptation. A higher $\eta$ makes it less costly for the deployer to achieve a given application quality. We combine four related dimensions: (i) \textit{access modality}, including whether access is limited to a remote API or extends to downloadable weights and inspectable architecture; (ii) \textit{documentation, tooling, and recipes}, including reference implementations, fine-tuning recipes, and evaluation harnesses; (iii) \textit{modularity and interface stability}, including standardized adapters, RAG connectors, and stable deployment interfaces; and (iv) \textit{licensing frictions around adaptation}, including restrictions on commercial fine-tuning, wrapping, and redistribution. Research on model openness distinguishes several such dimensions \citep{liesenfeld2023opening}, while the Foundation Model Transparency Index assesses disclosed information \citep{bommasani20242024}. Our aggregation is a modeling assumption: $\eta$ represents the resulting adaptation-cost benefit, rather than a measured transparency score.

\begin{table}[ht]
\centering
\scriptsize
\caption{Mapping Openness, License Fees, and Resource Costs to Practice}
\label{tab:openness-map}
\begin{tabular}{@{}p{2.8cm}p{5cm}p{3cm}p{4.85cm}@{}}
\toprule
\textbf{Observable feature} & \textbf{Representative artifact} & \textbf{Model object} & \textbf{Direct economic channel} \\
\midrule
Weight and architecture access &
API-only access versus downloadable weights and inspectable architecture &
Current model's openness $\eta_1$, $\eta_2$, or $\tilde{\eta}_2$ &
Lowers the deployer's cost of adapting the current model through its model-specific openness denominator in $C_1$, $C_{2I}$, or $C_{2E}$. \\
\addlinespace
Documentation, tooling, and recipes &
Model cards, reference code, fine-tuning recipes, and evaluation harnesses &
Current model's openness $\eta_1$, $\eta_2$, or $\tilde{\eta}_2$ &
Reduces the technical and organizational effort required to reach a target application quality. \\
\addlinespace
Modularity and interface stability &
Standardized adapters, RAG connectors, and stable deployment interfaces &
Current model's openness $\eta_1$, $\eta_2$, or $\tilde{\eta}_2$ &
Reduces integration and reconfiguration frictions in the deployer's adaptation process. \\
\addlinespace
Commercial adaptation and redistribution rights &
Permission to fine-tune, wrap, deploy, and redistribute a model commercially &
Current model's openness $\eta_1$, $\eta_2$, or $\tilde{\eta}_2$ &
Raises effective adaptability and reuse. It does not mechanically determine the separately chosen license fee $w_t$. \\
\addlinespace
Public ecosystem reuse across model families &
Public architecture information, reference code, and training or evaluation resources that a deployer can reuse when adapting a competing model &
Incumbent's period-1 openness $\eta_1$ &
Lowers the deployer's cost of adapting the entrant through the additional $(1+\eta_1)$ denominator in $C_{2E}$. \\
\addlinespace
Commercial license fee &
Usage-, token-, seat-, or managed-hosting charge &
License fee $w_t$ &
Changes the deployer's revenue retained per unit of usage, $\theta-w_t$, separately from technical openness $\eta_t$. \\
\addlinespace
Raw compute and data acquisition &
Accelerators, cloud-compute inputs, and purchased or curated training data &
Additional resource costs &
Outside the baseline openness channel; Appendix~\ref{appen-three k channels} examines common per-unit and fixed resource costs. The scalar $c$ continues to scale adaptation cost. \\
\bottomrule
\end{tabular}
\vspace{1mm}

\parbox{0.94\textwidth}{\scriptsize\textit{Note}: This table provides a conceptual mapping rather than an empirical measurement protocol. The first four rows identify observable components of model-specific adaptability; the fifth isolates the intertemporal public-spillover channel; and the final two separate openness from price and from additional resource costs.}
\end{table}

\textbf{Mapping Openness and Pricing to the Model.} Table~\ref{tab:openness-map} separates current-model adaptability, public reuse, license fees, and additional resource costs. The openness of the model being used enters its adaptation cost: $\eta_1$ in $C_1(Q_1)$, $\eta_2$ in $C_{2I}(Q_2)$, and $\tilde{\eta}_2$ in $C_{2E}(Q_2)$. The incumbent's earlier disclosure also helps the deployer adapt the entrant's model, generating the additional $(1+\eta_1)$ term in the denominator of $C_{2E}(Q_2)$. The license fee $w_t$ determines the revenue the deployer retains per unit of usage, $\theta-w_t$. The parameter $c$ scales adaptation costs; raw compute and data-acquisition costs are separate resource inputs. Appendix~\ref{appen-three k channels} examines common per-unit and fixed resource costs while holding these additional components independent of disclosure. A developer can therefore make its model easier to adapt while charging a positive license fee for its service, as the examples below illustrate.

\textbf{Developer-Operated Model Access}. The supplier configuration in our model places disclosure and license-fee setting within the same developer organization. Proprietary models offered through OpenAI's and Anthropic's direct APIs illustrate this arrangement: the developer determines the information and adaptation interfaces it makes available and charges for access to its own models.\footnote{See OpenAI's API pricing, \url{https://developers.openai.com/api/docs/pricing}, and Anthropic's API pricing, \url{https://platform.claude.com/docs/en/about-claude/pricing}, accessed October 22, 2026.} Access to weights is one component of $\eta$; documentation, tooling, and interfaces also shape the deployer's adaptation opportunities. The aggregate index therefore accommodates proprietary API services as well as services based on released weights, while $w$ separately captures the license fee.

Released-weight models illustrate why the dimensions should be considered separately. Llama 3.1 and DeepSeek-R1 both provide downloadable weights, but Llama 3.1 uses Meta's community license and DeepSeek-R1 uses the MIT license, with different conditions governing adaptation and redistribution.\footnote{See the Llama 3.1 community license, \url{https://huggingface.co/meta-llama/Llama-3.1-8B/blob/main/LICENSE}, and the DeepSeek-R1 model card and licensing statement, \url{https://huggingface.co/deepseek-ai/DeepSeek-R1}, accessed October 23, 2026.} Weight access and licensing flexibility are thus distinct components of $\eta$. A deployer that downloads and operates released weights independently does not purchase the developer's managed-access service. The positive fee $w$ in our model concerns the ongoing supplier service, whose joint disclosure and pricing decisions are illustrated by the following reference case.

Our main reference case is Alibaba's Qwen supplied through Alibaba Cloud Model Studio, which illustrates how released weights and substantial adaptability can coexist with paid managed access. For a specific model, Qwen3-32B is released under Apache 2.0, while Model Studio documents full-parameter and parameter-efficient supervised fine-tuning in its Beijing region and charges for dedicated deployment of that model.\footnote{See the Qwen3-32B model card, \url{https://huggingface.co/Qwen/Qwen3-32B}; Model Studio's supported fine-tuning models, \url{https://www.alibabacloud.com/help/en/model-studio/model-training-overview}; and its training and deployment pricing, \url{https://www.alibabacloud.com/help/en/model-studio/model-training-and-deployment-billing}, accessed October 22, 2026.} Model disclosure and managed-service pricing thus sit within Alibaba's organization. Common ownership gives the supplier a reason to consider how disclosure affects demand for its own services. To study this joint choice, the model represents the supplier as a single decision maker maximizing profit from the managed-service relationship. This objective is a modeling assumption motivated by the common ownership of the two activities.

The deployer is an independent application company that adapts the model and retains feedback from its own users. Developer-operated hosting is therefore compatible with an independently owned downstream application. Qwen's dedicated deployment is billed by capacity and time; the model represents this tariff as a constant per-unit license fee for the developer's service, distinct from a royalty on downloaded weights.

IBM Granite through IBM watsonx.ai provides supporting evidence for the same institutional configuration. IBM releases Granite models under Apache 2.0 and lists usage-based prices for managed access to selected models, including Granite 4.0 H Small.\footnote{See IBM's model card, \url{https://huggingface.co/ibm-granite/granite-4.0-h-small}, and watsonx.ai pricing, \url{https://www.ibm.com/products/watsonx-ai/pricing}, accessed October 22, 2026.} In both cases, the developer's own organization supplies paid services alongside adaptable released models.

These institutional features motivate the choice of players and instruments. The analysis takes the managed-service relationship as given, holding self-hosting, alternative hosts, supplier serving costs, and other supplier revenues outside the model. The binary price menu, normalized model capability, and cost and learning functions specify the economic environment within which the supplier makes its joint choice. Estimating the associated cost reductions and learning magnitudes would require additional empirical analysis.

\subsection{Adaptation Costs, Deployer Learning, and Policy Relevance}\label{appen-three k channels}
This subsection distinguishes three sources of adaptation-cost reductions and illustrates them in a deployment workflow. It then examines how common additional costs affect competition and welfare, reviews evidence on implementation costs, and describes measurements that would inform a policy comparison.

\textbf{Three Adaptation-Cost Channels.}
The cost functions \eqref{eq:cost_period1}, \eqref{eq:cost_entrant}, and \eqref{eq:cost_incumbent} separate public reuse, current-model adaptability, and accumulated deployment experience.
\begin{enumerate}[$(i)$]
\item \textbf{Public reuse when adapting the entrant's model ($\eta_1$)}. The incumbent's first-period openness enters the deployer's cost of adapting an entrant model through reusable public knowledge and infrastructure. An explicit example is provided by Qwen2.5-32B and DeepSeek-R1-Distill-Qwen-32B. DeepSeek identifies Qwen2.5-32B as the base of this distilled model, and both published configurations use the Qwen2 causal-language-model architecture.\footnote{See the base-model table in \url{https://huggingface.co/deepseek-ai/DeepSeek-R1} and the published configurations at \url{https://huggingface.co/Qwen/Qwen2.5-32B/blob/main/config.json} and \url{https://huggingface.co/deepseek-ai/DeepSeek-R1-Distill-Qwen-32B/blob/main/config.json}. This technical example concerns the Qwen2.5 lineage; the managed-service example in Appendix \ref{appen-openness-mapping} concerns Qwen3-32B.} Qwen's documentation describes fine-tuning with the ms-swift framework, whose supported-model registry includes both releases.\footnote{See \url{https://qwen.readthedocs.io/en/v2.5/training/SFT/ms_swift.html} and \url{https://swift.readthedocs.io/en/v3.4/Instruction/Supported-models-and-datasets.html}. The registry specifies model-dependent templates. DeepSeek's model card also instructs users to use its modified configurations and tokenizers.}

This compatibility gives the deployer a concrete reuse opportunity when evaluating the later supplier's model: it can retain the common training framework and compatible data-processing and evaluation routines, while changing the checkpoint, configuration, and model-specific preparation. The potential saving accrues to the deployer through reduced engineering work needed to adapt its application. Retraining adapters, preparing suitable domain data, paying for computation, and validating the resulting application remain part of the deployment process. The documented lineage and tool support establish the technical basis for this reuse; the implied reduction in engineering expenditure is an economic inference whose magnitude depends on the application. Closely related models can also share private deployment know-how, a channel allowed by the transferable-learning extension in Appendix \ref{appen-data flywheel on entrant}. The example therefore illustrates the public pathway alongside possible private transfer.

Reusable adaptation infrastructure also spans distinct model families. \citet{zheng2024llamafactory} describe LlamaFactory, a common fine-tuning framework supporting more than 100 language models through shared data-processing and training modules and model-specific configurations. Such infrastructure provides a practical route for reusing implementation work when adopting another model; the expenditure saving depends on the tasks and model-specific adjustments that remain.

\textit{Reduced-form pass-through}. Let $m(\eta_1)$ denote the maturity of the public entrant-facing ecosystem, with $m'(\eta_1)>0$, and let $\phi(m)>0$ be the deployer's adaptation-cost shifter, with $\phi'(m)<0$. The entrant-facing cost can then be written as
\[
C_{2E}(Q_2)=\frac{cQ_2^2}{1+\tilde{\eta}_2}\phi\!\left(m(\eta_1)\right).
\]
The baseline uses $\phi(m(\eta_1))=1/(1+\eta_1)$ to represent the downstream benefit of public compatibility in a tractable form. This specification describes the cost of adapting an application to the entrant; the entrant's own development expenditure is a separate economic object. The sign and size of the public pass-through depend on how much of the released knowledge is useful downstream. Appendix~\ref{appen-spillover-strength} parameterizes its strength and derives the conditions for openness restriction and a welfare-loss region within the stated competitive setting. The measurement design below describes how access to specific public artifacts could be varied to investigate this link.

\item \textbf{Current-model adaptability ($c$ and $\eta$)}. Openness lowers the cost of adapting the model currently used, while $c$ scales the adaptation component. The modeled work includes engineering, experimentation, and integration, such as adaptation scripts, evaluation routines, and model-specific debugging. Compute, data acquisition, and other resource expenditures can remain even when these tasks become easier. The common-cost calculation below separates those expenditures from the adaptation component.

\item \textbf{Model-specific deployment experience ($k$)}. Application usage can teach the deployer about a model family's behavior, failure modes, evaluation routines, and effective adaptation practices. In the baseline, the factor $(1+k\alpha_1)$ captures the resulting reduction in the cost of continuing with the incumbent; this benefit requires the deployer to retain and use application-layer feedback. Appendix \ref{appen-data flywheel on entrant} allows a weaker transferable component to lower the cost of adapting the entrant's model as well. This learning improves the deployment process while intrinsic model capability remains fixed. Developer-side learning would introduce an additional channel through changes in the base model itself. Research on recursive training with model-generated data studies a different mechanism \citep{shumailov2024ai}; it does not determine whether ordinary application feedback can improve a base model.
\end{enumerate}

These channels distinguish publicly reusable resources from experience accumulated by the deployer. The worked example illustrates their different roles; the subsequent cost calculation and measurement design examine their economic importance.

\textbf{Worked example: an independent application company using managed Qwen access through Alibaba Cloud Model Studio.} Consider an independent application company, the deployer, that customizes Qwen3-32B and uses Model Studio's paid deployment service, as described in Appendix \ref{appen-openness-mapping}. For this illustrative workflow, assume that the deployer retains and uses its own evaluation and user-feedback records. This application-level activity supplies the experience underlying the deployer's learning channel.

First, released model information and reusable adaptation libraries can make a later model family easier for the deployer to evaluate and adapt. The model represents this public-ecosystem pathway through $(1+\eta_1)$ in $C_{2E}$. Reuse operates through knowledge and infrastructure, allowing model-specific adapters to be retrained as needed when the deployer changes model families. Second, inspectable model artifacts and reusable tools can reduce the deployer's engineering and integration burden when adapting its current model, represented by the corresponding current-model openness term. Third, repeated application use can improve the deployer's model-specific evaluation routines, handling of failure modes, and adaptation practices. The baseline represents the part of this experience that remains specific to the incumbent family through $(1+k\alpha_1)$ in $C_{2I}$; Appendix \ref{appen-data flywheel on entrant} permits partial transfer.

The workflow distinguishes public knowledge available across suppliers from experience accumulated by the deployer. This distinction connects the illustrative workflow to the separate cost terms in the model. The feedback requirement concerns the deployer's retention and use of application records, so deployer learning can arise under either hosted or local deployment. When this feedback is unavailable or unused, the corresponding flywheel channel is absent.

\textbf{Common Additional Costs and Participation.}
Deployment also requires compute, suitable data, and infrastructure. To examine expenditures outside the adaptation component, suppose an active deployer incurs
\[
C^{\mathrm{tot}}(Q;D)=\frac{cQ^2}{D}+aQ+F,\qquad a,F\geq0,
\]
in each period, where $D_1=1+\eta_1$, $D_{2I}=(1+k\alpha_1)(1+\eta_2)$, and $D_{2E}=(1+\eta_1)(1+\tilde{\eta}_2)$. The parameter $a$ is a common per-unit resource cost, and $F$ is common fixed operating expenditure per active period. Both are independent of supplier, openness, and the mandate, and exclude expenditure already included in the license fee $w$. An inactive deployer receives zero and incurs neither component. This decomposition separates a cost that reduces the return to additional usage from expenditure that affects the decision to operate.

\textit{Equilibrium with common additional costs}.
Write $\vartheta=\theta-a$ and maintain $0<w_L<w_H\leq\vartheta/2$. The participation condition
\[
0\leq F<\frac{(\vartheta-w_H)^2}{4c}
\]
is sufficient to keep the deployer active in both periods at every feasible price--openness choice. We also impose the Section 4.1 competition restriction using revenue net of the additional per-unit resource cost:
\[
0\leq k< K_{\max}(\vartheta):=\min\left\{
\frac{2c\bar{\eta}}{(1+\bar{\eta})(\vartheta-w_L)},\,
\frac{2c(2\vartheta-w_H-w_L)(w_H-w_L)}{(\vartheta-w_H)^2(\vartheta-w_L)}
\right\}.
\]
Under these conditions, the baseline equilibrium and full-openness-mandate allocation are obtained by replacing $\theta$ with $\vartheta$ throughout the corresponding formulas. The candidate strategies remain Harvest, Defend, and Dominate, with cost-dependent selection thresholds. In particular, the retention thresholds become
\[
\bar{\eta}_j(a)=\frac{k(\vartheta-w_j)}{2c-k(\vartheta-w_j)},\qquad j\in\{H,L\}.
\]

To verify the result, the active deployer's operating problem gives
\[
Q^*(D,w)=\frac{D(\vartheta-w)}{2c},\qquad
\Pi^*(D,w)=\frac{D(\vartheta-w)^2}{4c}-F.
\]
Because $D\geq1$ and $w\leq w_H$, the stated bound on $F$ makes this profit strictly positive at every feasible choice. The common fixed expenditure cancels from the comparison between suppliers, and the usage and supplier-revenue functions coincide with those of the baseline evaluated at $\vartheta$. Backward induction therefore gives the same strategic optimization with revenue per unit reduced by the common variable cost. For either the equilibrium or the mandated allocation, summing profits and consumer surplus yields
\[
W_{a,F}(\theta)=W_{0,0}(\theta-a)-2F.
\]
Here each welfare term is evaluated at its corresponding allocation. Consequently, the welfare difference between the mandate and the unregulated equilibrium equals the baseline difference at $\theta-a$. A welfare reversal remains possible with positive additional costs. Under the participation condition, the per-unit cost $a$ changes the comparison through $\vartheta$, while the common fixed expenditure $F$ cancels. Applying Proposition \ref{b-p-welfare} at $\vartheta=\theta-a$, define $\bar{k}^{W}(\vartheta):=\max\{\bar{k}_2(\vartheta),k_W(\vartheta)\}$. The mandate lowers welfare exactly when $\bar{k}^{W}(\vartheta)<k<K_{\max}(\vartheta)$, and this region is nonempty if and only if $\max\{0,\bar{k}^{W}(\vartheta)\}<K_{\max}(\vartheta)$.

For illustration, set $\theta=5$, $c=1$, $w_H=2.25$, $w_L=0.5$, $\bar{\eta}=1.5$, and $F=0.2$. The following values show the two equilibrium thresholds, the welfare-reversal threshold $\bar{k}^{W}$, and the admissible upper bound on $k$:
\[
\begin{array}{c|cccc}
a & \bar{k}_1 & \bar{k}_2 & \bar{k}^{W} & K_{\max}\\ \hline
0 & 0.1719 & 0.2500 & 0.2580 & 0.2667\\
0.25 & 0.1778 & 0.2510 & 0.2726 & 0.2824\\
0.50 & 0.1822 & 0.2500 & 0.2889 & 0.3000
\end{array}
\]
In each row, the mandate lowers welfare for $\bar{k}^{W}<k<K_{\max}$, with both-period participation satisfied. The table also shows why the same strategic response can have different welfare consequences. At $k=0.26$, the unregulated incumbent chooses \textit{Dominate} in all three rows and the mandate induces \textit{Harvest}, but welfare falls only when $a=0$; it rises when $a=0.25$ or $a=0.50$. Variable resource costs change the return to deployment and hence the balance between the benefits of greater openness and the loss from reduced initial fine-tuning. Common fixed expenditure cancels from this comparison while the stated participation condition holds. These values illustrate sensitivity within the model; an empirical calibration would require application-level expenditure and learning data.

The result isolates how common additional costs affect the return to deployment and the resulting equilibrium. Supplier-specific costs, costs that depend on disclosure, and fixed costs large enough to change participation create additional competitive effects. Assessing those settings requires extending the corresponding supplier-choice and participation comparisons.

\textit{Economic relevance of adaptation}. Published application evidence distinguishes the computing bill from the work required to implement a model. In a headline-selection application, \citet[Online Appendix E, Table A4]{ye2025lola} report about \$3 of GPU expenditure for a two-hour fine-tuning replicate of Llama-3-8B using low-rank adaptation (LoRA), while assessing its implementation effort as high and explaining the engineering needed for fine-tuning and deployment. This task-specific comparison illustrates why a compute charge alone does not measure the adaptation burden; the study does not price the engineering labor.

Broader descriptive evidence also points to implementation costs as a consideration in technology choice. In a 2025 survey by McKinsey, the Mozilla Foundation, and the Patrick J.\ McGovern Foundation, 60\% of technology decision-makers reported lower implementation costs and 46\% lower maintenance costs with open-source AI, while 48\% of respondents favored proprietary tools for faster time to value \citep[pp.~8--9, Exhibit~6]{bisht2025opensource}. These percentages describe respondents' assessments, with the report grouping partially open and open technologies; they are not estimates of expenditure reductions or the effect of disclosure alone.

The economic role of the adaptation component in our model is to change the return to deployment and the relative payoff from continuing with the incumbent or adopting the entrant. Its share of total expenditure need not be the largest for those incentives to change. The common-cost result makes this distinction explicit: additional variable costs reduce the return to deployment and shift the thresholds, while common fixed expenditure cancels from comparisons between active alternatives under the participation condition. The evidence motivates studying competition through adaptation costs; the analysis characterizes the conditions under which these costs affect disclosure, pricing, and welfare.

\textbf{Measuring Learning and Public Reuse.} The cost functions suggest a design that separates accumulated application experience from contemporaneous model capability. Otherwise comparable application teams would be randomly assigned prescribed amounts of prior incumbent usage $\alpha$, including a zero-exposure group, and subsequently assigned fresh tasks adapting either the incumbent or an entrant to a common, feasible quality target $Q$. Model versions, available public artifacts, task difficulty, hardware, and access terms would be held fixed across exposure groups. Engineering time, compute, data preparation, and fixed operating expenditure would be recorded separately, allowing the adaptation component to be distinguished from the other resource costs.

Let $K_I(Q,\alpha)$ and $K_E(Q,\alpha)$ denote the adaptation costs on these matched tasks, after removing the separately measured additional components. Under the cost specification in the model and the transferable-learning extension, the implied moments are
\[
\frac{K_I(Q,0)}{K_I(Q,\alpha)}-1=k\alpha,\qquad
\frac{K_E(Q,0)}{K_E(Q,\alpha)}-1=\hat{k}\alpha.
\]
The ratios compare exposure groups within each model, so the two models may have different baseline adaptation costs. The first identifies the incumbent continuation benefit and the second its transferable component; the baseline corresponds to $\hat{k}=0$. Several assigned exposure levels would allow the shape of the learning relationship to be assessed as well as its magnitude. This interpretation requires prior exposure to affect the measured task cost through retained application learning, with the listed task and technology conditions held fixed. Random assignment addresses selection into incumbent use, while the entrant tasks reveal how much of the accumulated experience remains useful across suppliers.

Public reuse can be examined in a complementary comparison that varies access to specified adaptation scripts and evaluation tools across otherwise comparable entrant tasks while holding the entrant checkpoint and private exposure fixed. This comparison would measure the cost effect of the supplied artifacts, which is the relevant downstream link in the public-spillover mechanism. Its interpretation as an effect of broader openness depends on how the intervention changes the availability of those artifacts. Under the spillover-strength specification in Appendix~\ref{appen-spillover-strength}, a comparison of the no-artifact and artifact conditions corresponding to openness levels $0$ and $\eta_1$ identifies the cost-efficiency gain $\gamma\eta_1$, holding the other cost shifters fixed. Recovering $\gamma$ separately requires a common scale for $\eta_1$, such as one anchored to the incumbent-facing adaptation benefit in the model. A contrast between two positive openness levels instead identifies their relative cost effect.

The learning ratios first recover the dimensionless gains $k\alpha$ and $\hat{k}\alpha$. Mapping them into the model's $k$ axis requires a common usage normalization; locating the policy thresholds additionally requires estimates of monetization, adaptation efficiency, the price menu, additional costs, and the intervention's public pass-through. These inputs would support a comparison of deployment and welfare with and without the disclosure requirement, allowing the incumbent's choices to adjust in each case. The common-cost illustration above shows why both the learning benefit and the return to deployment matter for that comparison.

The measurements also indicate how much deployment experience transfers across suppliers. Appendix \ref{appen-data flywheel on entrant} examines this sensitivity separately: transferable learning can attenuate or eliminate the openness trap under the extension's conditions. Estimates of the transferable component would therefore inform the choice of competitive benchmark as well as its parameters. Uncertainty in the estimated inputs should be carried into the regime and welfare comparisons. This design supplies a route from the modeled mechanism to measurement; implementing it and estimating the parameters are empirical tasks.

\subsection{Capability Normalization and Scope of Baseline Competition}\label{appen-capacity normalization}
We normalize intrinsic base-model capability to study competition among models with sufficiently similar underlying capabilities for a given application. In this setting, adaptability, pricing, public spillovers, and accumulated deployment experience can determine model choice.
The baseline abstracts from endogenous R\&D and capability differences. DeepSeek-R1 illustrates the scope for a rival developer to narrow capability gaps \citep{guo2025deepseek}. More recent examples include Kimi K3, whose full weights Moonshot AI released, and GLM-5.2, which Z.ai released under an MIT license \citep{moonshot2026kimik3,zai2026glm52}. At release, independent evaluation placed Kimi K3 near GPT-5.5 and Claude Opus 4.8 and GLM-5.2 near selected effort levels of GPT-5.5 on a composite intelligence index, while performance varied across tasks and the strongest proprietary models retained advantages on some comparisons \citep{artificialanalysis2026kimik3,artificialanalysis2026glm52}.

These comparisons establish neither uniform parity across quality, cost, latency, safety, and deployment reliability nor a causal effect of releasing weights on capability. Even when base models perform similarly on benchmarks, the deployer may need to adapt its application when switching. Multi-prompt evaluation shows that semantically equivalent instruction templates can change both absolute performance and relative model rankings, while parameter-efficient adaptations trained on one base-model version can deteriorate after that model is updated \citep{mizrahi2024state,gu2025adapt}. A descriptive survey of 100 CIOs also reports that investments in prompts, guardrails, and agentic workflows can raise model-switching costs \citep{wang2025enterpriseai}. The UK Competition and Markets Authority identifies technical switching costs and limited interoperability as potential sources of persistence in foundation-model relationships \citep{cma2024fmupdate}. This evidence motivates treating model-specific deployment know-how as a source of switching costs.

\textbf{Intrinsic Capability and Deployed Application Quality.} Let $Q_2^*$ and $\tilde Q_2^*$ denote the deployer's optimal application qualities conditional on using the incumbent's and entrant's models, respectively. In the period-2 competitive comparison, both developers choose maximal openness $\bar{\eta}$ and charge $w_L$. The conditional qualities satisfy
\begin{equation*}
\frac{\tilde{Q}_{2}^*}{Q_{2}^*}=\frac{1+\eta_1}{1+kQ_1}.
\end{equation*}
The deployer therefore achieves higher application quality with the entrant when $\eta_1>kQ_1$ and weakly higher quality with the incumbent when $\eta_1\leq kQ_1$. At the common license fee, this ratio also ranks the deployer's optimized profits, yielding the retention condition $kQ_1\geq\eta_1$. Thus, equal intrinsic capability can coexist with different application qualities because the deployer's adaptation costs differ.

Proposition~\ref{b-p-equilibrium-k} links this comparison to the three strategies. In a Harvest outcome in which the incumbent strictly loses period 2, the public spillover exceeds the learning advantage, so $\tilde Q_2^*>Q_2^*$ and the deployer switches. Under Defend and Dominate, the incumbent chooses the greatest first-period openness consistent with retention. The constraint binds, $\eta_1=kQ_1$, so the two models yield equal conditional application quality; the deployer stays with the incumbent under the tie-breaking rule.

\newpage
\subsection{Additional Profit and Welfare Comparisons}\label{appen-baseline-welfare}
Figure~\ref{b-f-each party profit} reports developer 1's profit, the deployer's profit, consumer surplus, and social welfare across the three unregulated strategies for the calibration of Figure~\ref{b-f-equilibrium}.

\begin{figure}[ht]
	\centering \footnotesize
    \begin{subfigure}[b]{0.495\textwidth}
    \centering
        \includegraphics[width=\textwidth]{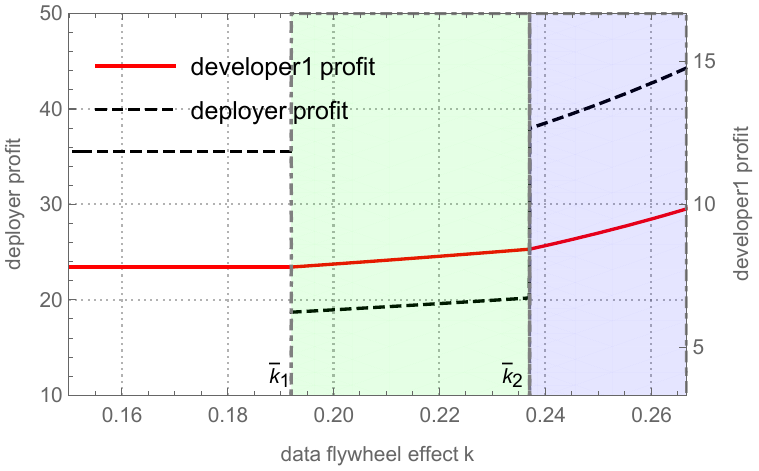}
       \caption{developer 1 \& deployer}\label{b-f-developer1 and deployer}
    \end{subfigure}
    ~~~
    \begin{subfigure}[b]{0.455\textwidth}
    \centering
        \includegraphics[width=\textwidth]{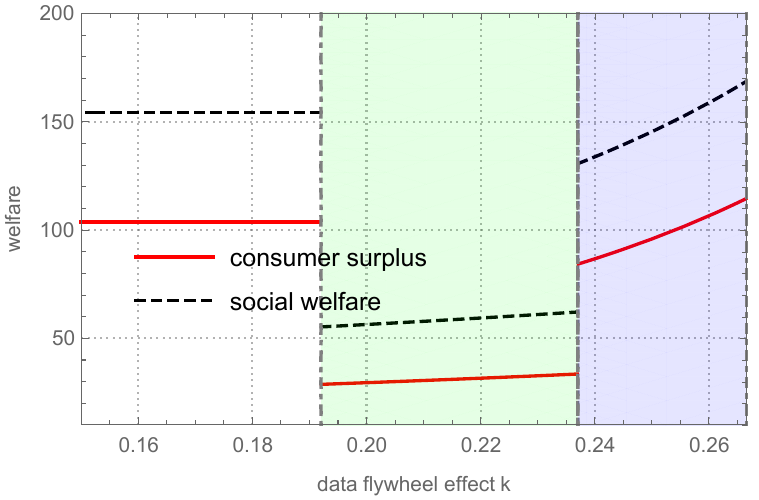}
       \caption{consumer surplus \& social welfare}\label{b-f-consumer and social welfare}
    \end{subfigure}
    	\caption{Welfare Implications of the Incumbent's Equilibrium Regimes\\($\theta=5$, $c=1$, $w_H=2.5$, $w_L=0.5$, $\bar{\eta}=1.5$)}\label{b-f-each party profit}
\end{figure}

Figure \ref{b-f-regulate-appendix} compares the entrant's profit and both developers' combined profit under the full-openness mandate and the unregulated equilibrium. These developer-profit comparisons complement the welfare analysis in Section \ref{Openness Trap}.

\begin{figure}[ht]
	\centering \footnotesize
    \begin{subfigure}[b]{0.464\textwidth}
    \centering
        \includegraphics[width=\textwidth]{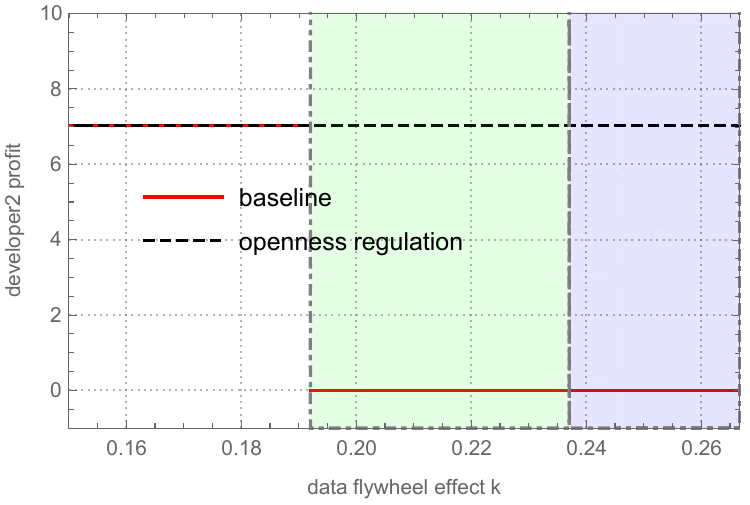}
       \caption{developer 2's profit}\label{b-f-regulate-developer2}
    \end{subfigure}
    ~~~
    \begin{subfigure}[b]{0.458\textwidth}
    \centering
        \includegraphics[width=\textwidth]{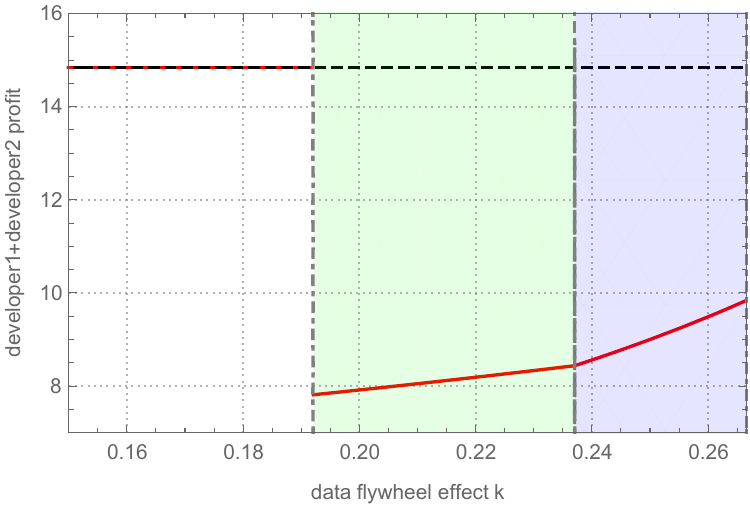}
       \caption{Both developers' combined profit}\label{b-f-regulate-developer1 and developer2}
    \end{subfigure}
    	\caption{Impacts of Openness Regulation on Developer 2's Profit and Both Developers' Combined Profit\\ ($\theta=5$, $c=1$, $w_H=2.5$, $w_L=0.5$, $\bar{\eta}=1.5$)}\label{b-f-regulate-appendix}
\end{figure}

\subsection{Equilibrium Derivation under Vertical Integration}\label{appen-vertical integration}
We retain full integration and the operating rule in Section \ref{vi}: the owner chooses openness over two periods, while the operating unit chooses effort to maximize current-period profit. Deployment remains with the incumbent, and there is no internal license fee. Given first-period effort $Q_{1v}$, second-period profit is
\begin{equation*}
    \pi_{2v}(\eta_{2v}, Q_{2v})=\theta Q_{2v}-\frac{cQ_{2v}^2}{(1+kQ_{1v})(1+ \eta_{2v})},
\end{equation*}
The factor $1+kQ_{1v}$ captures learning generated by first-period deployment. Greater second-period openness lowers adaptation costs, so the owner chooses $\eta_{2v}=\bar{\eta}$. The operating unit's effort response is
\begin{equation*}
    Q_{2v}(Q_{1v})=\frac{(1+kQ_{1v})(1+\bar{\eta}) \theta}{2c}.
\end{equation*}

In period 1, the operating unit takes $\eta_{1v}$ as given and maximizes current-period profit:
\begin{equation*}
\pi_{1v}(\eta_{1v},Q_{1v})=\theta Q_{1v}-\frac{cQ_{1v}^2}{1+\eta_{1v}}.
\end{equation*}
The owner chooses openness to maximize two-period profit, anticipating this effort response. Greater openness raises both current profit and continuation profit along the response, so the owner chooses $\eta_{1v}=\bar{\eta}$.
\begin{proposition}\label{v-p-equilibrium}
Under vertical integration with the operating rule above, $\eta_{1v}=\eta_{2v}=\bar{\eta}$, and the operating unit's efforts are
$$Q^*_{1v}=\frac{(1+\bar{\eta})\theta}{2c},\qquad Q^*_{2v}=\frac{(1+\bar{\eta})[2c+k\theta(1+\bar{\eta})]\theta}{4c^2}.$$
\end{proposition}

We provide a comparison of the fine-tuning efforts in both periods, the openness level at $t=1$, and social welfare between integration and the baseline in Figure \ref{v-f-appendix}.

\begin{figure}[ht]
    \centering \footnotesize
    \begin{subfigure}[b]{0.45\textwidth}
    \centering
        \includegraphics[width=\textwidth]{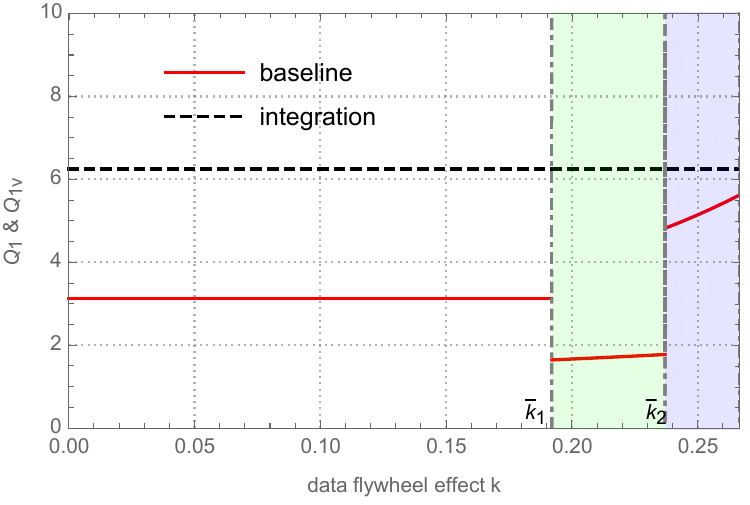}

      \caption{fine-tuning effort $Q_1$ and $Q_{1v}$}\label{v-f-q1}
    \end{subfigure}
    ~~~
    \begin{subfigure}[b]{0.45\textwidth}
    \centering
        \includegraphics[width=\textwidth]{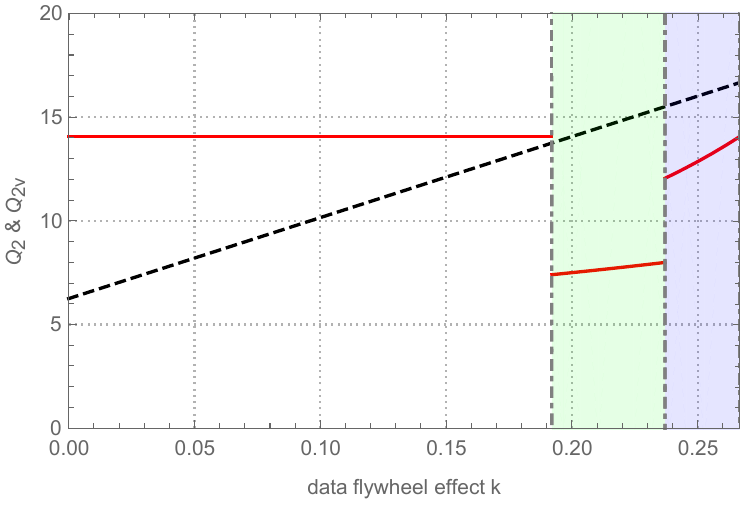}
       \caption{fine-tuning effort $Q_2$ and $Q_{2v}$}\label{v-f-q2}
    \end{subfigure}
	~~~
    \begin{subfigure}[b]{0.455\textwidth}
    \centering
        \includegraphics[width=\textwidth]{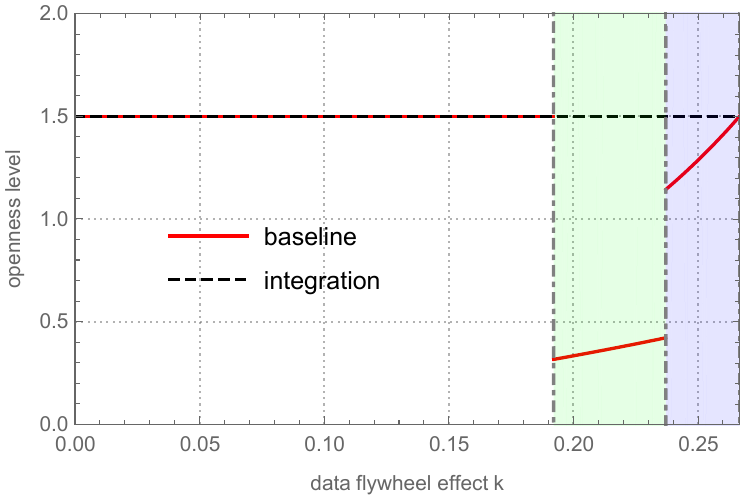}
       \caption{openness level $\eta_1$}\label{v-f-openness}
    \end{subfigure}
    ~~~
    \begin{subfigure}[b]{0.455\textwidth}
    \centering
        \includegraphics[width=\textwidth]{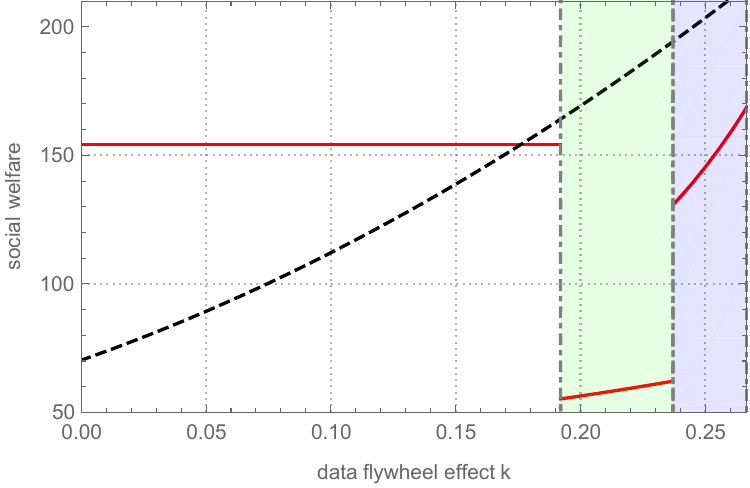}
       \caption{social welfare}\label{v-f-social welfare}
    \end{subfigure}
    	\caption{Impacts of Vertical Integration on Fine-tuning Effort, Openness Decision and Social Welfare \\($\theta=5$, $c=1$, $w_H=2.5$, $w_L=0.5$, $\bar{\eta}=1.5$)}\label{v-f-appendix}
\end{figure}

\subsection{Equilibrium Derivation under the Government Subsidy}\label{appen-free trial}
We use the per-unit subsidy and the period-by-period operating objective defined in Section \ref{subsidy}. In period 2, the deployer compares the profits available from the incumbent and entrant, taking their license fees $w_{2g}$ and $\Tilde{w}_{2g}$, openness levels $\eta_{2g}$ and $\Tilde{\eta}_{2g}$, and inherited usage $\alpha_{1g}$ as given. If it continues with the incumbent, its profit is
\begin{equation*}
\Pi_{2Ig} (Q_{2g},\alpha_{2g})= (\theta-w_{2g}+s) \alpha_{2g} - \frac{cQ_{2g}^2}{(1+k\alpha_{1g})(1+\eta_{2g})}.
\end{equation*}
The return per unit of usage is $\theta-w_{2g}+s$, after the subsidized license payment. Consumers maximize $Q_{2g} \alpha_{2g} - \alpha^2_{2g}/2$, giving $\alpha^*_{2g} = Q_{2g}$. Substituting this usage response and optimizing effort gives
$$Q^*_{2g} = \frac{(1+k \alpha_{1g})(1+\eta_{2g})(\theta-w_{2g}+s)}{2c},$$
and the resulting operating profit
$$\Pi_{2Ig} (w_{2g}, \eta_{2g}) = \frac{(1+k \alpha_{1g})(1+\eta_{2g})(\theta-w_{2g}+s)^2}{4c}.$$

If the deployer switches to the entrant, its profit is
\begin{equation*}
\Pi_{2Eg} (\Tilde{Q}_{2g},\Tilde{\alpha}_{2g})= (\theta-\Tilde{w}_{2g}+s) \Tilde{\alpha}_{2g} - \frac{c\Tilde{Q}_{2g}^2}{(1+\eta_{1g})(1+\Tilde{\eta}_{2g})}.
\end{equation*}
The factor $1+\eta_{1g}$ captures the public-reuse benefit for adapting the entrant's model. The same optimization gives
$$\Tilde{Q}^*_{2g} = \frac{(1+ \eta_{1g})(1+\Tilde{\eta}_{2g})(\theta-\Tilde{w}_{2g}+s)}{2c},$$
and operating profit
$$\Pi_{2Eg} (\Tilde{w}_{2g}, \Tilde{\eta}_{2g}) = \frac{(1+ \eta_{1g})(1+\Tilde{\eta}_{2g})(\theta-\Tilde{w}_{2g}+s)^2}{4c}.$$
The deployer selects the model that yields the larger of $\Pi_{2Ig} (w_{2g}, \eta_{2g})$ and $\Pi_{2Eg} (\Tilde{w}_{2g}, \Tilde{\eta}_{2g})$.

In the terminal period, greater openness lowers adaptation costs and expands usage if a developer's model is selected, without creating a subsequent competitive disadvantage. As in the baseline, we consider the equilibrium with $\eta_{2g} = \Tilde{\eta}_{2g} = \bar{\eta}$.
The entrant's license fee is fixed at $\Tilde{w}_{2g}=w_L$. We maintain the strict competition bound
$$k < \min\left\{\frac{2c\bar{\eta}}{(1+\bar{\eta})(\theta-w_L+s)},\frac{2c(2\theta+2s-w_H-w_L)(w_H-w_L)}{(\theta-w_H+s)^2(\theta-w_L+s)}\right\}.$$
The first component ensures that full first-period openness prevents retention even at the low first-period fee; the second prevents retention at the high second-period fee.

Within this range, we therefore consider $w_{2g}=w_L$. The deployer retains the incumbent when $\Pi_{2Ig}^* \geq \Pi_{2Eg}^*$, with ties resolved in the incumbent's favor. Substituting $\alpha_{1g}=Q_{1g}=(1+\eta_{1g})(\theta-w_{1g}+s)/(2c)$ yields the retention condition
\begin{equation}\label{thred_raw_subsidy}
\frac{2c(1+\eta_{1g})}{2c+k(1+\eta_{1g})(\theta-w_{1g}+s)} \leq 1.
\end{equation}

The incumbent now chooses its first-period license fee and openness, anticipating the deployer's period-2 model choice. Equation \ref{thred_raw_subsidy} is equivalent to $kQ_{1g}\geq\eta_{1g}$: model-specific learning must at least match the public-reuse benefit available when adapting the entrant. At a fixed fee, greater openness raises both current usage and public reuse. The maintained bound implies $k(\theta-w_{1g}+s)/(2c)<1$, so public reuse grows faster than the incumbent-specific learning benefit as openness rises. A higher first-period fee instead lowers effort and usage, reducing the experience accumulated with the incumbent's model. The next lemma states the resulting retention thresholds and how openness affects the incumbent's profit within each period-2 outcome.

\begin{lemma}\label{f-l-etabar}
\begin{enumerate}[$(a)$]
    \item \textbf{Retention thresholds.} Define $\bar{\eta}_{Hg} = \frac{k(\theta+s-w_H)}{2c-k(\theta+s-w_H)}$ and $\bar{\eta}_{Lg} = \frac{k(\theta+s-w_L)}{2c-k(\theta+s-w_L)}$.
    If the first-period fee is $w_H$ (respectively, $w_L$), the incumbent retains the deployer in period 2 if and only if $\eta_{1g}\leq\bar{\eta}_{Hg}$ (respectively, $\eta_{1g}\leq\bar{\eta}_{Lg}$).
    The thresholds satisfy $\bar{\eta}_{Hg}\leq\bar{\eta}_{Lg}$, with strict inequality when $k>0$ and $w_H>w_L$.

    \item \textbf{Profit within an outcome.} For a fixed first-period fee and a fixed period-2 model choice, the incumbent's two-period profit is weakly increasing in $\eta_{1g}$.
\end{enumerate}
\end{lemma}

For each first-period fee, Lemma \ref{f-l-etabar} makes the largest openness consistent with retention the relevant retaining candidate. If the incumbent instead gives up period-2 deployment, full openness maximizes current usage and the high fee maximizes current revenue within the maintained price menu. The first-period problem therefore reduces to comparing Harvest ($S_{0g}$), Defend ($S_{1g}$), and Dominate ($S_{2g}$).

\begin{proposition}\label{f-p-equilibrium-k}
Under the maintained competition bound above, there exist thresholds $\bar{k}_{1g}$ and $\bar{k}_{2g}$ determining the incumbent's first-period strategy.
In the equilibrium considered, both developers offer $w_{2g}=\Tilde{w}_{2g}=w_L$ and $\eta_{2g}=\Tilde{\eta}_{2g}=\bar{\eta}$ in period 2.
Let $Q_{2g}^*$ denote effort on the model selected by the deployer in period 2.
The equilibrium outcomes are:

\begin{enumerate}[$(a)$]
    \item If $k \leq \bar{k}_{1g}$, the incumbent adopts a \textbf{Harvest} strategy. The deployer selects the entrant in period 2, and the equilibrium outcomes are:
    \begin{equation*}
    \left(w_{1g}^*,\eta_{1g}^*,Q_{1g}^*,Q_{2g}^*\right) = \left(w_H,\bar{\eta},\frac{(1+\bar{\eta})(\theta-w_H+s)}{2c},\frac{(1+\bar{\eta})^2(\theta-w_L+s)}{2c} \right).
    \end{equation*}

    \item If $\bar{k}_{1g} < k \leq \bar{k}_{2g}$, the incumbent adopts a \textbf{Defend} strategy. The deployer selects the incumbent in period 2, and the equilibrium outcomes are:
    \begin{equation*}
    \left(w_{1g}^*,\eta_{1g}^*,Q_{1g}^*,Q_{2g}^*\right) = \left(w_H,\bar{\eta}_{Hg},\frac{\theta-w_H+s}{2c-k(\theta-w_H+s)},\frac{(1+\bar{\eta})(\theta-w_L+s)}{2c-k(\theta-w_H+s)} \right).
    \end{equation*}

    \item If $k>\bar{k}_{2g}$, the incumbent adopts a \textbf{Dominate} strategy. The deployer selects the incumbent in period 2, and the equilibrium outcomes are:
    \begin{equation*}
    \left(w_{1g}^*,\eta_{1g}^*,Q_{1g}^*,Q_{2g}^*\right) = \left(w_L,\bar{\eta}_{Lg},\frac{\theta-w_L+s}{2c-k(\theta-w_L+s)},\frac{(1+\bar{\eta})(\theta-w_L+s)}{2c-k(\theta-w_L+s)} \right).
    \end{equation*}

\end{enumerate}
Some regions may be empty within the maintained range; the threshold formulas are given in the proof.
\end{proposition}

Figure \ref{f-f-appendix} compares private payoffs and social welfare with and without the subsidy at the stated calibration. The welfare panel uses the convention in Section \ref{subsidy}: participant profits plus consumer surplus before deducting subsidy expenditure. The payoff comparisons in Proposition \ref{f-p-welfare comparison} additionally use its sufficient subsidy cutoff, which is separate from the competition bound imposed above.
\newpage
\begin{figure}[h!]
	\centering \footnotesize
    \begin{subfigure}[b]{0.45\textwidth}
    \centering
        \includegraphics[width=\textwidth]{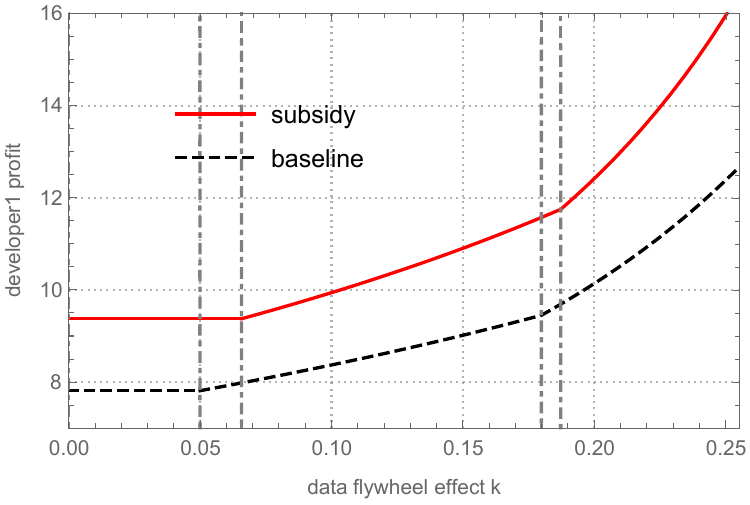}
       \caption{developer 1 profit}\label{f-f-developer1}
    \end{subfigure}
    ~~~
    \begin{subfigure}[b]{0.45\textwidth}
    \centering
        \includegraphics[width=\textwidth]{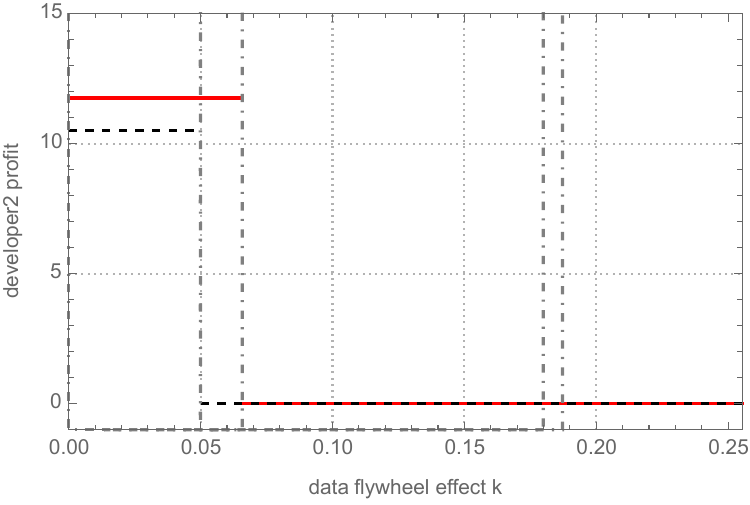}
       \caption{developer 2 profit}\label{f-f-developer2}
    \end{subfigure}

    \begin{subfigure}[b]{0.45\textwidth}
    \centering
        \includegraphics[width=\textwidth]{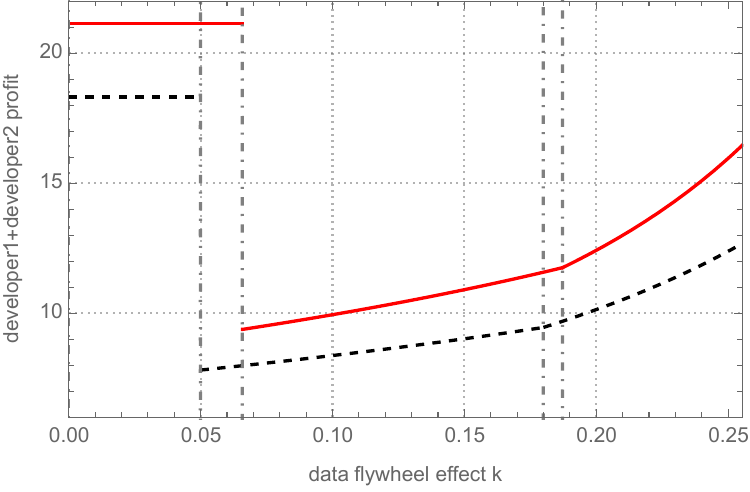}
       \caption{Both developers' combined profit}\label{f-f-developer1 and developer2}
    \end{subfigure}
    ~~~
    \begin{subfigure}[b]{0.45\textwidth}
    \centering
        \includegraphics[width=\textwidth]{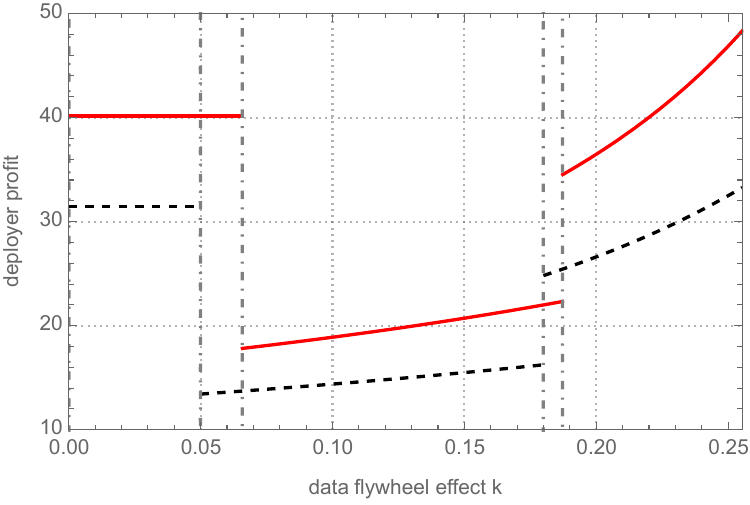}
       \caption{deployer profit}\label{f-f-deployer}
    \end{subfigure}

    \begin{subfigure}[b]{0.45\textwidth}
    \centering
        \includegraphics[width=\textwidth]{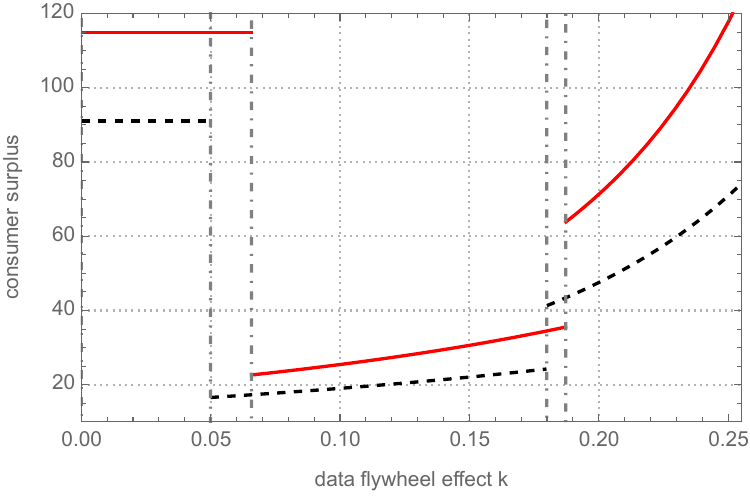}
       \caption{consumer surplus}\label{f-f-consumer surplus}
    \end{subfigure}
    ~~~
    \begin{subfigure}[b]{0.45\textwidth}
    \centering
        \includegraphics[width=\textwidth]{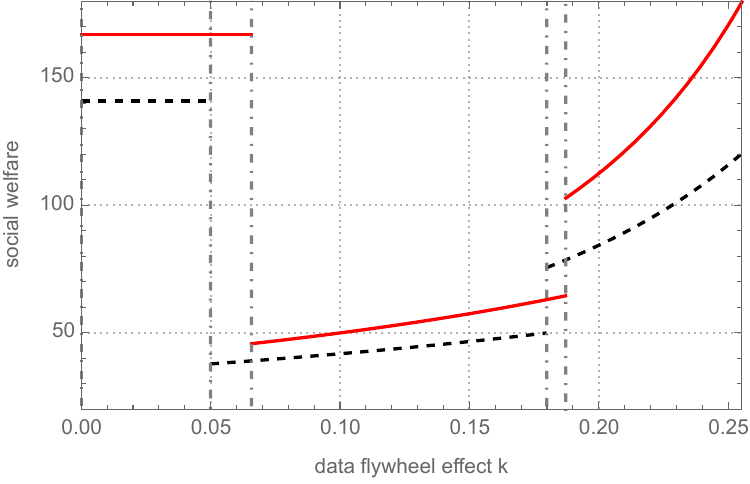}
       \caption{social welfare}\label{f-f-social welfare}
    \end{subfigure}
    	\caption{Impacts of Government Subsidy on the Developers' Profits, the Deployer's Profit, Consumer Surplus, and Social Welfare\\ ($\theta=5$, $c=1$, $w_H=2.5$, $w_L=0.8$, $\bar{\eta}=1.5$, $s=0.5$)}\label{f-f-appendix}
\end{figure}

\newpage

\section{Extensions}\label{sec:extensions}
We examine five extensions: transferable deployment experience, benefits of earlier openness when continuing with the incumbent, access to the entrant under integration, endogenous entry, and the strength of public reuse. All retain the current-period operating objective. Appendix~\ref{appendix-Partial Vertical Integration} starts from the integration benchmark of Section~\ref{vi}, and Appendix~\ref{appendix-Developer 2 Entry Cost} pairs endogenous entry with a binding posted period-2 license fee. Together, these analyses identify when the strategy and welfare conclusions persist and when they change, under the conditions stated in each subsection.

\subsection{Transferable Deployment Experience}\label{appen-data flywheel on entrant}
As discussed in the Introduction, a July 2026 coalition of model developers, cloud firms, and open-source organizations argues that open weights help organizations retain control over the knowledge and capabilities they build \citep{openweights2026letter}. Retaining access to those capabilities and being able to reuse them with another model are distinct benefits. In the baseline, greater openness lowers adaptation costs and encourages usage, through which the deployer accumulates model-specific experience. That experience can favor continued use of the incumbent even when open weights give the deployer greater control over its deployment. This extension examines how the incumbent's strategy and the welfare effects of openness change when some of that experience also remains useful after switching models.

The baseline separates experience accumulated by the deployer from publicly reusable resources. If the deployer continues with the incumbent developer 1's FM in period 2, its fine-tuning cost is $C_{2I}(Q_2) = \frac{cQ_2^2}{(1+k\alpha_1)(1+\eta_2)}$, where $1+k\alpha_1$ captures the adaptation benefit from prior engagement with that model. If it switches to the entrant developer 2's FM, its cost is $C_{2E}(Q_2) = \frac{cQ_2^2}{(1+\eta_1)(1+\tilde{\eta}_2)}$, where $1+\eta_1$ represents public resources created by the incumbent's first-period openness. The latter cost contains no benefit from the deployer's private experience in the baseline.

We now allow the deployer's experience with the incumbent's model to reduce its cost of adapting the entrant's model. If the deployer adopts developer 2's FM in period 2, its fine-tuning cost becomes $C_{2Ef}(Q_2) = \frac{cQ_2^2}{(1+\hat{k} Q_1)(1+\eta_1)(1+\tilde{\eta}_2)}$. The additional factor $1+\hat{k}Q_1$ captures the experience that remains useful after switching, alongside the public resources represented by $1+\eta_1$. Because this experience was acquired on the incumbent's model family, we assume $k \ge \hat{k}$: prior experience lowers adaptation costs at least as strongly when the deployer continues with the incumbent as when it switches to the entrant. The characterization below takes $\hat{k}>0$; setting $\hat{k}=0$ returns the baseline. For each first-period license fee, define the incumbent's retention frontier as follows:

\begin{equation*}
\left \{
\begin{aligned}
&\bar{\eta}_{Hf}  = \frac{-2c+(k-2\hat{k})(\theta-w_H)+\sqrt{[2c-k(\theta-w_H)]^2+8c\hat{k}(\theta-w_H)}}{2\hat{k}(\theta-w_H)},\\
&\bar{\eta}_{Lf}  = \frac{-2c+(k-2\hat{k})(\theta-w_L)+\sqrt{[2c-k(\theta-w_L)]^2+8c\hat{k}(\theta-w_L)}}{2\hat{k}(\theta-w_L)}.
\end{aligned}
\right.
\end{equation*}

For comparison with the baseline, we maintain the competition range $0\le k<K_{\max}$ defined in Section~\ref{benchmark}. Using backward induction as in the main paper, we obtain the following candidate characterization.

\begin{proposition}\label{ext-b-p-equilibrium-k}
Suppose $\hat{k}>0$, $k\ge\hat{k}$, and the maintained second-period pricing restriction of the baseline holds. Both developers choose maximal second-period openness and an active developer charges $w_L$ in head-to-head competition. The incumbent selects the most profitable of three candidates:
\begin{align*}
S_{0f}\;(\textbf{Harvest}):\quad
& (w_1,\eta_1)= (w_H,\bar{\eta}),\\[-1mm]
& \pi_{S_{0f}}=\frac{(1+\bar{\eta})(\theta-w_H)w_H}{2c};\\
S_{1f}\;(\textbf{Defend}):\quad
& (w_1,\eta_1)= (w_H,\bar{\eta}_{Hf}),\\[-1mm]
& \pi_{S_{1f}}=\frac{(1+\bar{\eta}_{Hf})(\theta-w_H)w_H}{2c}\\[-1mm]
& \hspace{20mm}+\frac{(1+\bar{\eta})(\theta-w_L)w_L[2c+k(1+\bar{\eta}_{Hf})(\theta-w_H)]}{4c^2};\\
S_{2f}\;(\textbf{Dominate}):\quad
& (w_1,\eta_1)= (w_L,\bar{\eta}_{Lf}),\\[-1mm]
& \pi_{S_{2f}}=\frac{(1+\bar{\eta}_{Lf})(\theta-w_L)w_L}{2c}\\[-1mm]
& \hspace{20mm}+\frac{(1+\bar{\eta})(\theta-w_L)w_L[2c+k(1+\bar{\eta}_{Lf})(\theta-w_L)]}{4c^2}.
\end{align*}
Thus the equilibrium is $\arg\max\{\pi_{S_{0f}},\pi_{S_{1f}},\pi_{S_{2f}}\}$.
Holding $\hat{k}$ fixed, the retention frontiers $\bar{\eta}_{Hf}$ and $\bar{\eta}_{Lf}$ increase with $k$.
The ordering of the profit-maximizing strategies as $k$ varies nevertheless depends on the other primitives and on $\hat{k}$.
Transferable deployment experience can shrink or eliminate the \textit{Dominate} region; the baseline three-regime ordering is recovered as $\hat{k}\to0$ whenever the baseline crossings are strict.
\end{proposition}

At given first-period choices, a larger $\hat{k}$ lowers the deployer's cost of adapting the entrant's model and improves its return from switching. The incumbent must therefore limit openness more tightly to retain the deployer, which can change which candidate is most profitable. Figure~\ref{ext-b-f-equilibrium} illustrates the resulting strategies and payoffs. At its displayed parameters and over the maintained range $k<K_{\max}\approx0.2667$, the equilibrium moves from \textit{Harvest} to \textit{Defend}; \textit{Dominate} does not arise.

To quantify this sensitivity, hold the other primitives at the values in Figure~\ref{ext-b-f-equilibrium} and vary $\hat{k}$ over the maintained range $\hat{k}\le k<K_{\max}\approx0.2667$. Numerical comparison of the candidates yields a nonempty \textit{Dominate} interval only for $\hat{k}$ below approximately 0.0180, and a welfare-loss region only for $\hat{k}$ below approximately 0.00481. Thus, at this calibration, the welfare reversal disappears before the \textit{Dominate} strategy does as more deployment experience transfers across models. These cutoffs are values of the learning coefficient $\hat{k}$, not estimated shares of knowledge that transfer.

Define social welfare as the sum of both developers' profits, the deployer's profit, and consumer surplus, and let a full-openness mandate fix $\eta_1=\bar{\eta}$.
\begin{proposition}\label{ext-b-p-welfare}
Suppose the conditions of Proposition~\ref{ext-b-p-equilibrium-k} hold and full openness lies above both of the incumbent's retention frontiers.
The mandate induces \textit{Harvest}, $S_{0f}$, whose welfare is independent of $k$ and increasing in $\hat{k}$.
Holding $\hat{k}$ fixed, welfare is strictly increasing in $k$ within any unregulated \textit{Dominate} interval.
An openness trap exists on such an interval if and only if unregulated \textit{Dominate} welfare exceeds mandated welfare for some admissible $k$.
Transferable deployment experience can therefore attenuate or eliminate the baseline welfare reversal.
\end{proposition}
For the calibration in Figure~\ref{ext-b-f-equilibrium}, the mandate is nonbinding in \textit{Harvest} and raises welfare after the equilibrium switches to \textit{Defend}, so no openness trap arises. The transferable component matters both for the incumbent's strategy and for the mandated outcome: experience accumulated in period 1 now lowers the deployer's adaptation cost even when the mandate induces a switch to the entrant. The extension therefore identifies how the baseline welfare comparison depends on the model specificity of deployment experience.

\begin{figure}[ht]
	\centering \footnotesize
    \begin{subfigure}[b]{0.464\textwidth}
    \centering
        \includegraphics[width=\textwidth]{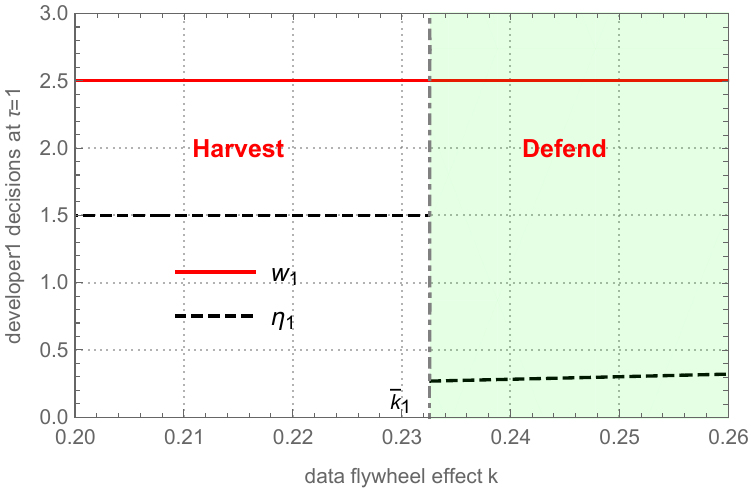}
       \caption{License fee $w_1$ and openness $\eta_1$}\label{ext-b-f-w1 and eta1}
    \end{subfigure}
    ~~~
    \begin{subfigure}[b]{0.458\textwidth}
    \centering
        \includegraphics[width=\textwidth]{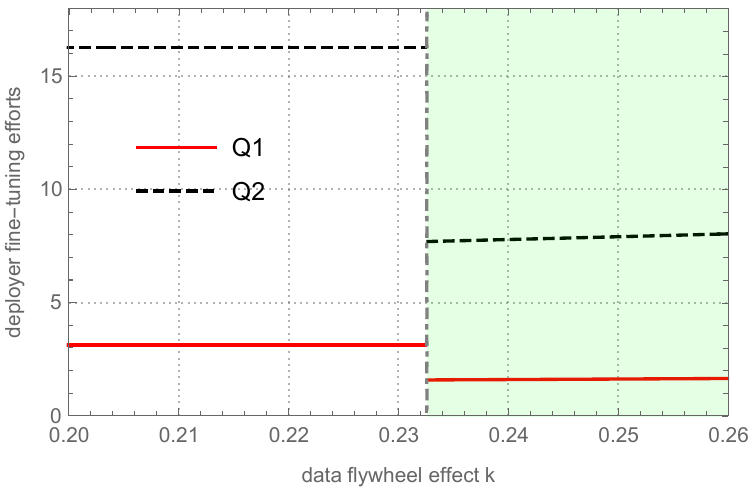}
       \caption{Deployer effort: $Q_1$ and $Q_2$}\label{ext-b-f-q1 and q2}
    \end{subfigure}
    ~~~
    \begin{subfigure}[b]{0.495\textwidth}
    \centering
        \includegraphics[width=\textwidth]{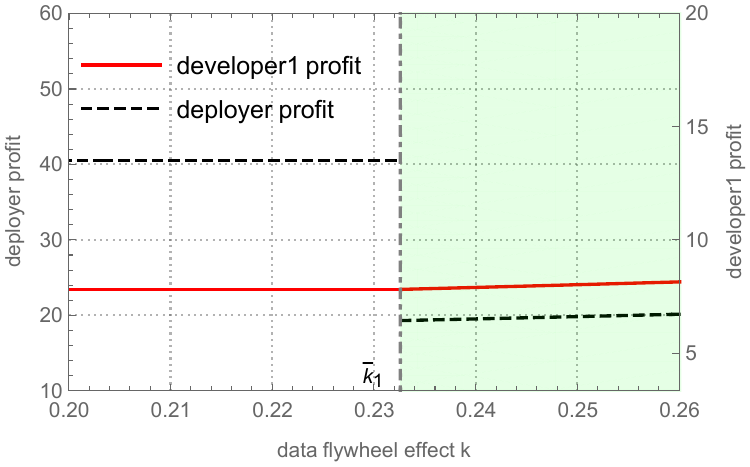}
       \caption{Incumbent and deployer profits}\label{ext-b-f-developer1 and deployer}
    \end{subfigure}
    ~~~
    \begin{subfigure}[b]{0.435\textwidth}
    \centering
        \includegraphics[width=\textwidth]{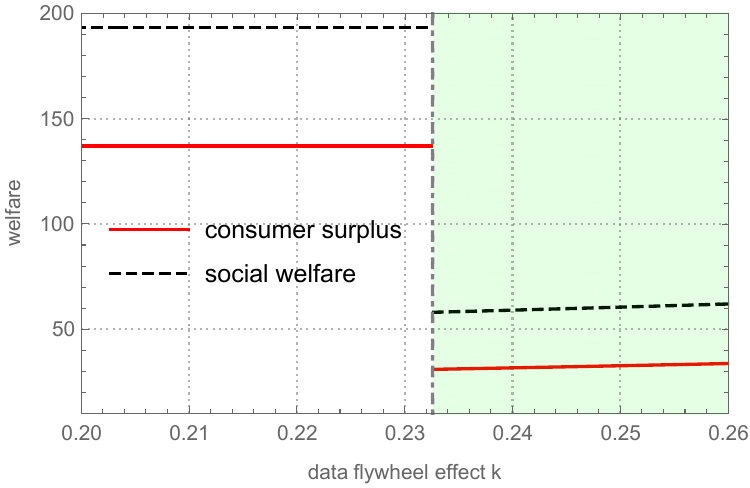}
       \caption{Consumer surplus and social welfare}\label{ext-b-f-consumer and social welfare}
    \end{subfigure}
    	\caption{Equilibrium Choices, Profits, and Welfare with Transferable Deployment Experience\\ ($\theta=5$, $c=1$, $w_H=2.5$, $w_L=0.5$, $\bar{\eta}=1.5$, $\hat{k}=0.05$, $k\le0.26$)}\label{ext-b-f-equilibrium}
\end{figure}

\subsection{Openness and the Cost of Continuing with the Incumbent}\label{appen-knowledge spillover on incumbent}
We extend the baseline by allowing the incumbent's first-period openness to lower the deployer's subsequent cost of adapting the incumbent's model. Public documentation, tooling, and modularity can remain useful in later adaptation, beyond the experience accumulated through first-period usage. In the baseline, the cost of adapting the entrant's FM is $C_{2E}(Q_2) = \frac{cQ_2^2}{(1+\eta_1)(1+\tilde{\eta}_2)}$, so the public resources represented by $1+\eta_1$ remain useful in period 2. The cost of continuing with the incumbent's FM is $C_{2I}(Q_2) = \frac{cQ_2^2}{(1+k\alpha_1)(1+\eta_2)}$. This cost captures private experience through $1+k\alpha_1$ but contains no separate persistent benefit from first-period openness.

We extend the incumbent-model cost to $C_{2Is}(Q_2) = \frac{cQ_2^2}{(1+k Q_1)(1+\beta\eta_1)(1+\eta_2)}$. The additional term $1+\beta\eta_1$ represents persistent benefits from public documentation, tooling, and modularity when the deployer continues adapting the incumbent's model.
These benefits arise from earlier openness and are distinct from the private learning captured by $1+kQ_1$. We study the strict parameter region
\[
0<\beta<\bar{\beta}(k)\equiv\frac{2c\bar{\eta}-k(1+\bar{\eta})(\theta-w_L)}{\bar{\eta}\,[2c+k(1+\bar{\eta})(\theta-w_L)]},
\]
which is nonempty only when the numerator of $\bar{\beta}(k)$ is positive. Within this region, the deployer prefers the entrant at full first-period openness under either first-period license fee. At $\beta=\bar{\beta}(k)$, the deployer is indifferent when the incumbent chose the low first-period fee and stays with the incumbent under the tie-breaking rule. A larger $\beta$ can therefore permit retention at full openness. Define the incumbent's retention frontiers as follows:

\begin{equation*}
\left \{
\begin{aligned}
&\bar{\eta}_{Hs}  = \frac{2(1-\beta)c-(1+\beta)k(\theta-w_H)-\sqrt{(1-\beta)^2[2c-k(\theta-w_H)]^2-8\beta(1-\beta)ck(\theta-w_H)}}{2\beta k(\theta-w_H)},\\
&\bar{\eta}_{Ls}  = \frac{2(1-\beta)c-(1+\beta)k(\theta-w_L)-\sqrt{(1-\beta)^2[2c-k(\theta-w_L)]^2-8\beta(1-\beta)ck(\theta-w_L)}}{2\beta k(\theta-w_L)}.
\end{aligned}
\right.
\end{equation*}

The displayed expressions apply for $k>0$; at $k=0$, both frontiers are zero, as obtained by continuity. Using backward induction as in the main paper, we obtain the following candidate characterization.

\begin{proposition}\label{ext2-b-p-equilibrium}
Suppose $0<\beta<\bar{\beta}(k)$ and the maintained second-period pricing restriction holds. Both developers choose maximal second-period openness and an active developer charges $w_L$ in head-to-head competition. The incumbent selects the most profitable of:
\begin{align*}
S_0\;(\textbf{Harvest}):\quad
& (w_1,\eta_1)=(w_H,\bar{\eta}),&
\pi_{S_0}&=\frac{(1+\bar{\eta})(\theta-w_H)w_H}{2c},\\
S_1\;(\textbf{Defend}):\quad
& (w_1,\eta_1)=(w_H,\bar{\eta}_{Hs}),&
\pi_{S_1}&=\frac{(1+\bar{\eta}_{Hs})[(\theta-w_H)w_H+(1+\bar{\eta})(\theta-w_L)w_L]}{2c},\\
S_2\;(\textbf{Dominate}):\quad
& (w_1,\eta_1)=(w_L,\bar{\eta}_{Ls}),&
\pi_{S_2}&=\frac{(1+\bar{\eta}_{Ls})(2+\bar{\eta})(\theta-w_L)w_L}{2c}.
\end{align*}
Thus the equilibrium is $\arg\max\{\pi_{S_0},\pi_{S_1},\pi_{S_2}\}$.
Holding $\beta$ fixed, both retention frontiers increase with $k$.
A sequence from \textit{Harvest} to \textit{Defend} to \textit{Dominate} obtains when the candidate-profit crossings exist, are unique, and are ordered.
These conditions hold for the displayed calibration in Figure~\ref{ext2-b-f-equilibrium}; the sequence is not asserted outside that parameter region.
\end{proposition}

\begin{figure}[ht]
	\centering \footnotesize
    \begin{subfigure}[b]{0.464\textwidth}
    \centering
        \includegraphics[width=\textwidth]{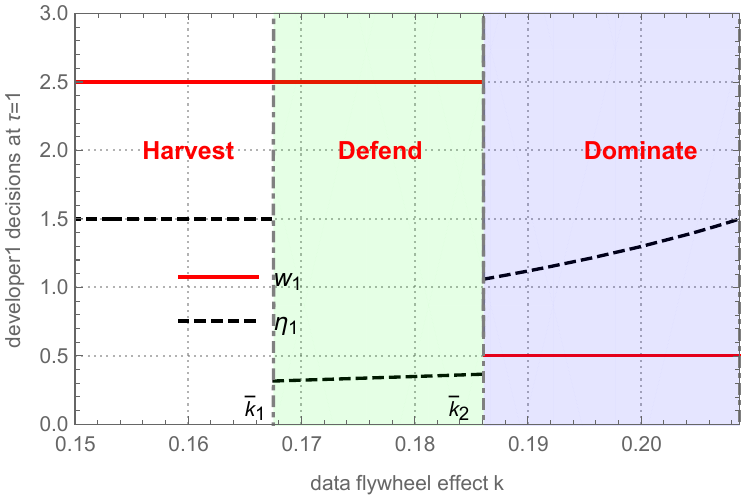}
       \caption{License fee $w_1$ and openness $\eta_1$}\label{ext2-b-f-w1 and eta1}
    \end{subfigure}
    ~~~
    \begin{subfigure}[b]{0.458\textwidth}
    \centering
        \includegraphics[width=\textwidth]{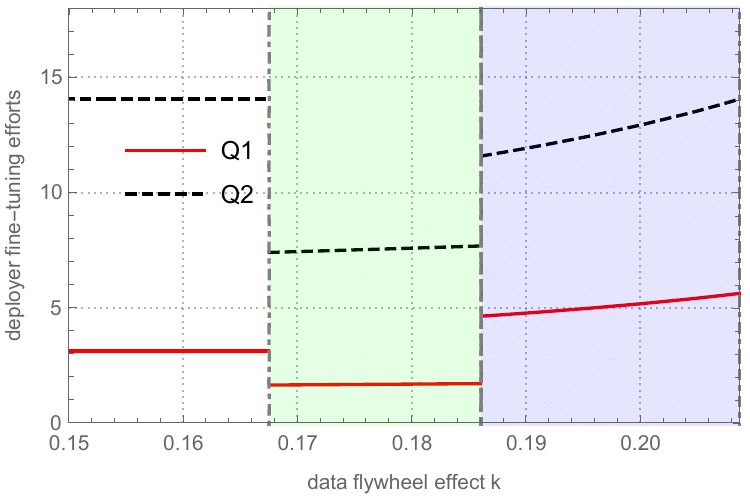}
       \caption{Deployer effort: $Q_1$ and $Q_2$}\label{ext2-b-f-q1 and q2}
    \end{subfigure}
    ~~~
    \begin{subfigure}[b]{0.495\textwidth}
    \centering
        \includegraphics[width=\textwidth]{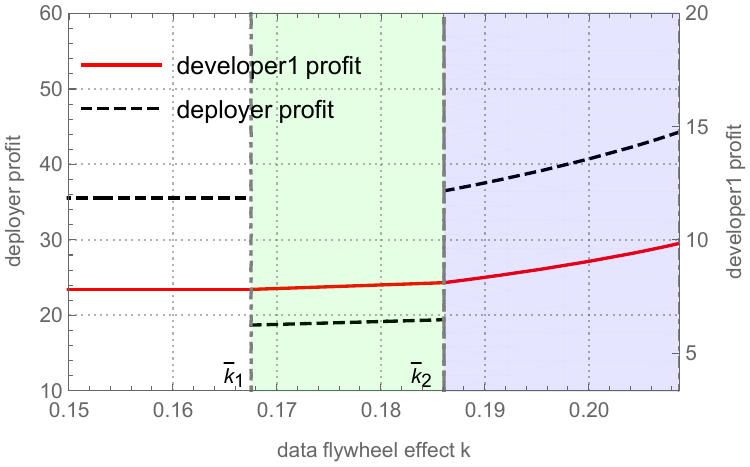}
       \caption{Incumbent and deployer profits}\label{ext2-b-f-developer1 and deployer}
    \end{subfigure}
    ~~~
    \begin{subfigure}[b]{0.435\textwidth}
    \centering
        \includegraphics[width=\textwidth]{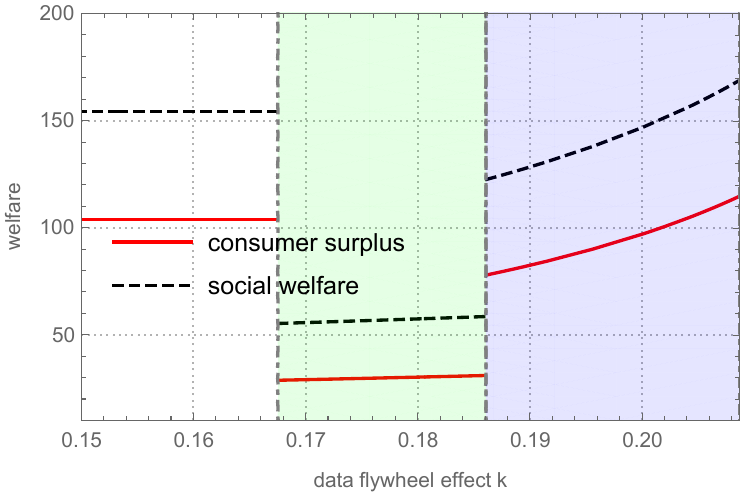}
       \caption{Consumer surplus and social welfare}\label{ext2-b-f-consumer and social welfare}
    \end{subfigure}
    	\caption{Equilibrium and Welfare with Benefits of Earlier Openness on the Incumbent's Model\\ ($\theta=5$, $c=1$, $w_H=2.5$, $w_L=0.5$, $\bar{\eta}=1.5$, $\beta=0.1$)}\label{ext2-b-f-equilibrium}
\end{figure}

At a fixed $k>0$ within the strict admissible region, a larger $\beta$ lowers the deployer's continuation cost at any positive first-period openness. The incumbent can consequently sustain greater openness while retaining the deployer under either first-period license fee. For the winning condition $\Phi$ defined in the proof of Proposition~\ref{ext2-b-p-equilibrium}, $\partial\Phi/\partial\beta=\eta_1[2c+k(1+\eta_1)(\theta-w_j)]>0$ and $\partial\Phi/\partial\eta_1<0$ at the smaller root, so both $\bar{\eta}_{Hs}$ and $\bar{\eta}_{Ls}$ increase with $\beta$. This raises openness within a given \textit{Defend} or \textit{Dominate} regime; the equilibrium effect also depends on whether the incumbent changes strategies.

Figure~\ref{ext2-b-f-equilibrium} illustrates the equilibrium choices and payoffs. For this calibration, the strict admissible range defined by $\beta<\bar{\beta}(k)$ has an upper bound of approximately 0.2087. The equilibrium switches from \textit{Harvest} to \textit{Defend} at $\bar{k}_1\approx0.1675$ and from \textit{Defend} to \textit{Dominate} at $\bar{k}_2\approx0.1861$, preserving the baseline non-monotonicity of openness. Both thresholds lie below their baseline values of 0.1920 and 0.2370.

We next ask when this persistence preserves the openness-trap welfare reversal, using the same social-welfare definition and full-openness mandate as above.
\begin{proposition}\label{ext2-b-p-welfare}
Suppose $0<\beta<\bar{\beta}(k)$ and the unregulated equilibrium is \textit{Dominate} on an interval $\mathcal{K}_D$. The mandate induces \textit{Harvest}, whose welfare $W_{S_0}$ is independent of $k$, whereas unregulated Dominate welfare $W_{S_2}(k)$ is strictly increasing on $\mathcal{K}_D$. If $W_{S_2}(k)$ is below $W_{S_0}$ at the lower endpoint of $\mathcal{K}_D$ and above it at some admissible $k\in\mathcal{K}_D$, there is a unique crossing $\bar{k}_s$ on that interval, and mandating full openness reduces welfare exactly for $k>\bar{k}_s$ within $\mathcal{K}_D$.
\end{proposition}
Persistent openness benefits lower the deployer's cost of continuing with the incumbent and can support greater openness within a retention strategy. They do not by themselves guarantee that a full-openness mandate reduces welfare. Proposition~\ref{ext2-b-p-welfare} states the sufficient crossing conditions; they hold for the calibration in Figure~\ref{ext2-b-f-equilibrium}, where \textit{Dominate} welfare crosses mandated \textit{Harvest} welfare $W_{S_0}\approx154.15$ at $\bar{k}_s\approx0.2033$. The mandate lowers welfare between this crossing and the admissible upper bound $k\approx0.2087$. At the excluded boundary $\beta=\bar{\beta}(k)$, full openness permits retention under the low first-period license fee, so the mandate no longer forces \textit{Harvest}.

\subsection{Partial Vertical Integration}\label{appendix-Partial Vertical Integration}
The full-integration benchmark ties all deployment to the incumbent and removes the internal license fee. We retain common ownership of developer 1 and the deployer but allow some operations to use the entrant in period 2. As in Section \ref{vi}, the operating unit chooses effort to maximize current-period profit, while the owner chooses openness to maximize two-period profit, anticipating the effort response.

A fixed share $\rho\in[0,1]$ of the deployer's operations must remain with the incumbent in period 2; the remaining share $1-\rho$ can use either developer's model. Operations using the incumbent pay no internal license fee, while those using the entrant pay its license fee $w_L$.
We refer to continued use of the incumbent by all operations as full integration, or complete foreclosure, and to use of the entrant by the switchable share as partial integration. At $\rho=1$, all operations are tied to the incumbent; at $\rho=0$, all can switch, although common ownership still removes the internal license fee when the incumbent is used. Each share serves the corresponding share of a unit mass of otherwise identical users. The quantities $Q_{2p}$ and $\tilde{Q}_{2p}$ are effort and engagement intensities on the incumbent and entrant branches, respectively; aggregate effort and profit weight these intensities and the associated profits by the shares using each model. Consumer surplus is summed across the same user groups, so it equals $[\rho(Q_{2p})^2+(1-\rho)(\tilde{Q}_{2p})^2]/2$ in period 2 when the switchable share uses the entrant.

We solve backward from period 2, taking first-period effort $Q_{1p}$ as given. If all operations remain with the incumbent, the integrated firm's period-2 profit is
\begin{equation*}
    \pi_{2Ip}(\eta_{2p}, Q_{2p})=\theta Q_{2p}-\frac{cQ_{2p}^2}{(1+kQ_{1p})(1+ \eta_{2p})},
\end{equation*}
where $1+kQ_{1p}$ captures the learning generated by first-period deployment. Greater openness lowers adaptation costs, so the owner chooses $\eta_{2p}=\bar{\eta}$.
The operating unit's effort response is
\begin{equation*}
    Q_{2p}=\frac{(1+kQ_{1p})(1+\bar{\eta}) \theta}{2c}.
\end{equation*}
If the switchable share instead uses the entrant, the tied share $\rho$ remains with the incumbent and the remaining share pays $w_L$ per unit of engagement to developer 2. The integrated firm's period-2 profit, which combines developer 1's and the deployer's profits, is then
\begin{equation*}
    \pi_{2Ep}(\tilde{\eta}_{2p}, \tilde{Q}_{2p})=\rho \pi_{2Ip}(\eta_{2p}, Q_{2p}) +(1-\rho) \left[ (\theta-w_L) \tilde{Q}_{2p}-\frac{c\tilde{Q}_{2p}^2}{(1+\eta_{1p})(1+ \tilde{\eta}_{2p})} \right],
\end{equation*}
For $0\le\rho<1$, greater entrant openness lowers adaptation costs on its branch and raises the effort response. This increases developer 2's license revenue whenever that branch is used and $w_L>0$, so the entrant chooses $\tilde{\eta}_{2p}=\bar{\eta}$; full openness is also optimal when $w_L=0$. At $\rho=1$, the entrant branch carries no operations and its openness does not affect the allocation. For $0\le\rho<1$, the operating unit's effort response on the entrant branch is
\begin{equation*}
    \tilde{Q}_{2p}=\frac{(1+\eta_{1p})(1+\bar{\eta}) (\theta-w_L)}{2c}.
\end{equation*}

For $0\le\rho<1$, the firm keeps all operations with the incumbent when $\pi_{2Ip}^*\geq\pi_{2Ep}^*$. Substituting the period-2 responses and the first-period operating response $Q_{1p}=(1+\eta_{1p})\theta/(2c)$, and cancelling the positive factor $1-\rho$, gives
\begin{equation}
\frac{2c(1+\eta_{1p})}{2c+k(1+\eta_{1p})\theta} \leq \left( \frac{\theta}{\theta-w_L} \right)^2.
\end{equation}

The next lemma identifies when the incumbent retains all operations and how first-period openness affects the integrated firm's profit.

\begin{lemma}\label{ext2-b-l-etabar}
\begin{enumerate}[$(a)$]
\item \textbf{Retention of all operations.} For $0\le\rho<1$, define $\hat{k}_p \equiv 2c(\theta-w_L)^2/\theta^3$.
If $k\ge\hat{k}_p$, all operations remain with the incumbent for every $\eta_{1p}\in[0,\bar{\eta}]$.
If $k<\hat{k}_p$, define $\bar{\eta}_p = \min \{ \bar{\eta},\frac{2c\theta^2 -2c(\theta-w_L)^2+k\theta^3}{2c(\theta-w_L)^2 - k\theta^3}\} \ge 0$.
All operations remain with the incumbent if and only if $\eta_{1p}\le\bar{\eta}_p$.
At $\rho=1$, the entrant is foreclosed for every $\eta_{1p}$ and $k$.

\item \textbf{Return to openness.} Conditional on either second-period allocation, the integrated firm's two-period profit is monotonically increasing in first-period openness $\eta_{1p}$.
\end{enumerate}
\end{lemma}

It remains to choose first-period openness. The operating unit takes $\eta_{1p}$ as given and maximizes current-period profit:
\begin{equation*}
\pi_{1p}(\eta_{1p},Q_{1p})=\theta Q_{1p}-\frac{cQ_{1p}^2}{1+\eta_{1p}}.
\end{equation*}
The owner chooses openness to maximize two-period profit, anticipating this effort response and the subsequent model allocation. The next proposition characterizes the resulting equilibrium.

\begin{proposition}\label{ext2-v-p-equilibrium}

Under partial vertical integration with the maintained operating rule, the owner chooses first-period openness $\eta_{1p}=\bar{\eta}$. In period 2, the owner sets $\eta_{2p}=\bar{\eta}$ for the incumbent and the entrant sets $\tilde{\eta}_{2p}=\bar{\eta}$ for its own model; choices on unused branches do not affect the allocation. For $0\le\rho<1$, there exists a threshold $\bar{k}_p$ such that all operations remain with the incumbent if and only if $k\geq\bar{k}_p$. At $\rho=1$, all operations remain with the incumbent for every $k$. For $0\le\rho<1$, the operating unit's effort intensities are
\begin{equation*}
		(Q^*_{1p}, Q^*_{2p}, \tilde{Q}^*_{2p})=\left \{
		\begin{aligned}
 		&\left(\frac{(1+\bar{\eta})\theta}{2c}, \frac{(1+\bar{\eta})[2c+k\theta(1+\bar{\eta})]\theta}{4c^2}, N.A. \right) &&\text{if}~~k \geq \bar{k}_p,\\
		&\left(\frac{(1+\bar{\eta})\theta}{2c}, \frac{(1+\bar{\eta})[2c+k\theta(1+\bar{\eta})]\theta}{4c^2},\frac{(1+\bar{\eta})^2 (\theta-w_L)}{2c} \right) &&\text{if} ~~k < \bar{k}_p;
		\end{aligned}
		\right.
\end{equation*}

\end{proposition}

\begin{figure}[ht]
	\centering
    \begin{subfigure}[b]{0.55\textwidth}
    \centering
        \includegraphics[width=\textwidth]{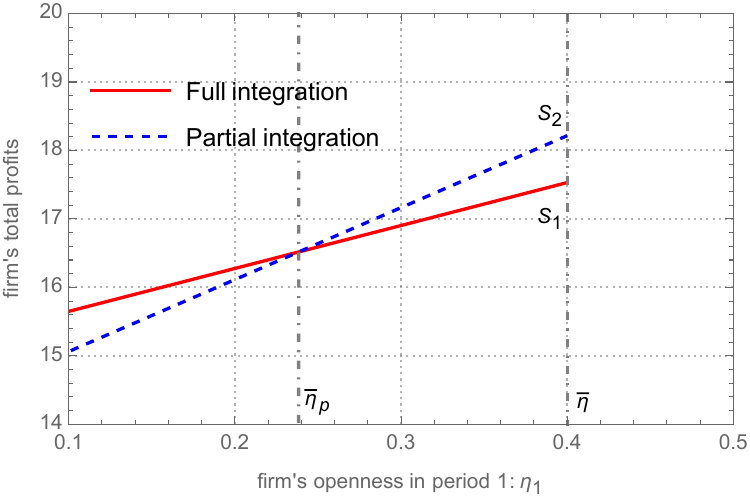}
    \end{subfigure}~~~
\caption{Joint Profit under Complete Foreclosure and Partial Integration\\ ($\theta=5$, $c=1$, $w_H=2.5$, $w_L=0.5$, $\bar{\eta}=0.4$, $\rho=0.4$, $k=0.001$)}\label{ext2-b-f-etahbar and etalbar}
\end{figure}

Greater first-period openness raises current operating profit and weakly increases continuation profit on either allocation branch. At $\bar{\eta}_p$, whenever this threshold lies below $\bar{\eta}$, the two continuation profits coincide, so allowing the switchable share to use the entrant creates no downward jump in the firm's profit. The owner therefore chooses full openness even when the switchable share uses the entrant under partial integration. The distinction from decentralization is whose profit the owner values: a separate incumbent loses period-2 license revenue when the deployer switches, whereas the integrated owner continues to receive the deployer's operating profit, including the adaptation gains from using the entrant. At the calibration in Figure \ref{ext2-b-f-etahbar and etalbar}, the partial-integration candidate $S_2$ consequently yields higher profit than the complete-foreclosure candidate $S_1$ at full openness.

For $0\le\rho<1$, $\bar{k}_p$ is the learning strength at which continued use of the incumbent becomes optimal for all operations at full openness. Below that threshold, the share $1-\rho$ uses the entrant; at and above it, stronger model-specific learning sustains complete foreclosure. If $\bar{k}_p\le0$, all admissible values of $k$ yield complete foreclosure. At $\rho=1$, the allocation remains fully tied to the incumbent for every $k$.

\noindent\textbf{Welfare under partial integration.}
For $0\le\rho<1$, second-period model choice compares operating profit per unit of operations, so the common factor $1-\rho$ cancels. The owner chooses full openness for every such $\rho$, and the effort intensities and switching threshold $\bar{k}_p$ are also independent of $\rho$; the shares using each model determine aggregate effort and welfare. Let $W_v$ denote welfare under full integration (Proposition~\ref{v-p-equilibrium}). For $k<\bar{k}_p$, first-period outcomes coincide with full integration, and summing welfare across the two user groups gives
\[
W_p(\rho)=W_v+(1-\rho)\Delta(k),\qquad
\Delta(k)=\frac{(\theta+w_L)\tilde{Q}^*_{2p}+(\tilde{Q}^*_{2p})^2}{2}-\frac{\theta Q^*_{2p}+(Q^*_{2p})^2}{2}>0.
\]
The firm switches operations only when the entrant branch yields strictly higher operating profit. At the operating optimum, profit is $(\theta-w)Q/2$, so switching implies $\tilde{Q}^*_{2p}>\theta Q^*_{2p}/(\theta-w_L)\ge Q^*_{2p}$. The switched operations therefore generate both higher profit for the integrated firm and higher consumer surplus, while developer 2 gains license revenue when $w_L>0$. Relative to complete foreclosure, the welfare gain is proportional to the share $1-\rho$ that can switch. Against the decentralized benchmark in Proposition~\ref{v-p-profit comparison}, this improvement reduces any welfare loss from integration and increases any welfare gain.

For $k\ge\bar{k}_p$, all operations remain with the incumbent and welfare coincides with full integration; the same is true for every $k$ when $\rho=1$. For $0\le\rho<1$, if $0<\bar{k}_p<K_{\max}$ and $w_L>0$, welfare falls by $(1-\rho)\Delta(\bar{k}_p)>0$ at the switch to complete foreclosure. Here $\Delta(\bar{k}_p)$ is evaluated using the two continuation branches at the threshold, where the firm is indifferent and retains the incumbent. This private indifference does not imply equal welfare: the switch reduces effort and consumer surplus on the previously switchable share and eliminates the entrant's license revenue. The integrated firm does not internalize the loss of consumer surplus or the entrant's license revenue. The welfare ranking relative to decentralization therefore need not retain the single-threshold form of the full-integration comparison.

\subsection{Endogenous Entry}\label{appendix-Developer 2 Entry Cost}
In the baseline, developer~2 is available at the beginning of period~2 and has no separate participation decision. Here, it must pay a fixed cost $e$ to enter after observing the incumbent's binding posted period-2 license fee. Entry therefore requires both that developer~2 would win the deployer and that the resulting operating profit would justify the cost. The model-choice comparison is unchanged: when both developers charge the low period-2 fee, the incumbent retains the deployer if and only if
\begin{equation*}
\frac{2c(1+\eta_1)}{2c+k(1+\eta_1)(\theta-w_1)} \leq 1.
\end{equation*}
Lemma~\ref{b-l-etabar} expresses this retention condition as an upper bound on first-period openness. Greater $\eta_1$ makes the entrant more attractive, while the low first-period fee $w_L$ increases usage and the incumbent-specific learning that favors retention.

\noindent\textbf{Timing, information, and commitment.} In period~1, the incumbent chooses $(w_1,\eta_1)$, the deployer chooses $Q_1$, and user engagement is $\alpha_1=Q_1$, as in the baseline. At the beginning of period~2, the incumbent publicly posts $w_2\in\{w_H,w_L\}$, which is contractually fixed for that period. Developer~2 observes $(w_1,\eta_1,Q_1,w_2)$ and decides whether to pay $e>0$ and enter. If it enters, it charges the fixed fee $\tilde{w}_2=w_L$; both active developers choose maximal second-period openness, $\eta_2=\tilde{\eta}_2=\bar{\eta}$. The deployer selects the available model that delivers the higher second-period operating profit and chooses $Q_2$. An entrant exactly indifferent between entering and staying out stays out; as in the baseline, a deployer exactly indifferent between the two models stays with the incumbent. All other primitives and the maintained competition restrictions are unchanged, so $w_2=w_H$ cannot retain the deployer against an active entrant.

The commitment concerns the period-2 fee after period-1 deployment has occurred; it requires no period-1 promise about a future fee. A public API or marketplace tariff, an enterprise licensing agreement, or another posted rate fixed over a contracting or billing period could provide this commitment. It makes $w_2=w_L$ a credible limit price: the incumbent cannot raise the fee after observing non-entry. Without that restriction, it would prefer the monopoly fee $w_H$ once the entrant stayed out, and the thresholds below would have to be re-derived. Proposition~\ref{ext4-p-entry} is conditional on this binding posted-tariff timing.

If $w_L=0$, an entrant facing $e>0$ earns no license revenue and never enters, leaving the incumbent an unconstrained two-period monopolist. The entry-cutoff formulas below consider $w_L>0$. If developer~2 wins, the deployer's optimal effort is $\tilde{Q}_{2}^* = \frac{(1+\eta_1)(1+\tilde{\eta}_2)(\theta-\tilde{w}_2)}{2c}$, with $\tilde{\eta}_2 = \bar{\eta}$ and the fixed fee $\tilde{w}_2 = w_L$. Developer~2 enters only if it would win and its gross operating profit strictly exceeds the entry cost, $w_L \tilde{Q}_{2}^* > e$. Because $w_L\tilde{Q}_2^*=\frac{w_L(1+\eta_1)(1+\bar{\eta})(\theta-w_L)}{2c}$ increases in $\eta_1$, the profitability requirement conditional on winning is equivalent to $\eta_1>\bar{\eta}_e$, where
\begin{equation*}
\bar{\eta}_e = \frac{2ce}{w_L(1+\bar{\eta})(\theta-w_L)}-1 .
\end{equation*}
The incumbent can retain the deployer in period~2 in two ways. \textit{Channel~A (openness deterrence)} sets $\eta_1\le\bar{\eta}_e$, so entry is unprofitable even at the monopoly fee $w_2=w_H$. Anticipating non-entry, the incumbent optimally posts $w_H$. \textit{Channel~B (limit pricing)} applies when winning would make entry profitable: the incumbent posts the binding low fee $w_2=w_L$, so the entrant would lose the deployer and stays out. This requires $\eta_1\le\bar{\eta}_H$ when $w_1=w_H$ or $\eta_1\le\bar{\eta}_L$ when $w_1=w_L$, and reproduces the Defend and Dominate strategies. Thus both channels deter entry, but only Channel~A permits the incumbent to charge the period-2 monopoly fee.

Channel~A yields the new \textbf{Exclude} candidate. The sufficient regularity conditions below make Exclude preferable to the feasible Defend and Dominate strategies in the intermediate-entry-cost region. For other admissible primitives, the incumbent compares the profits of the feasible candidates. Define
\begin{equation*}
\left\{
\begin{aligned}
&\bar{e}_1 = \frac{w_L(1+\bar{\eta})(\theta-w_L)}{2c}, \qquad \bar{e}_2 = (1+\bar{\eta})\,\bar{e}_1 = \frac{w_L(1+\bar{\eta})^2(\theta-w_L)}{2c},\\
&\bar{k}_{eH} = \frac{2c(\theta-w_H-w_L)}{(1+\bar{\eta})\,w_H(\theta-w_H)} .
\end{aligned}
\right.
\end{equation*}
The threshold $\bar{e}_1$ satisfies $\bar{\eta}_e=0$: below it, even zero openness cannot deter profitable entry at the high period-2 fee. The threshold $\bar{e}_2$ is the entrant's maximal attainable gross operating profit and satisfies $\bar{\eta}_e=\bar{\eta}$: at or above it, entry is never strictly profitable. The threshold $\bar{k}_{eH}$ determines when the incumbent switches its first-period fee from $w_H$ to $w_L$ within Exclude and is independent of the entry cost.

For the intermediate-cost region, define the incumbent's Exclude profit at first-period fee $w_1$ and its Dominate profit by
\begin{align*}
\pi_X(w_1)
&=\frac{(\theta-w_1)w_1(1+\bar{\eta}_e)}{2c}
+\frac{w_H(\theta-w_H)(1+\bar{\eta})[2c+k(1+\bar{\eta}_e)(\theta-w_1)]}{4c^2},\\
\pi_{\mathrm{Dominate}}
&=\frac{(2+\bar{\eta})(\theta-w_L)w_L}{2c-k(\theta-w_L)}.
\end{align*}
We call the primitives \textit{regular} if, throughout the maintained admissible range of $k$, (i) $\bar{\eta}_H\le\bar{\eta}_e$ whenever Defend is feasible and (ii) $\pi_X(w_L)\ge\pi_{\mathrm{Dominate}}$. These restrictions on the primitives and entry cost are sufficient to eliminate, respectively, a residual Defend interval and a return to Dominate in the intermediate-cost region. For a given admissible entry cost satisfying them, Figure~\ref{appendix-entry} depicts the resulting equilibrium ordering schematically; its threshold locations are uncalibrated.

\begin{proposition}\label{ext4-p-entry}
Under the binding posted-tariff timing above, every active developer chooses maximal openness $\bar{\eta}$ in period~2. Conditional on entry, developer~2 charges $\tilde{w}_2=w_L$; the incumbent can retain the deployer against an active entrant only at $w_2=w_L$. When openness makes entry unprofitable at either fee, the incumbent posts the monopoly fee $w_2=w_H$. Let $\bar{k}_1,\bar{k}_2$ be the baseline thresholds of Proposition~\ref{b-p-equilibrium-k}. The equilibrium has the following three entry-cost cases, with case~(II) subject to the stated regularity conditions.

\smallskip
\noindent\textbf{(I) Small entry cost, $e < \bar{e}_1$ (openness deterrence infeasible, $\bar{\eta}_e< 0$).} The allocation and incumbent profit coincide with the baseline, although entry occurs only under Harvest:
\begin{enumerate}[$(a)$]
    \item If $k \leq \bar{k}_1$ (\textbf{Harvest}): $(w_1^*,\eta_1^*,w_2^*,Q_1^*,Q_2^*)=\big(w_H,\bar{\eta},-,\frac{(1+\bar{\eta})(\theta-w_H)}{2c},\frac{(1+\bar{\eta})^2(\theta-w_L)}{2c}\big)$; the deployer selects the entrant in period~2.
    \item If $\bar{k}_1 < k \leq \bar{k}_2$ (\textbf{Defend}): $\big(w_H,\bar{\eta}_H,w_L,\frac{\theta-w_H}{2c-k(\theta-w_H)},\frac{(1+\bar{\eta})(\theta-w_L)}{2c-k(\theta-w_H)}\big)$; the deployer selects the incumbent.
    \item If $k>\bar{k}_2$ (\textbf{Dominate}): $\big(w_L,\bar{\eta}_L,w_L,\frac{\theta-w_L}{2c-k(\theta-w_L)},\frac{(1+\bar{\eta})(\theta-w_L)}{2c-k(\theta-w_L)}\big)$; the deployer selects the incumbent.
\end{enumerate}

\smallskip
\noindent\textbf{(II) Intermediate entry cost, $\bar{e}_1 \le e < \bar{e}_2$ (openness deterrence feasible, $0\le\bar{\eta}_e<\bar{\eta}$).} Under the regularity conditions above, Exclude is optimal: the incumbent limits openness so that entry is unprofitable and posts the period-2 monopoly fee $w_2=w_H$. Its first-period fee switches at $\bar{k}_{eH}$, the same threshold that governs case~(III):
\begin{enumerate}[$(a)$]
    \item If $k \leq \bar{k}_{eH}$ (\textbf{Exclude}, high-fee sub-case): $\big(w_H,\bar{\eta}_e,w_H,\frac{(1+\bar{\eta}_e)(\theta-w_H)}{2c},\frac{(1+\bar{\eta})(\theta-w_H)[2c+k(1+\bar{\eta}_e)(\theta-w_H)]}{4c^2}\big)$; the deployer stays with the incumbent, which monopolizes period~2 at $w_2=w_H$.
    \item If $k > \bar{k}_{eH}$ (\textbf{Exclude}, low-fee sub-case): $\big(w_L,\bar{\eta}_e,w_H,\frac{(1+\bar{\eta}_e)(\theta-w_L)}{2c},\frac{(1+\bar{\eta})(\theta-w_H)[2c+k(1+\bar{\eta}_e)(\theta-w_L)]}{4c^2}\big)$; the incumbent lowers $w_1$ to strengthen the flywheel but still monopolizes period~2 at $w_2=w_H$.
\end{enumerate}
In both sub-cases $\eta_1^*=\bar{\eta}_e<\bar{\eta}$: developer~2's gross operating profit equals $e$, so the entrant stays out under the stated tie-breaking rule. Under regularity condition~(i), Exclude permits weakly greater openness than Defend and retains the deployer at the monopoly fee $w_H$ instead of the low fee $w_L$; condition~(ii) also rules out a return to Dominate within the admissible range. Hence Exclude is the equilibrium throughout region~(II), as depicted in Figure~\ref{appendix-entry}. For other admissible primitives, Exclude remains an additional equilibrium candidate, but a Defend interval or an eventual return to Dominate can survive.

\smallskip
\noindent\textbf{(III) Large entry cost, $e \ge \bar{e}_2$ ($\bar{\eta}_e\ge\bar{\eta}$).} Developer~2 never enters, including at zero net profit under the tie-breaking rule. The incumbent is therefore an unconstrained two-period monopolist with full openness (\textbf{Monopoly}): $\eta_1^*=\bar{\eta}$, $w_2^*=w_H$, and $w_1^*=w_H$ if $k\le\bar{k}_{eH}$, $w_1^*=w_L$ if $k>\bar{k}_{eH}$, with $Q_1^*=\frac{(1+\bar{\eta})(\theta-w_1^*)}{2c}$ and $Q_2^*=\frac{(1+\bar{\eta})(\theta-w_H)[2c+k(1+\bar{\eta})(\theta-w_1^*)]}{4c^2}$.
\end{proposition}

Figure~\ref{appendix-entry} summarizes the three entry-cost cases schematically. In the intermediate-cost case, the stated regularity restrictions make Exclude the preferred strategy: limiting openness deters entry and lets the incumbent charge the monopoly fee in period 2. At $e\geq\bar{e}_2$, entry is unprofitable even at maximal openness, so the incumbent can choose full openness without inducing entry.

\noindent\textbf{An openness mandate with endogenous entry.}
The participation decision also changes the effect of a full-openness mandate, $\eta_1=\bar{\eta}$.
We measure social welfare as in Section~\ref{Openness Trap}, net of the entry cost $e$ whenever developer~2 enters.
Under the maintained bound $0\le k<K_{\max}$, full openness lies above both retention frontiers, so limit pricing cannot retain the deployer.
At full openness, developer~2 earns gross operating profit $w_L\tilde{Q}_2^*=\bar{e}_2$; hence it enters whenever $e<\bar{e}_2$.
In regions~(I) and~(II), the mandate therefore induces Harvest with entry and welfare $W_H-e$.

In region~(I), the unregulated allocation is the baseline one.
Harvest is unaffected by the mandate, while Defend and Dominate deter entry through limit pricing and do not incur $e$.
Write $W_F(k)$ and $W_D(k)$ for welfare under these two candidates, as derived in the proof of Proposition~\ref{b-p-welfare}.
Within Dominate, the mandate lowers welfare exactly when $W_D(k)>W_H-e$.
Define the candidate crossing $k_W(e)$ by $W_D(k_W(e))=W_H-e$ on the positive-denominator domain of $W_D$.
Because $e>0$ and $W_D$ is strictly increasing there, $k_W(e)<k_W$, but this crossing need not lie in the admissible equilibrium Dominate region.
Combining the welfare comparison with strategy selection gives a strict loss in Dominate exactly when
\[
\max\{\bar{k}_2,k_W(e)\}<k<K_{\max},\qquad k\ge0.
\]
This region is nonempty if and only if $\max\{0,\bar{k}_2,k_W(e)\}<K_{\max}$.
Within Defend, the mandate raises welfare if $e<W_H-W_F(k)$, leaves it unchanged at equality, and lowers it if $e>W_H-W_F(k)$.
Thus, in the small-entry-cost case, entry costs weakly enlarge the welfare-loss region relative to the baseline: the mandate incurs a real cost that the unregulated Defend and Dominate strategies avoid.

In region~(II), Exclude is the unregulated equilibrium under the stated sufficient regularity conditions.
The mandate removes openness deterrence, so the entrant enters and wins the deployer.
Greater public reuse and the lower period-2 fee raise second-period welfare before deducting the entry cost.
To establish this comparison, let $Q_{X2}$ be second-period effort under Exclude and $Q_{H2}=(1+\bar{\eta})^2(\theta-w_L)/(2c)$ the mandated effort.
The first component of $K_{\max}$, together with $\bar{\eta}_e<\bar{\eta}$ and $w_1\ge w_L$, gives
\[
Q_{X2}<\frac{\theta-w_H}{\theta-w_L}Q_{H2}<Q_{H2}.
\]
Since $(\theta+w_H)(\theta-w_H)<(\theta+w_L)(\theta-w_L)$, both the linear and quadratic terms in per-period welfare, $[(\theta+w)Q+Q^2]/2$, are higher under the mandate.
First-period welfare also rises in the high-fee Exclude case, where the fee is unchanged and openness increases from $\bar{\eta}_e$ to $\bar{\eta}$.
These gains do not determine the total effect: net welfare also depends on $e$ and, in the low-fee Exclude case, on the first-period comparison. The total welfare effect can therefore be negative for other primitives.
The calibration of Figure~\ref{b-f-equilibrium} illustrates a positive net effect: $\bar{e}_1=2.8125$ and $\bar{e}_2\approx7.031$.
Direct comparison of the candidate profits selects Exclude throughout this intermediate-cost region over the admissible range of $k$, including where the sufficient regularity conditions do not hold, and the mandate raises welfare net of the entry cost throughout that region.

In region~(III), the incumbent already chooses $\eta_1^*=\bar{\eta}$, so the mandate does not bind.

Endogenous entry makes the mechanism sustaining non-entry central to the policy comparison.
Under Defend or Dominate, limited openness and the binding low period-2 fee keep the entrant from winning; a mandate can induce the baseline price response and also incur an entry cost.
Under Exclude, limited openness makes entry unprofitable even at the monopoly fee, so a mandate replaces period-2 monopoly with entry and a low-fee supplier.
The calibrated example shows that this gain can outweigh the entry cost.

\begin{figure}[ht]
\centering \footnotesize
\begin{tikzpicture}[scale=0.9]
\foreach \x in {0,3,9}{\draw (\x cm,5pt) -- (\x cm,-5pt);}
\node[align=center] at (3,0.5) {$\bar{k}_{1}$};
\node[align=center] at (9,0.5) {$\bar{k}_{2}$};
\draw [thick,->] (0,0) -- (12,0)node[below]{~~$k$};
\node at (-2,0) {~~~~~~$e < \bar{e}_1$};
\node[align=center] at (0,0.5) {0};
\draw [decorate,decoration={brace,amplitude=5pt,mirror,raise=2ex}] (0.05,0) -- (2.95,0) node[midway,yshift=-2em]{Harvest};
\draw [decorate,decoration={brace,amplitude=5pt,mirror,raise=2ex}] (3.05,0) -- (8.95,0) node[midway,yshift=-2em]{Defend};
\draw [decorate,decoration={brace,amplitude=5pt,mirror,raise=2ex}] (9.05,0) -- (11.8,0) node[midway,yshift=-2em]{Dominate};
\end{tikzpicture}

\begin{tikzpicture}[scale=0.9]
\fill[yellow!60] (0.01,-0.16) rectangle (11.79,0.16);
\foreach \x in {0,2.5}{\draw (\x cm,5pt) -- (\x cm,-5pt);}
\node[align=center] at (2.5,0.5) {$\bar{k}_{eH}$};
\node[align=center] at (0,0.5) {0};
\draw [thick,->] (0,0) -- (12,0)node[below]{~~$k$};
\node at (-1.9,0) {$\bar{e}_1 \leq e < \bar{e}_2$};
\draw [decorate,decoration={brace,amplitude=5pt,mirror,raise=2ex}] (0.05,0) -- (2.45,0) node[midway,yshift=-2em]{Exclude ($w_H$)};
\draw [decorate,decoration={brace,amplitude=5pt,mirror,raise=2ex}] (2.55,0) -- (11.8,0) node[midway,yshift=-2em]{Exclude ($w_L$)};
\end{tikzpicture}

\begin{tikzpicture}[scale=0.9]
\fill[green!60] (0.01,-0.16) rectangle (11.79,0.16);
\foreach \x in {0,2.5}{\draw (\x cm,5pt) -- (\x cm,-5pt);}
\node[align=center] at (2.5,0.5) {$\bar{k}_{eH}$};
\node[align=center] at (0,0.5) {0};
\draw [thick,->] (0,0) -- (12,0)node[below]{~~$k$};
\node at (-1.9,0) {~~~~~~$e \ge \bar{e}_2$};
\draw [decorate,decoration={brace,amplitude=5pt,mirror,raise=2ex}] (0.05,0) -- (2.45,0) node[midway,yshift=-2em]{Monopoly ($w_H$)};
\draw [decorate,decoration={brace,amplitude=5pt,mirror,raise=2ex}] (2.55,0) -- (11.8,0) node[midway,yshift=-2em]{Monopoly ($w_L$)};
\end{tikzpicture}

\caption{Schematic Equilibrium Ordering by Entry Cost. The intermediate-cost row applies under the stated sufficient regularity inequalities. Parenthetical fees are first-period license fees. Each row is read within the maintained admissible range of $k$; threshold locations are uncalibrated, and some intervals may be empty in that range.}\label{appendix-entry}

\end{figure}

\subsection{Strength of the Public Spillover}\label{appen-spillover-strength}
Public disclosure can make adaptation work reusable across models, but the benefit depends on compatibility and on how much work remains model-specific.
We vary this benefit by letting $\gamma>0$ scale the entrant-facing public spillover, while retaining the baseline specifications for current-model adaptability and incumbent-specific learning:
\[
C_{2E}^{\gamma}(Q_2)=\frac{cQ_2^2}{(1+\gamma\eta_1)(1+\tilde{\eta}_2)}.
\]
The baseline is $\gamma=1$, and $0<\gamma<1$ represents weaker public reuse.
Thus the same disclosure level can produce a smaller cost-efficiency gain when the entrant's model is less compatible with the publicly available adaptation resources.
The parameter scales this adaptation-efficiency gain; it does not measure a percentage reduction in total deployment expenditure.
In the reduced form of Appendix~\ref{appen-three k channels}, this specification corresponds to $\phi(m(\eta_1))=1/(1+\gamma\eta_1)$.

With both developers at full second-period openness and the incumbent charging $w_L$, the incumbent retains the deployer if and only if $kQ_1\geq\gamma\eta_1$.
Thus a weaker public spillover makes retention possible at greater openness for a given learning rate.
We examine the competitive region
\[
0\leq k<K_{\max}(\gamma):=\min\{\gamma K_1,K_2\},\qquad
K_1=\frac{2c\bar{\eta}}{(1+\bar{\eta})(\theta-w_L)},\qquad
K_2=\frac{2c(2\theta-w_H-w_L)(w_H-w_L)}{(\theta-w_H)^2(\theta-w_L)}.
\]
Here full openness prevents retention at either first-period fee, and a high second-period fee also loses the deployer.
A full-openness mandate therefore induces Harvest.
This region extends the baseline competition restriction, which is recovered at $\gamma=1$.
Within it, the largest openness compatible with retention after first-period fee $w_j$ is
\[
\bar{\eta}_j(\gamma)=\frac{k(\theta-w_j)}{2c\gamma-k(\theta-w_j)},\qquad j\in\{H,L\}.
\]
For exact existence conditions, the following result focuses on $\gamma K_1\leq K_2$, so the upper bound is $\gamma K_1$.
It asks whether any learning rates below this bound induce openness restriction and whether any produce an openness trap.
Let $\bar{\eta}_H^{*}$ denote the limiting Defend openness as learning approaches the bound from below, and let $r$ normalize current high-fee revenue relative to the scale of future low-fee revenue:
\[
\bar{\eta}_H^{*}=\frac{\bar{\eta}(\theta-w_H)}{(1+\bar{\eta})(\theta-w_L)-\bar{\eta}(\theta-w_H)},\qquad
r=\frac{w_H(\theta-w_H)}{(1+\bar{\eta})w_L(\theta-w_L)},
\]
The corresponding comparison thresholds are
\[
\gamma_R=\frac{(\bar{\eta}-\bar{\eta}_H^{*})r-1}{\bar{\eta}_H^{*}},\qquad
\gamma_D=\frac{(1+\bar{\eta})r-2}{\bar{\eta}},\qquad
\gamma_F=\frac{(1+\bar{\eta}_H^{*})r-1}{\bar{\eta}-\bar{\eta}_H^{*}}.
\]

\begin{proposition}\label{ext5-p-spillover}
Suppose $c>0$, $\bar{\eta}>0$, $0<w_L<w_H\leq\theta/2$, $\gamma>0$, and $\gamma K_1\leq K_2$.
For $0\leq k<\gamma K_1$, retain the baseline tie-breaking rule: Harvest is selected whenever it maximizes profit, and Defend otherwise at a profit tie.
\begin{enumerate}[$(a)$]
\item \textbf{Strategies and the mandate.} The incumbent selects among Harvest $(w_H,\bar{\eta})$, Defend $(w_H,\bar{\eta}_H(\gamma))$, and Dominate $(w_L,\bar{\eta}_L(\gamma))$.
Their respective profits are
\begin{align*}
\pi^{H}&=\frac{(1+\bar{\eta})(\theta-w_H)w_H}{2c},\\
\pi^{F}&=\frac{(1+\bar{\eta}_H(\gamma))(\theta-w_H)w_H+w_L(1+\gamma\bar{\eta}_H(\gamma))(1+\bar{\eta})(\theta-w_L)}{2c},\\
\pi^{D}&=\frac{(1+\bar{\eta}_L(\gamma))(\theta-w_L)w_L+w_L(1+\gamma\bar{\eta}_L(\gamma))(1+\bar{\eta})(\theta-w_L)}{2c}.
\end{align*}
A full-openness mandate induces Harvest.
Relative to the unregulated allocation, it leaves welfare unchanged in Harvest and strictly raises welfare in Defend.
\item \textbf{Existence of openness restriction.} A nonempty interval of learning rates with openness restriction exists if and only if $\gamma>\min\{\gamma_R,\gamma_D\}$.
When this condition holds, restriction occurs on an interval adjacent to $\gamma K_1$.
\item \textbf{Existence of an openness trap.} A nonempty Dominate interval, and a nonempty interval on which the mandate lowers welfare, exist if and only if $\gamma>\max\{\gamma_D,\gamma_F\}$.
When this condition holds, Dominate and a welfare loss occur on an interval adjacent to $\gamma K_1$.
Otherwise, the mandate weakly increases welfare throughout the stated range.
\end{enumerate}
\end{proposition}

The restriction condition compares the future business protected by restricting disclosure with the current revenue forgone.
Harvest profit is constant in $k$, whereas the profits from Defend and Dominate increase with learning.
Their limiting profits at $\gamma K_1$ therefore determine whether either strategy can overtake Harvest before the upper bound is reached.
An openness-trap region additionally requires Dominate to beat Defend: the low first-period fee must generate enough deployment and learning to make retention worth that price cut.

As learning approaches the upper bound, Dominate combines almost full openness with the low first-period fee.
The mandate instead induces Harvest and the high first-period fee.
Second-period effort converges to the same level under the two allocations, so the gain from additional public disclosure vanishes while the first-period loss from the higher fee remains.
This comparison gives a welfare-loss interval near the bound when Dominate is optimal there.

At the illustrative parameters of Figure~\ref{b-f-equilibrium}, $\gamma_R=2/9$, $\gamma_D=14/27$, and $\gamma_F=2/3$.
Within the stated competitive region, some learning rates induce restriction exactly when $\gamma>2/9$, and a welfare-loss region exists exactly when $\gamma>2/3$.
For $2/9<\gamma\leq2/3$, some learning rates induce openness restriction, but the mandate never lowers welfare within the stated region.
Above $2/3$, the welfare effect still depends on the particular learning rate; the threshold establishes existence of a loss region, not a loss at every $k$.
Across other primitives the two existence thresholds can coincide or lie outside $0<\gamma\leq1$.
The policy question is therefore whether public reuse is sufficiently strong relative to the application's learning and pricing incentives.
Appendix~\ref{appen-three k channels} separately examines common additional costs: per-unit resource costs change deployment incentives, while common fixed expenditure cancels from the welfare comparison under the stated sufficient participation condition.

\newpage
\section{Proofs of Statements}
We present the proofs of the Lemmas and Propositions in this appendix.\\

\noindent\textbf{Proof of Lemma \ref{b-l-etabar}:}\\
Given the fine-tuning effort $Q_t$ in period $t$, The consumer's utility at period $t$ is:
\begin{equation*}
    u_t(Q_t,\alpha_t)= Q_t\alpha_t - \alpha^2_t/2,
\end{equation*}
from which we derive that the consumer's optimal usage level at period $t$ is $\alpha_t(Q_t)=Q_t$.

Given developer 1's openness level $\eta_1$ and the license fee decision $w_1$ at $t=1$, the deployer's objective function in the first period is:
\begin{equation*}
\Pi_1 (Q_1,\alpha_1)= (\theta-w_1) \alpha_1 - \frac{cQ_1^2}{1+\eta_1},
\end{equation*}
from which we derive the deployer's best response:
\begin{equation*}
    Q_1(w_1,\eta_1)=\frac{(1+\eta_1)(\theta-w_1)}{2c}.
\end{equation*}

Given developer 1's openness level $\eta_1$ and the license fee decision $w_1$ at $t=1$, and developer 1's openness level $\eta_2$ and the license fee decision $w_2$ at $t=2$, the deployer's profit of choosing developer 1 in the second period is:
\begin{equation*}
\Pi_{2I} (Q_2,\alpha_2)= (\theta-w_2) \alpha_2 - \frac{cQ_2^2}{(1+k\alpha_1)(1+\eta_2)}.
\end{equation*}
Plugging $\alpha_t(Q_t)$ and $Q_1(w_1,\eta_1)$ in the deployer's objective function $\Pi_{2I} (Q_2,\alpha_2)$, we have:
\begin{equation*}
    (\theta-w_2)Q_2-\frac{2c^2Q^2_2}{[2c+k(1+\eta_1)(\theta-w_1)](1+\eta_2)}.
\end{equation*}
We derive the deployer's best response:
\begin{equation*}
    Q_2(w_1,w_2,\eta_1,\eta_2)=\frac{(\theta-w_2)[2c+k(1+\eta_1)(\theta-w_1)](1+\eta_2)}{4c^2}.
\end{equation*}
Plugging $Q_2(w_1,w_2,\eta_1,\eta_2)$ into the deployer's profit function, we have:
\begin{equation*}
\Pi_{2I} (w_1,w_2,\eta_1,\eta_2)=\frac{(\theta-w_2)^2[2c+k(1+\eta_1)(\theta-w_1)](1+\eta_2)}{8c^2}.
\end{equation*}

Consider next the deployer's profit of choosing developer 2 in the second period. Given developer 2's openness level $\Tilde{\eta}_2$ and the license fee decision $\Tilde{w}_2$, and developer 1's openness level $\eta_1$ chosen in the first period, the deployer's profit at $t=2$ is:
\begin{equation*}
\Pi_{2E} (\Tilde{Q}_2,\Tilde{\alpha}_2)= (\theta-\Tilde{w}_2) \Tilde{\alpha}_2 - \frac{c\Tilde{Q}_2^2}{(1+\eta_1)(1+\Tilde{\eta}_2)}.
\end{equation*}
Plugging $\Tilde{\alpha}_2=\Tilde{Q}_2$ into the deployer's profit function $\Pi_{2E} (\Tilde{Q}_2,\Tilde{\alpha}_2)$ and taking the first order derivative to $\Tilde{Q}_2$, we derive the deployer's optimal fine-tuning effort:
\begin{equation*}
    Q_2(\Tilde{w}_2,\eta_1,\Tilde{\eta}_2)=\frac{(1+\eta_1)(1+\Tilde{\eta}_2)(\theta-\Tilde{w}_2)}{2c}.
\end{equation*}
Plugging $Q_2(\Tilde{w}_2,\eta_1,\Tilde{\eta}_2)$ into the deployer's profit function, we have:
\begin{equation*}
\Pi_{2E} (\Tilde{w}_2,\eta_1,\Tilde{\eta}_2)= \frac{(1+\eta_1)(1+\Tilde{\eta}_2)(\theta-\Tilde{w}_2)^2}{4c}.
\end{equation*}

Both developer 1 and developer 2 choose $\eta_2=\bar{\eta}_2= \bar{\eta}$, which increases developer's profit and competitiveness. In addition, by our assumption, developer 2 chooses the low license fee $\Tilde{w}_2=w_L$. Developer 1 wins in the second period iff $\Pi_{2I} (w_1,w_2,\eta_1,\eta_2) \geq \Pi_{2E} (\Tilde{w}_2,\eta_1,\Tilde{\eta}_2)$. Plugging in $\eta_2=\bar{\eta}_2= \bar{\eta}$ and $\Tilde{w}_2=w_L$, we simplify $\Pi_{2I} (w_1,w_2,\eta_1,\eta_2) \geq \Pi_{2E} (\Tilde{w}_2,\eta_1,\Tilde{\eta}_2)$ as:
\begin{equation*}
    \left(\frac{\theta-w_2}{\theta-w_L}\right)^2 \geq \frac{2c(1+\eta_1)}{2c+k(1+\eta_1)(\theta-w_1)}.
\end{equation*}
By assumption $k \leq \min\{\frac{2c\bar{\eta}}{(1+\bar{\eta})(\theta-w_L)},\frac{2c(2\theta-w_H-w_L)(w_H-w_L)}{(\theta-w_H)^2(\theta-w_L)}\}$, developer 1 loses the deployer for sure if developer 1 charges the high license fee $w_2=w_H$ in the second period. Thus, we plug in $w_2=w_L$ and developer 1's winning condition becomes:
\begin{equation*}
    \frac{2c(1+\eta_1)}{2c+k(1+\eta_1)(\theta-w_1)} \leq 1,
\end{equation*}
which depends on developer 1's openness level $\eta_1$ and the license fee decision $w_1$ in the first period.

If developer 1 chooses the high license fee $w_1=w_H$ in the first period, $\eta_1 \leq \frac{k(\theta-w_H)}{2c-k(\theta-w_H)}$ guarantees that developer 1 can win in the second period. We denote $\bar{\eta}_H=\frac{k(\theta-w_H)}{2c-k(\theta-w_H)}$. If developer 1 chooses the low license fee $w_1=w_L$ in the first period, $\eta_1 \leq \frac{k(\theta-w_L)}{2c-k(\theta-w_L)}$ guarantees that developer 1 can win in the second period. We denote $\bar{\eta}_L=\frac{k(\theta-w_L)}{2c-k(\theta-w_L)}$. It can be verified that $\bar{\eta}_H \leq \bar{\eta}_L$.

Next, we investigate developer 1's profit. Note that if developer 1 wins in the second period, $w_2=w_L$ and $\eta_2=\bar{\eta}$. Given developer 1's $w_1$ and $\eta_1$ in the first period, winning in the second period, developer 1's profit over two periods is:
\begin{equation*}
    \pi_{win} (w_1,\eta_1) = w_1 \alpha_1 + w_L \alpha_2.
\end{equation*}
For each fixed binary fee $w_1\in\{w_H,w_L\}$, substituting $\alpha_1$, $\alpha_2$, $Q_1$, and $Q_2$ into $\pi_{win}(w_1,\eta_1)$ shows that developer~1's profit increases in $\eta_1$. The sign of the comparison between $w_H$ and $w_L$ is not globally positive; the two fee candidates must be compared directly. Throughout the proof, developer~1's winning-period revenue is evaluated at the winning fee $w_2=w_L$, so that $\pi_{win}=w_1\alpha_1+w_L\alpha_2$ as defined above.

Given developer 1's $w_1$ and $\eta_1$ in the first period, losing in the second period, developer 1's profit only in the first period is:
\begin{equation*}
    \pi_{lose} (w_1,\eta_1) = w_1 \alpha_1.
\end{equation*}
Plugging $\alpha_1$, $\alpha_2$, $Q_1$, $Q_2$ into $ \pi_{lose} (w_1,\eta_1)$, it can be verified that developer 1's profit increases in both $w_1$ and $\eta_1$ as well.\Halmos\\[3mm]

\noindent\textbf{Proof of Proposition \ref{b-p-equilibrium-k}:}\\
Lemma \ref{b-l-etabar} implies that developer 1 selects $w_2=w_L$ and $\eta_2=\bar{\eta}$ in the second period. So developer 1's decision $\eta_1$ and $w_1$ in the first period determines if developer 1 can win in the second period and the corresponding profit.

Let us suppose developer 1 loses in the second period. Plugging $\alpha_1$, $\alpha_2$, $Q_1$, $Q_2$ into $ \pi_{lose} (w_1,\eta_1)$, we have:
\begin{equation*}
    \pi_{lose} (w_1,\eta_1) = \frac{(1+\eta_1)(\theta-w_1)w_1}{2c}.
\end{equation*}
Given that developer 1 loses for sure, it is optimal for developer 1 to select $\eta_1 = \bar{\eta}$ to maximize the profit. In addition, under the assumption $\theta/2 \geq w_H \geq w_L \geq 0$, it can be verified that $\pi_{lose} (w_1,\eta_1)$ also increases in $w_1$. Therefore, $\pi_{lose} (w_H,\bar{\eta}) \geq \pi_{lose} (w_L,\bar{\eta})$. Given the scenario that developer 1 loses, developer 1 chooses $w_1 = w_H$ and $\eta_1 = \bar{\eta}$, which is denoted as scenario $S_0$ (Harvest strategy). The corresponding profit for developer 1:
\begin{equation*}
    \pi_{S_0} = \frac{(1+\bar{\eta})(\theta-w_H)w_H}{2c}.
\end{equation*}

Let us suppose developer 1 wins in the second period. Plugging $\alpha_1$, $\alpha_2$, $Q_1$, $Q_2$ into $ \pi_{win} (w_1,\eta_1)$, we have:
\begin{equation*}
    \pi_{win} (w_1,\eta_1) = \frac{2c(1+\eta_1)(\theta-w_1)w_1+(1+\bar{\eta})[2c+(1+\eta_1)k(\theta-w_1)](\theta-w_L)w_L}{4c^2}.
\end{equation*}
Given that developer 1 is certain to win, it is optimal for developer 1 to choose a higher openness level $\eta_1$. However, Lemma \ref{b-l-etabar} shows that developer 1 will lose in the second period if choosing a sufficiently large openness level. Specifically, to guarantee winning in the second period while achieving the largest possible openness level, developer 1 can choose from two scenarios. That is, scenario $S_1$ (Defend strategy), where $w_1=w_H$ and $\eta_1 = \bar{\eta}_H$; scenario $S_2$ (Dominate strategy), where $w_1=w_L$ and $\eta_1 = \bar{\eta}_L$. We can derive developer 1's profit for the two scenarios:
\begin{align*}
    \pi_{S_1} &=  \frac{2c(1+\bar{\eta}_H)(\theta-w_H)w_H+(1+\bar{\eta})[2c+(1+\bar{\eta}_H)k(\theta-w_H)](\theta-w_L)w_L}{4c^2}.\\
    \pi_{S_2} &=  \frac{2c(1+\bar{\eta}_L)(\theta-w_L)w_L+(1+\bar{\eta})[2c+(1+\bar{\eta}_L)k(\theta-w_L)](\theta-w_L)w_L}{4c^2}.
\end{align*}

Develope 1 chooses from the three scenarios $\pi_{S_0}$, $\pi_{S_1}$, and $\pi_{S_2}$ to generate the largest profit. We first compare $\pi_{S_1}$ and $\pi_{S_2}$. We solve that $\pi_{S_1} \geq \pi_{S_2}$ iff:
\begin{equation*}
    k \leq \frac{2c(\theta-w_L-w_H)}{(\theta-w_L)(\theta-w_H+w_L+\bar{\eta} w_L)}.
\end{equation*}
We denote $\bar{k}_{12} = \frac{2c(\theta-w_L-w_H)}{(\theta-w_L)(\theta-w_H+w_L+\bar{\eta} w_L)}$. Then we compare $\pi_{S_1}$ and $\pi_{S_0}$. We solve that $\pi_{S_1} \geq \pi_{S_0}$ iff:
\begin{equation*}
    k \geq \frac{2c[\bar{\eta}(\theta-w_H)w_H-(1+\bar{\eta})\theta w_L+(1+\bar{\eta})w^2_L]}{(1+\bar{\eta})(\theta-w_H)^2 w_H}.
\end{equation*}
We denote $\bar{k}_{13} = \frac{2c[\bar{\eta}(\theta-w_H)w_H-(1+\bar{\eta})\theta w_L+(1+\bar{\eta})w^2_L]}{(1+\bar{\eta})(\theta-w_H)^2 w_H}$. Finally, we compare $\pi_{S_2}$ and $\pi_{S_0}$. We solve that $\pi_{S_2} \geq \pi_{S_0}$ iff:
\begin{equation*}
    k \geq 2c \left(\frac{1}{\theta-w_L}-\frac{(2+\bar{\eta})w_L}{(1+\bar{\eta})(\theta-w_H)w_H}\right).
\end{equation*}

\begin{figure}[ht]
	\centering \footnotesize
\begin{tikzpicture}[scale=0.9]
\linespread{0.8}
\foreach \x in {0,3,6,9}{
    \draw (\x cm,5pt) -- (\x cm,-5pt);
}
\node[align=center] at (3,0.5) {$\bar{k}_{13}$};
\node[align=center] at (6,0.5) {$\bar{k}_{23}$};
\node[align=center] at (9,0.5) {$\bar{k}_{12}$};

\draw [thick,->] (0,0) -- (11,0)node[below]{~~$k$};
\node at (-2,0) {$\bar{\eta} \leq \eta'$};
\node[align=center] at (0,0.5) {0};

\draw [decorate,decoration={brace,amplitude=5pt,mirror,raise=2ex}] (0.05,0) -- (2.95,0) node[midway,yshift=-2em]{$S_0$};
\draw [decorate,decoration={brace,amplitude=5pt,mirror,raise=2ex}] (3.05,0) -- (8.95,0) node[midway,yshift=-2em]{$S_1$};
\draw [decorate,decoration={brace,amplitude=5pt,mirror,raise=2ex}] (9.05,0) -- (10.8,0) node[midway,yshift=-2em]{\textcolor{black}{$S_2$}};
\end{tikzpicture}
\begin{tikzpicture}[scale=0.9]
\linespread{0.8}
\foreach \x in {0,3,6,9}{
    \draw (\x cm,5pt) -- (\x cm,-5pt);
}
\node[align=center] at (3,0.5) {$\bar{k}_{12}$};
\node[align=center] at (6,0.5) {$\bar{k}_{23}$};
\node[align=center] at (9,0.5) {$\bar{k}_{13}$};
\node[align=center] at (0,0.5) {0};

\draw [thick,->] (0,0) -- (11,0)node[below]{~~$k$};
\node at (-1.9,0) {$\bar{\eta} \geq \eta'$};

\draw [decorate,decoration={brace,amplitude=5pt,mirror,raise=2ex}] (0,0) -- (5.95,0)  node[midway,yshift=-2em]{$S_0$};
\draw [decorate,decoration={brace,amplitude=5pt,mirror,raise=2ex}] (6.05,0) -- (10.8,0) node[midway,yshift=-2em]{\textcolor{black}{$S_2$}};
\end{tikzpicture}

\caption{Developer 1's Scenario Selection Decision}\label{b-f-appendix}
\end{figure}

We analyze the relationship among $\bar{k}_{12}$, $\bar{k}_{13}$, and $\bar{k}_{23}$. There exists a threshold:
\begin{equation*}
    \eta' = \frac{(\theta-w_H)^2+w_L(w_H-w_L)}{(\theta-w_H-w_L)(w_H-w_L)},
\end{equation*}
such that $\bar{k}_{13} \leq \bar{k}_{23} \leq \bar{k}_{12}$ when $\bar{\eta} \leq \eta'$, and $\bar{k}_{12} \leq \bar{k}_{23} \leq \bar{k}_{13}$ when $\bar{\eta} \geq \eta'$. The results are summarized in Figure \ref{b-f-appendix}.

We denote $\bar{k}_1 = \min\{\bar{k}_{13},\bar{k}_{23}\}$, and $\bar{k}_2 = \max\{\bar{k}_{12},\bar{k}_{23}\}$. The proposition is proved.\Halmos\\[3mm]

\noindent\textbf{Proof of Proposition \ref{b-p-welfare}:}\\
We categorize developer 1's profit, developer 2's profit, the deployer's profit, and consumer surplus under the baseline (no openness regulation). Proposition \ref{b-p-equilibrium-k} shows developer 1 manipulates the license fee $w_1$ and the openness level $\eta_1$ in the first period to choose among three scenarios $S_1$ (Defend strategy), $S_2$ (Dominate strategy), and $S_0$ (Harvest strategy). Developer 1's profit under three scenarios is provided below:
\begin{equation*}
\left \{
\begin{aligned}
\pi_{S_1} &= \frac{2c(1+\bar{\eta}_H)(\theta-w_H)w_H+(1+\bar{\eta})[2c+(1+\bar{\eta}_H)k(\theta-w_H)](\theta-w_L)w_L}{4c^2},\\
\pi_{S_2} &= \frac{2c(1+\bar{\eta}_L)(\theta-w_L)w_L+(1+\bar{\eta})[2c+(1+\bar{\eta}_L)k(\theta-w_L)](\theta-w_L)w_L}{4c^2},\\
\pi_{S_0} &= \frac{(1+\bar{\eta})(\theta-w_H)w_H}{2c}.
\end{aligned}
\right.
\end{equation*}

Next, consider developer 2's profit. Developer 2 only makes a profit at $t=2$ under the scenario $S_0$ where developer 1 loses in the second period. Developer 2's profit under three scenarios is provided below:
\begin{equation*}
\left \{
\begin{aligned}
\Tilde{\pi}_{S_1} &= 0,\\
\Tilde{\pi}_{S_2} &= 0,\\
\Tilde{\pi}_{S_0} &= \frac{(1+\bar{\eta})^2(\theta-w_L)w_L}{2c}.
\end{aligned}
\right.
\end{equation*}

Similarly, we analyze the deployer's profit. Write $\Pi_1=(\theta-w_1)\alpha_1-C_1(Q_1)$. If developer 1 is chosen in the second period, the deployer's total profit is $\Pi=\Pi_1+\Pi_{2I}(w_1,w_2,\eta_1,\eta_2)$; otherwise it is $\Pi=\Pi_1+\Pi_{2E}(\tilde{w}_2,\eta_1,\tilde{\eta}_2)$. Substituting the corresponding equilibrium choices gives:
\begin{equation*}
\left \{
\begin{aligned}
\Pi_{S_1} &= \frac{(2+\bar{\eta})\theta^2+w^2_H+(1+\bar{\eta})w^2_L-2\theta(w_H+w_L+\bar{\eta} w_L)}{4c-2k(\theta-w_H)},\\
\Pi_{S_2} &= \frac{(2+\bar{\eta})(\theta-w_L)^2}{4c-2k(\theta-w_L)},\\
\Pi_{S_0} &= \frac{(1+\bar{\eta})[(2+\bar{\eta})\theta^2 +w^2_H+ (1+\bar{\eta})w^2_L-2\theta(w_H+w_L+\bar{\eta} w_L)]}{4c}.
\end{aligned}
\right.
\end{equation*}

Then we turn to consumer surplus. Following a similar analysis, we derive:
\begin{equation*}
\left \{
\begin{aligned}
u_{S_1} &= \frac{[2+\bar{\eta}(2+\bar{\eta})]\theta^2+w^2_H+(1+\bar{\eta})^2 w^2_L-2\theta[w_H+(1+\bar{\eta})^2 w_L]}{2[2c-k(\theta-w_H)]^2},\\
u_{S_2} &= \frac{[2+\bar{\eta}(2+\bar{\eta})](\theta-w_L)^2}{2[2c-k(\theta-w_L)]^2},\\
u_{S_0} &= \frac{(1+\bar{\eta})^2[(\theta-w_H)^2+(1+\bar{\eta})^2(\theta-w_L)^2]}{8c^2}.
\end{aligned}
\right.
\end{equation*}

Combining each party's profit, we derive social welfare under the baseline where there exists no openness regulation.

Under the maintained bound $0\leq k<K_{\max}$, full openness lies strictly above both winning frontiers. A full-openness mandate therefore induces Harvest $S_0$, with
\[
Q_{H1}=\frac{(1+\bar{\eta})(\theta-w_H)}{2c},\qquad
Q_{H2}=\frac{(1+\bar{\eta})^2(\theta-w_L)}{2c}.
\]
At the deployer's operating optimum, the adaptation cost in a period with fee $w$ and effort $Q$ is $(\theta-w)Q/2$. Summing developer and deployer profits cancels the license payments, and consumer surplus is $Q^2/2$. Hence the period's welfare is $[(\theta+w)Q+Q^2]/2$, and mandated welfare is the constant
\[
W_H=\frac{(\theta+w_H)Q_{H1}+(\theta+w_L)Q_{H2}+Q_{H1}^2+Q_{H2}^2}{2}.
\]

The mandate leaves an unregulated Harvest outcome unchanged. Under Defend, the fee sequence is also $(w_H,w_L)$, while
\[
Q_{F1}=\frac{\theta-w_H}{2c-k(\theta-w_H)},\qquad
Q_{F2}=\frac{(1+\bar{\eta})(\theta-w_L)}{2c-k(\theta-w_H)}.
\]
The first component of the bound $K_{\max}$ implies $2c-k(\theta-w_H)>2c/(1+\bar{\eta})$, so $Q_{F1}<Q_{H1}$ and $Q_{F2}<Q_{H2}$. Welfare is strictly increasing in each quantity at the common fee sequence, giving $W_F(k)<W_H$.

For Dominate, let $d(k)=2c-k(\theta-w_L)>0$. Its quantities are
\[
Q_{D1}=\frac{\theta-w_L}{d(k)},\qquad
Q_{D2}=\frac{(1+\bar{\eta})(\theta-w_L)}{d(k)}.
\]
Both periods have fee $w_L$, so its welfare simplifies to
\begin{align*}
W_D(k)&=\frac{A}{d(k)}+\frac{B}{d(k)^2},\\
A&=\frac{(2+\bar{\eta})(\theta^2-w_L^2)}{2},\qquad
B=\frac{[1+(1+\bar{\eta})^2](\theta-w_L)^2}{2}.
\end{align*}
Since $A,B,W_H>0$, the equation $W_Hd^2-Ad-B=0$ has exactly one positive root. Define
\[
d_W=\frac{A+\sqrt{A^2+4W_HB}}{2W_H},\qquad
k_W=\frac{2c-d_W}{\theta-w_L}.
\]
The Dominate candidate's welfare is strictly increasing in $k$ on its positive-denominator domain, and
\[
W_D(k)>W_H\quad\Longleftrightarrow\quad d(k)<d_W
\quad\Longleftrightarrow\quad k>k_W.
\]
The value $k_W$ is a crossing of the candidate welfare functions; it need not belong to the admissible equilibrium Dominate region. Under the strategy-selection convention used in Proposition~\ref{b-p-equilibrium-k}, Dominate is selected exactly when $k>\bar{k}_2$. Combining selection and welfare gives a strict welfare loss exactly for
\[
\max\{\bar{k}_2,k_W\}<k<K_{\max},\qquad k\geq0.
\]
This set is nonempty if and only if $\max\{0,\bar{k}_2,k_W\}<K_{\max}$. At $k=\bar{k}_2$, the convention selects Defend or, if the two regime thresholds coincide, Harvest; at $k=k_W>\bar{k}_2$, welfare is equal. Equilibrium welfare can jump across $W_H$ at a strategy switch, so equality at the policy threshold is not required. Finally, if $w_L=0$, retaining the deployer earns no second-period access revenue and requires less first-period openness than Harvest. Harvest is therefore optimal throughout the maintained domain; equivalently, $\bar{k}_2=2c/\theta>K_{\max}$ and the welfare-loss region is empty.\Halmos\\[3mm]

\noindent\textbf{Proof of Proposition \ref{v-p-equilibrium}:}\\
Under vertical integration, developer 1 and the deployer are integrated as one firm. Thus, in the second period, the new entrant developer 2's FM will not be adopted. In the second period, the integrated firm needs to determine both the openness level $\eta_{2v}$ and the fine-tuning effort $Q_{2v}$. The integrated firm's profit is:
\begin{equation*}
    \pi_{2v}(\eta_{2v}, Q_{2v})=\theta Q_{2v}-\frac{cQ_{2v}^2}{(1+kQ_{1v})(1+ \eta_{2v})}.
\end{equation*}
Note that the integrated firm's profit increases in the openness level $\eta_{2v}$. So the integrated firm chooses $\eta_{2v}=\bar{\eta}$. In addition, we derive the integrated firm's best response $Q_{2v}$ to the first-period user engagement:
\begin{equation*}
    Q_{2v}(Q_{1v})=\frac{(1+kQ_{1v})(1+\eta_{2v}) \theta}{2c}.
\end{equation*}

In period 1, the operating unit maximizes current-period profit at fixed $\eta_{1v}$:
\begin{equation*}
\pi_{1v}(\eta_{1v},Q_{1v})=\theta Q_{1v}-\frac{cQ_{1v}^2}{1+\eta_{1v}}.
\end{equation*}
Along the operating unit's unique effort response, both maximized current profit and continuation profit increase in $\eta_{1v}$, so the owner chooses $\eta_{1v}=\bar{\eta}$.
Strict concavity of current-period profit yields the effort response:

\begin{equation*}
    Q_{1v}=\frac{(1+\eta_{1v}) \theta}{2c}.
\end{equation*}
We solve the equilibrium.\Halmos\\[3mm]

\noindent\textbf{Proof of Proposition \ref{v-p-profit comparison}:}\\
Under the baseline (decentralized) model (Proposition \ref{b-p-equilibrium-k}), the equilibrium is summarized in three scenarios $S_0$, $S_1$, and $S_2$, which depend on the data flywheel effect $k$. Developer 1's profit under the three scenarios is:
\begin{equation*}
		\pi=\left \{
		\begin{aligned}
 		&\frac{(1+\bar{\eta})(\theta-w_H)w_H}{2c} &&\text{if}~~k \leq \bar{k}_1,\\
		&\frac{2c(1+\bar{\eta}_H)(\theta-w_H)w_H+(1+\bar{\eta})[2c+(1+\bar{\eta}_H)k(\theta-w_H)](\theta-w_L)w_L}{4c^2} &&\text{if} ~~ \bar{k}_1 <k \leq \bar{k}_2,\\
        &\frac{[2c(2+\bar{\eta}_L+\bar{\eta})+(1+\bar{\eta}_L)(1+\bar{\eta})k(\theta-w_L)](\theta-w_L)w_L}{4c^2} &&\text{if} ~\bar{k}_2 < k.
		\end{aligned}
		\right.
\end{equation*}

Developer 2's profit under the decentralized model is:
\begin{equation*}
		\Tilde{\pi}=\left \{
		\begin{aligned}
 		&\frac{(1+\bar{\eta})^2(\theta-w_L)w_L}{2c} &&\text{if}~~k \leq \bar{k}_1,\\
		&0 &&\text{if} ~~ \bar{k}_1 <k \leq \bar{k}_2,\\
            &0 &&\text{if} ~\bar{k}_2 < k.
		\end{aligned}
		\right.
\end{equation*}

The deployer's profit under the decentralized model is:
\begin{equation*}
		\Pi=\left \{
		\begin{aligned}
 		&\frac{(1+\bar{\eta})[(2+\bar{\eta})\theta^2 +w^2_H+ (1+\bar{\eta})w^2_L-2\theta(w_H+w_L+\bar{\eta} w_L)]}{4c} &&\text{if}~~k \leq \bar{k}_1,\\
		&\frac{(2+\bar{\eta})\theta^2+w^2_H+(1+\bar{\eta})w^2_L-2\theta(w_H+w_L+\bar{\eta} w_L)}{4c-2k(\theta-w_H)} &&\text{if} ~~ \bar{k}_1 <k \leq \bar{k}_2,\\
            &\frac{(2+\bar{\eta})(\theta-w_L)^2}{4c-2k(\theta-w_L)} &&\text{if} ~\bar{k}_2 < k.
		\end{aligned}
		\right.
\end{equation*}

The consumer surplus under the decentralized model is:
\begin{equation*}
		u=\left \{
		\begin{aligned}
 		&\frac{(1+\bar{\eta})^2[(\theta-w_H)^2+(1+\bar{\eta})^2(\theta-w_L)^2]}{8c^2} &&\text{if}~~k \leq \bar{k}_1,\\
		&\frac{[2+\bar{\eta}(2+\bar{\eta})]\theta^2+w^2_H+(1+\bar{\eta})^2 w^2_L-2\theta[w_H+(1+\bar{\eta})^2 w_L]}{2[2c-k(\theta-w_H)]^2} &&\text{if} ~~ \bar{k}_1 <k \leq \bar{k}_2,\\
            &\frac{[2+\bar{\eta}(2+\bar{\eta})](\theta-w_L)^2}{2[2c-k(\theta-w_L)]^2} &&\text{if} ~\bar{k}_2 < k.
		\end{aligned}
		\right.
\end{equation*}

Under vertical integration, the equilibrium is summarized in Proposition \ref{v-p-equilibrium}. The integrated firm's profit is:
\begin{equation*}
    \pi_v=\frac{(1+\bar{\eta})[4c+(1+\bar{\eta})k\theta]\theta^2}{8c^2}.
\end{equation*}

The consumer surplus under the centralized model is:
\begin{equation*}
    u_v=\frac{(1+\bar{\eta})^2\theta^2[1+(1+(1+\bar{\eta})k\theta/2/c)^2]}{8c^2}.
\end{equation*}

We first analyze joint profit, the combined profit of developer 1 and the deployer. Let us start from the scenario $S_1$ (Defend strategy), where $\bar{k}_1 <k \leq \bar{k}_2$. Under the decentralized model, joint profit is provided as follows:
\begin{align*}
    \pi_{S_1} + \Pi_{S_1}= &\frac{2c(1+\bar{\eta}_H)(\theta-w_H)w_H+(1+\bar{\eta})[2c+(1+\bar{\eta}_H)k(\theta-w_H)](\theta-w_L)w_L}{4c^2}\\
    + &\frac{(2+\bar{\eta})\theta^2+w^2_H+(1+\bar{\eta})w^2_L-2\theta(w_H+w_L+\bar{\eta} w_L)}{4c-2k(\theta-w_H)}.
\end{align*}
It is clear that joint profit under the decentralized model is monotone increasing in $k$ when $\bar{k}_1 <k \leq \bar{k}_2$.

Next, we analyze the scenario $S_2$ (Dominate strategy), where $k > \bar{k}_2$. Under the decentralized model, joint profit is provided as follows:
\begin{align*}
    \pi_{S_2} + \Pi_{S_2}= \frac{[2c(2+\bar{\eta}_L+\bar{\eta})+(1+\bar{\eta}_L)(1+\bar{\eta})k(\theta-w_L)](\theta-w_L)w_L}{4c^2} + \frac{(2+\bar{\eta})(\theta-w_L)^2}{4c-2k(\theta-w_L)}.
\end{align*}
Joint profit under the decentralized model is also monotone increasing in $k$ when $k > \bar{k}_2$. In addition, combining $S_1$ and $S_2$, it can be verified that joint profit $\pi_{S_1+S_2} + \Pi_{S_1+S_2}$ is monotone increasing in $k$ ($k > \bar{k}_1$).

Lastly, we analyze the scenario $S_0$ (Harvest strategy), where $k \leq \bar{k}_1$. Under the decentralized model, joint profit is provided as follows:
\begin{align*}
    \pi_{S_0} + \Pi_{S_0}= \frac{(1+\bar{\eta})(\theta-w_H)w_H}{2c} + \frac{(1+\bar{\eta})[(2+\bar{\eta})\theta^2 +w^2_H+ (1+\bar{\eta})w^2_L-2\theta(w_H+w_L+\bar{\eta} w_L)]}{4c}.
\end{align*}
Joint profit under the decentralized model does not change in $k$.

The decentralized joint profit is therefore constant on the Harvest region. Step 1 below establishes that the integrated firm's profit $\pi_v$ exceeds it outside that region. We complete the three comparisons after collecting the consumer-surplus expressions.

For consumer surplus, begin with $S_1$, where $\bar{k}_1<k\leq\bar{k}_2$:
\begin{equation*}
    u_{S_1}=\frac{[2+\bar{\eta}(2+\bar{\eta})]\theta^2+w^2_H+(1+\bar{\eta})^2 w^2_L-2\theta[w_H+(1+\bar{\eta})^2 w_L]}{2[2c-k(\theta-w_H)]^2}.
\end{equation*}
Clearly, the consumer surplus under the decentralized model is monotone increasing in $k$ when $\bar{k}_1 <k \leq \bar{k}_2$.

Next, we analyze the scenario $S_2$, where $k > \bar{k}_2$. Under the decentralized model, the consumer surplus is provided as follows:
\begin{align*}
    u_{S_2}= \frac{[2+\bar{\eta}(2+\bar{\eta})](\theta-w_L)^2}{2[2c-k(\theta-w_L)]^2}.
\end{align*}
The consumer surplus under the decentralized model increases in $k$ when $k > \bar{k}_2$. In addition, combining $S_1$ and $S_2$, it can be verified that the consumer surplus $u_{S_1+S_2}$ is monotone increasing in $k$ ($k > \bar{k}_1$).

Lastly, we analyze the scenario $S_0$, where $k \leq \bar{k}_1$. Under the decentralized model, the consumer surplus is provided as follows:
\begin{align*}
    u_{S_0}=\frac{(1+\bar{\eta})^2[(\theta-w_H)^2+(1+\bar{\eta})^2(\theta-w_L)^2]}{8c^2}.
\end{align*}
The consumer surplus under the decentralized model does not change in $k$.

For $j\in\{d,c,s\}$, let $\Delta_j(k)$ be the integrated value minus the decentralized value of joint profit ($d$), consumer surplus ($c$), and social welfare ($s$).
Decentralized social welfare is $\pi+\tilde{\pi}+\Pi+u$, whereas integrated social welfare is $\pi_v+u_v$.
We compare these measures on $0\leq k<K_{\max}$.

\textit{Step 1: Defend and Dominate.}
At any admissible $k>\bar{k}_1$, the deployer stays with developer 1 under both structures, so developer 2 earns zero and license payments are transfers within the incumbent--deployer pair.
The pair's decentralized profit in each period is $\theta Q_t-C_t(Q_t)$ at the deployer's chosen effort.
The maintained bound on $k$ implies $\bar{\eta}_H\leq\bar{\eta}_L<\bar{\eta}$, so integration strictly lowers the first-period adaptation cost at any positive effort.
The integrated operating unit maximizes current-period profit with revenue $\theta$ per unit of engagement and no internal license fee; evaluating that objective at the decentralized effort already yields a strict gain, and its optimum yields at least as much.
Its first-period effort is also strictly higher because it has weakly higher net revenue per unit of engagement and strictly higher openness.
Consequently, $1+kQ_{1v}^*\geq1+kQ_1$.
Both structures have period-2 openness $\bar{\eta}$, so integration weakly lowers the second-period adaptation cost at any given effort and raises net revenue per unit of engagement from $\theta-w_L$ to $\theta$.
The same comparison gives weakly higher second-period joint profit and $Q_{2v}^*\geq Q_2$.
The strict first-period gain implies $\Delta_d(k)>0$ and $\Delta_c(k)>0$, and hence $\Delta_s(k)=\Delta_d(k)+\Delta_c(k)>0$.

\textit{Step 2: Harvest.}
On the admissible Harvest region, $k\leq\bar{k}_1$, decentralized efforts, joint profit, consumer surplus, and entrant profit are all independent of $k$.
Decentralized social welfare is therefore also constant there.
Under integration, $Q_{1v}^*$ is constant and $Q_{2v}^*$ strictly increases with $k$; thus $\pi_v$, $u_v$, and $\pi_v+u_v$ strictly increase.
Each $\Delta_j(k)$ is consequently continuous and strictly increasing within Harvest.

\textit{Step 3: Cutoffs and boundary cases.}
Define
\[
\bar{k}_{jv}:=\inf\{k\in[0,K_{\max}):\Delta_j(k)>0\},
\]
with $\bar{k}_{jv}=K_{\max}$ when the set is empty.
Proposition \ref{b-p-equilibrium-k} makes Harvest an initial interval, possibly empty or consisting only of $k=0$.
By Steps 1 and 2, a positive gap cannot be followed by a nonpositive gap: within Harvest the gap strictly increases, and outside Harvest it is strictly positive.
For $k<\bar{k}_{jv}$ the gap must be strictly negative, since a zero in that range would be followed by a positive gap before the cutoff.
For $k>\bar{k}_{jv}$ it is strictly positive.
If Harvest is empty, Step 1 gives $\bar{k}_{jv}=0$.
Otherwise, the cutoff is no higher than the upper endpoint $\min\{\bar{k}_1,K_{\max}\}$ of the admissible Harvest region.

A crossing inside Harvest gives $\Delta_j(\bar{k}_{jv})=0$.
If the cutoff instead occurs when the strategy leaves Harvest, the value at that point is evaluated under the Harvest-favoring tie convention and need not be zero.
A zero cutoff likewise does not determine the sign at $k=0$: the gap may already be positive, equal zero there, or be negative when Harvest is selected only at $k=0$ and the gain begins immediately afterward.
The convention $\bar{k}_{jv}=K_{\max}$ covers an empty gain region; that endpoint lies outside the maintained domain.
Parts $(b)$ and $(c)$ follow by applying the two relevant strict comparisons together.\Halmos\\[3mm]

\noindent\textbf{Proof of Lemma \ref{f-l-etabar}:}\\
Given the fine-tuning effort $Q_{tg}$ in period $t$, The consumer's utility at period $t$ is:
\begin{equation*}
    u_{tg}(Q_{tg},\alpha_{tg})= Q_{tg}\alpha_{tg} - \alpha^2_{tg}/2,
\end{equation*}
from which we derive that the consumer's optimal usage level at period $t$ is $\alpha_{tg}(Q_{tg})=Q_{tg}$.

Given developer 1's openness level $\eta_{1g}$ and the license fee decision $w_{1g}$ at $t=1$, the deployer's objective function in the first period is:
\begin{equation*}
\Pi_{1g} (Q_{1g},\alpha_{1g})= (\theta-w_{1g}+s) \alpha_{1g} - \frac{cQ_{1g}^2}{1+\eta_{1g}},
\end{equation*}
from which we derive the deployer's best response:
\begin{equation*}
    Q_{1g}(w_{1g},\eta_{1g})=\frac{(1+\eta_{1g})(\theta-w_{1g}+s)}{2c}.
\end{equation*}

Given developer 1's openness level $\eta_{1g}$ and the license fee decision $w_{1g}$ at $t=1$, and developer 1's openness level $\eta_{2g}$ and the license fee decision $w_{2g}$ at $t=2$, the deployer's profit of choosing developer 1 in the second period is:
\begin{equation*}
\Pi_{2Ig} (Q_{2g},\alpha_{2g})= (\theta-w_{2g}+s) \alpha_{2g} - \frac{cQ_{2g}^2}{(1+k\alpha_{1g})(1+\eta_{2g})}.
\end{equation*}
Plugging $\alpha_{tg}(Q_{tg})$ and $Q_{1g}(w_{1g},\eta_{1g})$ in the deployer's objective function $\Pi_{2Ig} (Q_{2g},\alpha_{2g})$, we have:
\begin{equation*}
    (\theta-w_{2g}+s)Q_{2g}-\frac{2c^2Q^2_{2g}}{[2c+k(1+\eta_{1g})(\theta-w_{1g}+s)](1+\eta_{2g})}.
\end{equation*}
We derive the deployer's best response:
\begin{equation*}
    Q_{2g}(w_{1g},w_{2g},\eta_{1g},\eta_{2g})=\frac{(\theta-w_{2g}+s)[2c+k(1+\eta_{1g})(\theta-w_{1g}+s)](1+\eta_{2g})}{4c^2}.
\end{equation*}
Plugging $Q_{2g}(w_{1g},w_{2g},\eta_{1g},\eta_{2g})$ into the deployer's profit function, we have:
\begin{equation*}
\Pi_{2Ig} (w_{1g},w_{2g},\eta_{1g},\eta_{2g})=\frac{(\theta-w_{2g}+s)^2[2c+k(1+\eta_{1g})(\theta-w_{1g}+s)](1+\eta_{2g})}{8c^2}.
\end{equation*}

Consider next the deployer's profit of choosing developer 2 in the second period. Given developer 2's openness level $\Tilde{\eta}_{2g}$ and the license fee decision $\Tilde{w}_{2g}$, and developer 1's openness level $\eta_{1g}$ chosen in the first period, the deployer's profit at $t=2$ is:
\begin{equation*}
\Pi_{2Eg} (\Tilde{Q}_{2g},\Tilde{\alpha}_{2g})= (\theta-\Tilde{w}_{2g}+s) \Tilde{\alpha}_{2g} - \frac{c\Tilde{Q}_{2g}^2}{(1+\eta_{1g})(1+\Tilde{\eta}_{2g})}.
\end{equation*}
Plugging $\Tilde{\alpha}_{2g}=\Tilde{Q}_{2g}$ into the deployer's profit function $\Pi_{2Eg} (\Tilde{Q}_{2g},\Tilde{\alpha}_{2g})$ and taking the first order derivative to $\Tilde{Q}_{2g}$, we derive the deployer's optimal fine-tuning effort:
\begin{equation*}
    Q_{2g}(\Tilde{w}_{2g},\eta_{1g},\Tilde{\eta}_{2g})=\frac{(1+\eta_{1g})(1+\Tilde{\eta}_{2g})(\theta-\Tilde{w}_{2g}+s)}{2c}.
\end{equation*}
Plugging $Q_{2g}(\Tilde{w}_{2g},\eta_{1g},\Tilde{\eta}_{2g})$ into the deployer's profit function, we have:
\begin{equation*}
\Pi_{2Eg} (\Tilde{w}_{2g},\eta_{1g},\Tilde{\eta}_{2g})= \frac{(1+\eta_{1g})(1+\Tilde{\eta}_{2g})(\theta-\Tilde{w}_{2g}+s)^2}{4c}.
\end{equation*}

Both developer 1 and developer 2 choose $\eta_{2g}=\bar{\eta}_{2g}= \bar{\eta}$, which increases developer's profit and competitiveness. In addition, by our assumption, developer 2 chooses the low license fee $\Tilde{w}_{2g}=w_L$. Developer 1 wins in the second period iff $\Pi_{2Ig} (w_{1g},w_{2g},\eta_{1g},\eta_{2g}) \geq \Pi_{2Eg} (\Tilde{w}_{2g},\eta_{1g},\Tilde{\eta}_{2g})$. Plugging in $\eta_{2g}=\bar{\eta}_{2g}= \bar{\eta}$ and $\Tilde{w}_{2g}=w_L$, we simplify $\Pi_{2Ig} (w_{1g},w_{2g},\eta_{1g},\eta_{2g}) \geq \Pi_{2Eg} (\Tilde{w}_{2g},\eta_{1g},\Tilde{\eta}_{2g})$ as:
\begin{equation*}
    \left(\frac{\theta-w_{2g}+s}{\theta-w_L+s}\right)^2 \geq \frac{2c(1+\eta_{1g})}{2c+k(1+\eta_{1g})(\theta-w_{1g}+s)}.
\end{equation*}
Within the maintained strict competition range, the incumbent cannot retain the deployer at the high second-period fee $w_{2g}=w_H$.
We therefore consider $w_{2g}=w_L$, for which the retention condition becomes:
\begin{equation*}
    1 \geq \frac{2c(1+\eta_{1g})}{2c+k(1+\eta_{1g})(\theta-w_{1g}+s)},
\end{equation*}
which depends on developer 1's openness level $\eta_{1g}$ and the license fee decision $w_{1g}$ in the first period.

If developer 1 chooses the high license fee $w_{1g}=w_H$ in the first period, $\eta_{1g} \leq \frac{k(\theta+s-w_H)}{2c-k(\theta+s-w_H)}$ guarantees that developer 1 can win in the second period. We denote $\bar{\eta}_{Hg}=\frac{k(\theta+s-w_H)}{2c-k(\theta+s-w_H)}$. If developer 1 chooses the low license fee $w_{1g}=w_L$ in the first period, $\eta_{1g} \leq \frac{k(\theta+s-w_L)}{2c-k(\theta+s-w_L)}$ guarantees that developer 1 can win in the second period. We denote $\bar{\eta}_{Lg}=\frac{k(\theta+s-w_L)}{2c-k(\theta+s-w_L)}$. It can be verified that $\bar{\eta}_{Hg} \leq \bar{\eta}_{Lg}$.

Next, we investigate developer 1's profit. Note that if developer 1 wins in the second period, $w_{2g}=w_L$ and $\eta_{2g}=\bar{\eta}$. Given developer 1's $w_{1g}$ and $\eta_{1g}$ in the first period, winning in the second period, developer 1's profit over two periods is:
\begin{equation*}
    \pi_{wing} (w_{1g},\eta_{1g}) = w_{1g} \alpha_{1g} + w_L \alpha_{2g}.
\end{equation*}
Plugging $\alpha_{1g}$, $\alpha_{2g}$, $Q_{1g}$, $Q_{2g}$ into $ \pi_{wing} (w_{1g},\eta_{1g})$, it can be verified that developer 1's profit increases in $\eta_{1g}$.

Given developer 1's $w_{1g}$ and $\eta_{1g}$ in the first period, losing in the second period, developer 1's profit only in the first period is:
\begin{equation*}
    \pi_{loseg} (w_{1g},\eta_{1g}) = w_{1g} \alpha_{1g}.
\end{equation*}
Plugging $\alpha_{1g}$, $\alpha_{2g}$, $Q_{1g}$, $Q_{2g}$ into $ \pi_{loseg} (w_{1g},\eta_{1g})$, it can be verified that developer 1's profit increases in both $w_{1g}$ and $\eta_{1g}$ as well.\Halmos\\[3mm]

\noindent\textbf{Proof of Proposition \ref{f-p-equilibrium-k}:}\\
Lemma \ref{f-l-etabar} implies that developer 1 selects $w_{2g}=w_L$ and $\eta_{2g}=\bar{\eta}$ in the second period. So developer 1's decision $\eta_{1g}$ and $w_{1g}$ in the first period determines if developer 1 can win in the second period and the corresponding profit.

Let us suppose developer 1 loses in the second period. Plugging $\alpha_{1g}$, $\alpha_{2g}$, $Q_{1g}$, $Q_{2g}$ into $ \pi_{loseg} (w_{1g},\eta_{1g})$, we have:
\begin{equation*}
    \pi_{loseg} (w_{1g},\eta_{1g}) = \frac{(1+\eta_{1g})(\theta+s-w_{1g})w_{1g}}{2c}.
\end{equation*}
Given that developer 1 loses for sure, it is optimal for developer 1 to select $\eta_{1g} = \bar{\eta}$ to maximize the profit. In addition, under the assumption $\theta/2 \geq w_H \geq w_L \geq 0$, it can be verified that $\pi_{loseg} (w_{1g},\eta_{1g})$ also increases in $w_{1g}$. Therefore, $\pi_{loseg} (w_H,\bar{\eta}) \geq \pi_{loseg} (w_L,\bar{\eta})$. Given the scenario that developer 1 loses, developer 1 chooses $w_{1g} = w_H$ and $\eta_{1g} = \bar{\eta}$, which is denoted as scenario $S_{0g}$ (Harvest strategy). The corresponding profit for developer 1:
\begin{equation*}
    \pi_{S_{0g}} = \frac{(1+\bar{\eta})(\theta+s-w_H)w_H}{2c}.
\end{equation*}

Let us suppose developer 1 wins in the second period. Plugging $\alpha_{1g}$, $\alpha_{2g}$, $Q_{1g}$, $Q_{2g}$ into $ \pi_{wing} (w_{1g},\eta_{1g})$, we have:
\begin{equation*}
    \pi_{wing} (w_{1g},\eta_{1g}) = \frac{2c(1+\eta_{1g})(\theta+s-w_{1g})w_{1g}+(1+\bar{\eta})[2c+(1+\eta_{1g})k(\theta+s-w_{1g})](\theta+s-w_L)w_L}{4c^2}.
\end{equation*}

For either fixed first-period fee, winning profit increases with openness, so the best winning choice uses the largest openness consistent with retention in Lemma \ref{f-l-etabar}.
The two candidates are $S_{1g}$ (Defend), with $(w_{1g},\eta_{1g})=(w_H,\bar{\eta}_{Hg})$, and $S_{2g}$ (Dominate), with $(w_{1g},\eta_{1g})=(w_L,\bar{\eta}_{Lg})$.
Comparing their profits determines the incumbent's preferred winning strategy.
Substituting the two candidates gives:

\begin{align*}
    \pi_{S_{1g}} &= \frac{2c(1+\bar{\eta}_{Hg})(\theta+s-w_H)w_H+(1+\bar{\eta})[2c+(1+\bar{\eta}_{Hg})k(\theta+s-w_H)](\theta+s-w_L)w_L}{4c^2}.\\
    \pi_{S_{2g}} &= \frac{[2c(2+\bar{\eta}_{Lg}+\bar{\eta})+(1+\bar{\eta}_{Lg})(1+\bar{\eta})k(\theta+s-w_L)](\theta+s-w_L)w_L}{4c^2}.
\end{align*}

Develope 1 chooses from the three scenarios $\pi_{S_{0g}}$, $\pi_{S_{1g}}$, and $\pi_{S_{2g}}$ to generate the largest profit. We first compare $\pi_{S_{1g}}$ and $\pi_{S_{2g}}$. We solve that $\pi_{S_{1g}} \geq \pi_{S_{2g}}$ iff:
\begin{equation*}
    k \leq \frac{2c(\theta+s-w_L-w_H)}{(\theta+s-w_L)(\theta+s-w_H+w_L+\bar{\eta} w_L)}.
\end{equation*}
We denote $\bar{k}_{12g} = \frac{2c(\theta+s-w_L-w_H)}{(\theta+s-w_L)(\theta+s-w_H+w_L+\bar{\eta} w_L)}$. Then we compare $\pi_{S_{1g}}$ and $\pi_{S_{0g}}$. We solve that $\pi_{S_{1g}} \geq \pi_{S_{0g}}$ iff:
\begin{equation*}
    k \geq \frac{2c[\bar{\eta}(\theta+s-w_H)w_H-(1+\bar{\eta})(\theta+s) w_L+(1+\bar{\eta})w^2_L]}{(1+\bar{\eta})(\theta+s-w_H)^2 w_H}.
\end{equation*}
We denote $\bar{k}_{13g} = \frac{2c[\bar{\eta}(\theta+s-w_H)w_H-(1+\bar{\eta})(\theta+s) w_L+(1+\bar{\eta})w^2_L]}{(1+\bar{\eta})(\theta+s-w_H)^2 w_H}$. Finally, we compare $\pi_{S_{2g}}$ and $\pi_{S_{0g}}$. We solve that $\pi_{S_{2g}} \geq \pi_{S_{0g}}$ iff:
\begin{equation*}
    k \geq 2c \left(\frac{1}{\theta+s-w_L}-\frac{(2+\bar{\eta})w_L}{(1+\bar{\eta})(\theta+s-w_H)w_H}\right).
\end{equation*}

\begin{figure}[ht]
	\centering \footnotesize
\begin{tikzpicture}[scale=0.9]
\linespread{0.8}
\foreach \x in {0,3,6,9}{
    \draw (\x cm,5pt) -- (\x cm,-5pt);
}
\node[align=center] at (3,0.5) {$\bar{k}_{13g}$};
\node[align=center] at (6,0.5) {$\bar{k}_{23g}$};
\node[align=center] at (9,0.5) {$\bar{k}_{12g}$};

\draw [thick,->] (0,0) -- (11,0)node[below]{~~$k$};
\node at (-2,0) {$\bar{\eta} \leq \eta'_g$};
\node[align=center] at (0,0.5) {0};

\draw [decorate,decoration={brace,amplitude=5pt,mirror,raise=2ex}] (0.05,0) -- (2.95,0) node[midway,yshift=-2em]{$S_{0g}$};
\draw [decorate,decoration={brace,amplitude=5pt,mirror,raise=2ex}] (3.05,0) -- (8.95,0) node[midway,yshift=-2em]{$S_{1g}$};
\draw [decorate,decoration={brace,amplitude=5pt,mirror,raise=2ex}] (9.05,0) -- (10.8,0) node[midway,yshift=-2em]{\textcolor{black}{$S_{2g}$}};
\end{tikzpicture}
\begin{tikzpicture}[scale=0.9]
\linespread{0.8}
\foreach \x in {0,3,6,9}{
    \draw (\x cm,5pt) -- (\x cm,-5pt);
}
\node[align=center] at (3,0.5) {$\bar{k}_{12g}$};
\node[align=center] at (6,0.5) {$\bar{k}_{23g}$};
\node[align=center] at (9,0.5) {$\bar{k}_{13g}$};
\node[align=center] at (0,0.5) {0};

\draw [thick,->] (0,0) -- (11,0)node[below]{~~$k$};
\node at (-1.9,0) {$\bar{\eta} \geq \eta'_g$};

\draw [decorate,decoration={brace,amplitude=5pt,mirror,raise=2ex}] (0,0) -- (5.95,0)  node[midway,yshift=-2em]{$S_{0g}$};
\draw [decorate,decoration={brace,amplitude=5pt,mirror,raise=2ex}] (6.05,0) -- (10.8,0) node[midway,yshift=-2em]{\textcolor{black}{$S_{2g}$}};
\end{tikzpicture}

\caption{Developer 1's Scenario Selection Decision under Government Subsidy}\label{f-f-appendix-kbar}
\end{figure}

We analyze the relationship among $\bar{k}_{12g}$, $\bar{k}_{13g}$, and $\bar{k}_{23g}$. There exists a threshold:

\begin{equation*}
    \eta'_g = \frac{(\theta+s-w_H)^2+w_L(w_H-w_L)}{(\theta+s-w_H-w_L)(w_H-w_L)},
\end{equation*}
such that $\bar{k}_{13g} \leq \bar{k}_{23g} \leq \bar{k}_{12g}$ when $\bar{\eta} \leq \eta'_g$, and $\bar{k}_{12g} \leq \bar{k}_{23g} \leq \bar{k}_{13g}$ when $\bar{\eta} \geq \eta'_g$. The results are summarized in Figure \ref{f-f-appendix-kbar}.

We denote $\bar{k}_{1g} = \min\{\bar{k}_{13g},\bar{k}_{23g}\}$, and $\bar{k}_{2g} = \max\{\bar{k}_{12g},\bar{k}_{23g}\}$. The proposition is proved.\Halmos\\[3mm]

\noindent\textbf{Proof of Proposition \ref{f-p-welfare comparison}:}\\
We retain the price restrictions and compare the two equilibria within the common competition range maintained in Section \ref{benchmark} and Appendix \ref{appen-free trial}. Set $\bar{s}=0$ unless $0<w_L<w_H\leq\theta/2$ and the baseline has $0<\bar{k}_1<\bar{k}_2$. Otherwise, its two transition thresholds are
$$\begin{aligned}
\bar{k}_1 &=\frac{2c[\bar{\eta}w_H(\theta-w_H) -(1+\bar{\eta})w_L(\theta-w_L)]}{(1+\bar{\eta})w_H(\theta-w_H)^2},\\
\bar{k}_2 &=\frac{2c(\theta-w_H-w_L)}{(\theta-w_L)[\theta-w_H+(1+\bar{\eta})w_L]}.
\end{aligned}$$

For these primitives, define the sufficient subsidy cutoff by
$$\bar{s}=\max\left\{0,\min\left\{
\begin{gathered}
w_L,\\[2pt]
\frac{2(1+\bar{\eta})w_L(w_H-w_L)}
 {\bar{\eta}w_H-(1+\bar{\eta})w_L}-(\theta-w_H),\\[2pt]
\sqrt{(2+\bar{\eta})w_Hw_L}-(\theta-w_H-w_L),\\[2pt]
\frac{(\bar{k}_2-\bar{k}_1)w_H(\theta-w_H)^3}
 {4cw_L(w_H-w_L)},\\[2pt]
\frac{\bar{k}_2(\theta-w_L)(w_H-w_L)}{2c},\\[2pt]
\frac{\theta(w_H-w_L)(\theta-w_H)}
 {\theta^2+2(2+\bar{\eta})\theta w_L
 -w_H(\theta+w_L)+(1+\bar{\eta})w_L^2}
\end{gathered}
\right\}\right\}.$$

A positive subsidy satisfying $s<\bar{s}$ therefore selects a restricted parameter region; the cutoff can be zero for other primitives. In particular, $\bar{k}_1>0$ ensures $\bar{\eta}w_H-(1+\bar{\eta})w_L>0$, so its denominator above is positive. The conditions are sufficient, not necessary.

\medskip
\noindent\textit{Threshold ordering.}
The candidate crossings in the proof of Proposition \ref{f-p-equilibrium-k} are
$$\begin{aligned}
\bar{k}_{13g} &=\frac{2c[\bar{\eta}w_H(\theta+s-w_H) -(1+\bar{\eta})w_L(\theta+s-w_L)]}{(1+\bar{\eta})w_H(\theta+s-w_H)^2},\\
\bar{k}_{12g} &=\frac{2c(\theta+s-w_H-w_L)}{(\theta+s-w_L)[\theta+s-w_H+(1+\bar{\eta})w_L]}.
\end{aligned}$$
The second and third bounds defining $\bar{s}$ imply that both crossings increase throughout the interval from zero to $s$. Specifically,
$$\begin{aligned}
\frac{\partial\bar{k}_{13g}}{\partial s} &=\frac{2c\bigl\{2(1+\bar{\eta})w_L(w_H-w_L) -[\bar{\eta}w_H-(1+\bar{\eta})w_L](\theta+s-w_H)\bigr\}}{(1+\bar{\eta})w_H(\theta+s-w_H)^3}>0,\\
\frac{\partial\bar{k}_{12g}}{\partial s} &=\frac{2c[(2+\bar{\eta})w_Hw_L-(\theta+s-w_H-w_L)^2]}{(\theta+s-w_L)^2[\theta+s-w_H+(1+\bar{\eta})w_L]^2}>0.
\end{aligned}$$
Moreover,
$$\frac{\partial\bar{k}_{13g}}{\partial s} \leq \frac{4cw_L(w_H-w_L)}{w_H(\theta-w_H)^3}.$$
The fourth bound defining $\bar{s}$ consequently gives
$$0<\bar{k}_{13g}-\bar{k}_1 <\frac{4cw_L(w_H-w_L)}{w_H(\theta-w_H)^3}s <\bar{k}_2-\bar{k}_1.$$
It follows that $0<\bar{k}_1<\bar{k}_{13g}<\bar{k}_2<\bar{k}_{12g}$. Hence the subsidized equilibrium retains the three-regime ordering, with $\bar{k}_{1g}=\bar{k}_{13g}$ and $\bar{k}_{2g}=\bar{k}_{12g}$.

\medskip
\noindent\textit{Part (a).}
For $\bar{k}_1<k<\bar{k}_{1g}$, the baseline selects Defend and the subsidized equilibrium selects Harvest. Their efforts satisfy
$$\begin{aligned}
Q_1^*&=\frac{\theta-w_H}{2c-k(\theta-w_H)},& Q_2^*&=\frac{(1+\bar{\eta})(\theta-w_L)}{2c-k(\theta-w_H)},\\
Q_{1g}^*&=\frac{(1+\bar{\eta})(\theta+s-w_H)}{2c},& Q_{2g}^*&=\frac{(1+\bar{\eta})^2(\theta+s-w_L)}{2c}.
\end{aligned}$$
Because baseline Defend chooses $\bar{\eta}_H\leq\bar{\eta}$, while subsidized Harvest chooses maximal openness and retains more per unit of engagement after accounting for license payments and the subsidy, $Q_{1g}^*>Q_1^*$ and $Q_{2g}^*>Q_2^*$. At the deployer's optimum, its period-specific profit equals half of the amount remaining after license payments are deducted from revenue and subsidy receipts. Thus, its profit increases in both periods, and consumer surplus $(Q_1^2+Q_2^2)/2$ increases. Developer 2's profit rises from zero to $w_LQ_{2g}^*>0$. The incumbent's subsidized Defend candidate is more profitable than baseline Defend. Optimality of subsidized Harvest therefore implies
$$\pi_{S_{0g}}\geq\pi_{S_{1g}}>\pi_{S_1}.$$
This proves part~(a).

\medskip
\noindent\textit{Part (b).}
For $\bar{k}_2<k<\bar{k}_{2g}$, the baseline selects Dominate and the subsidized equilibrium selects Defend. Their efforts are
$$\begin{aligned}
Q_1^*&=\frac{\theta-w_L}{2c-k(\theta-w_L)},& Q_2^*&=\frac{(1+\bar{\eta})(\theta-w_L)}{2c-k(\theta-w_L)},\\
Q_{1g}^*&=\frac{\theta+s-w_H}{2c-k(\theta+s-w_H)},& Q_{2g}^*&=\frac{(1+\bar{\eta})(\theta+s-w_L)}{2c-k(\theta+s-w_H)}.
\end{aligned}$$
The fifth bound defining $\bar{s}$ satisfies
$$\frac{\bar{k}_2(\theta-w_L)(w_H-w_L)}{2c} =\frac{(\theta-w_H-w_L)(w_H-w_L)}{\theta-w_H+(1+\bar{\eta})w_L} <w_H-w_L.$$
Consequently, $s<w_H-w_L$ and $Q_{1g}^*<Q_1^*$. Furthermore,
$$Q_{2g}^*<Q_2^* \quad\Longleftrightarrow\quad k(\theta-w_L)(w_H-w_L)>2cs,$$
which holds because $k>\bar{k}_2$ and $s<\bar{s}$. Consumer surplus therefore decreases.

The joint profit of the incumbent and deployer is
$$\begin{aligned}
\pi_{S_2}+\Pi_{S_2} &=\frac{(2+\bar{\eta})(\theta+w_L)(\theta-w_L)}{2[2c-k(\theta-w_L)]},\\
\pi_{S_{1g}}+\Pi_{S_{1g}} &=\frac{(\theta+s+w_H)(\theta+s-w_H) +(1+\bar{\eta})(\theta+s+w_L)(\theta+s-w_L)}{2[2c-k(\theta+s-w_H)]}.
\end{aligned}$$
Since $s<w_H-w_L$, the ratio of the second expression to the first strictly decreases in $k$. At $k=\bar{k}_2$, the second expression is smaller precisely when
$$(2+\bar{\eta})w_Ls^2 + [\theta^2+2(2+\bar{\eta})\theta w_L -w_H(\theta+w_L)-w_L^2]s <\theta(w_H-w_L)(\theta-w_H).$$
This inequality follows from the last bound defining $\bar{s}$ and $s<w_L$, because its left-hand side is no greater than
$$ [\theta^2+2(2+\bar{\eta})\theta w_L -w_H(\theta+w_L)+(1+\bar{\eta})w_L^2]s.$$
Thus,
$$\pi_{S_{1g}}+\Pi_{S_{1g}} <\pi_{S_2}+\Pi_{S_2} \qquad\text{for every }k>\bar{k}_2$$
in the maintained range. The incumbent's subsidized Dominate candidate is more profitable than baseline Dominate. Optimality of subsidized Defend implies
$$\pi_{S_{1g}}\geq\pi_{S_{2g}}>\pi_{S_2}.$$
Its profit rises while joint profit falls, so $\Pi_{S_{1g}}<\Pi_{S_2}$. Developer 2 earns zero in both outcomes. Under the manuscript's welfare convention, social welfare is the sum of participant profits and consumer surplus; both components decline. This proves part~(b).\Halmos\\[3mm]

\noindent\textbf{Proof of Proposition \ref{ext-b-p-equilibrium-k}:}\\
We follow the backward-induction logic of the proof of Proposition \ref{b-p-equilibrium-k}, modifying only the second-period competition to accommodate the entrant-side data flywheel.

\emph{Second period.} If the deployer stays with the incumbent, its problem is unchanged from the baseline; setting $\eta_2=\bar{\eta}$ and $w_2=w_L$,
\begin{equation*}
Q_2=\frac{(1+\bar{\eta})(\theta-w_L)[2c+k(1+\eta_1)(\theta-w_1)]}{4c^2},\qquad
\Pi_{2I}=\frac{(\theta-w_L)^2\,[2c+k(1+\eta_1)(\theta-w_1)](1+\bar{\eta})}{8c^2}.
\end{equation*}
If the deployer instead switches to the entrant, its period-2 cost is $C_{2Ef}=\frac{c\tilde{Q}_2^2}{(1+\hat{k}Q_1)(1+\eta_1)(1+\tilde{\eta}_2)}$. Maximizing $(\theta-\tilde{w}_2)\tilde{Q}_2-C_{2Ef}$ over $\tilde{Q}_2$ (with $\tilde{\alpha}_2=\tilde{Q}_2$) gives
\begin{equation*}
\tilde{Q}_2=\frac{(1+\hat{k}Q_1)(1+\eta_1)(1+\tilde{\eta}_2)(\theta-\tilde{w}_2)}{2c},\qquad \Pi_{2Ef}=\frac{(1+\hat{k}Q_1)(1+\eta_1)(1+\tilde{\eta}_2)(\theta-\tilde{w}_2)^2}{4c}.
\end{equation*}
Both period-2 developers set $\eta_2=\tilde{\eta}_2=\bar{\eta}$ (profit is increasing in openness) and the entrant sets $\tilde{w}_2=w_L$. Substituting $1+\hat{k}Q_1=\frac{2c+\hat{k}(1+\eta_1)(\theta-w_1)}{2c}$,
\begin{equation*}
\Pi_{2Ef}=\frac{[2c+\hat{k}(1+\eta_1)(\theta-w_1)](1+\eta_1)(1+\bar{\eta})(\theta-w_L)^2}{8c^2}.
\end{equation*}

\emph{Winning condition.} The incumbent retains the deployer in period 2 iff $\Pi_{2I}\ge\Pi_{2Ef}$. Cancelling the common positive factor $\frac{(1+\bar{\eta})(\theta-w_L)^2}{8c^2}$ and writing $x\equiv 1+\eta_1\ge 1$ and $A\equiv\theta-w_1>0$, this becomes $2c+kxA\ge x[2c+\hat{k}xA]$, i.e.
\begin{equation*}
\hat{k}A\,x^2+(2c-kA)\,x-2c\le 0.
\end{equation*}
The left-hand side is a strictly convex quadratic in $x$ that equals $-2c<0$ at $x=0$; hence it has exactly one positive root
\begin{equation*}
x^{+}=\frac{(kA-2c)+\sqrt{(2c-kA)^2+8c\hat{k}A}}{2\hat{k}A},
\end{equation*}
and the inequality holds iff $x\le x^{+}$. Therefore the incumbent wins iff $\eta_1\le x^{+}-1$. Setting $w_1=w_H$ (so $A=\theta-w_H$) yields $x^{+}-1=\bar{\eta}_{Hf}$, and setting $w_1=w_L$ yields $x^{+}-1=\bar{\eta}_{Lf}$, exactly the closed forms defined preceding the proposition; as $\hat{k}\to 0$ they collapse to the baseline $\bar{\eta}_H,\bar{\eta}_L$. Because the quadratic changes sign from negative to positive as $x$ increases through $x^{+}$, we have $\partial(\Pi_{2I}-\Pi_{2Ef})/\partial\eta_1<0$ at the binding openness, while $\partial(\Pi_{2I}-\Pi_{2Ef})/\partial k=\frac{(1+\bar{\eta})(\theta-w_L)^2}{8c^2}(1+\eta_1)(\theta-w_1)>0$; by the implicit function theorem $\bar{\eta}_{Hf}$ and $\bar{\eta}_{Lf}$ are strictly increasing in $k$. The retention frontiers satisfy $\bar{\eta}_{Hf}\le\bar{\eta}_{Lf}$: pricing low permits weakly higher openness while still winning.
The ordering is strict when $k>\hat{k}$ and $w_L<w_H$; at $k=\hat{k}>0$, both frontiers equal zero.

\emph{First period.} Conditional on either second-period outcome and a fixed first-period fee, the incumbent's profit increases in $\eta_1$. If it concedes period 2, it earns only $w_1\alpha_1=\frac{(1+\eta_1)(\theta-w_1)w_1}{2c}$, which under $\theta/2\ge w_H\ge w_L\ge 0$ is maximized at $\eta_1=\bar{\eta}$, $w_1=w_H$. If it competes, it pushes openness to the relevant winning frontier. Hence the incumbent selects among $S_{0f}$ (\textbf{Harvest}), $S_{1f}$ (\textbf{Defend}), and $S_{2f}$ (\textbf{Dominate}). The corresponding incumbent profits are
\begin{align*}
\pi_{S_{0f}} &= \frac{(1+\bar{\eta})(\theta-w_H)w_H}{2c},\\
\pi_{S_{1f}} &= \frac{(1+\bar{\eta}_{Hf})(\theta-w_H)w_H}{2c}+\frac{(1+\bar{\eta})(\theta-w_L)w_L\,[2c+k(1+\bar{\eta}_{Hf})(\theta-w_H)]}{4c^2},\\
\pi_{S_{2f}} &= \frac{(1+\bar{\eta}_{Lf})(\theta-w_L)w_L}{2c}+\frac{(1+\bar{\eta})(\theta-w_L)w_L\,[2c+k(1+\bar{\eta}_{Lf})(\theta-w_L)]}{4c^2},
\end{align*}
where the winning profits use $\pi_{win}=w_1\alpha_1+w_L\alpha_2$ with $\alpha_1=Q_1$, $\alpha_2=Q_2$.

\emph{Candidate selection.} The incumbent therefore chooses $\arg\max\{\pi_{S_{0f}},\pi_{S_{1f}},\pi_{S_{2f}}\}$. The first candidate is constant in $k$, while the two winning candidates are continuous and increasing. This does not by itself imply that their difference is monotone, so the proposition does not impose a universal ordering or uniqueness claim. As $\hat{k}\to0$, the winning frontiers and candidate profits converge to their baseline counterparts; whenever the baseline crossings are strict, the baseline ordering is recovered locally by continuity. For the parameters in Figure~\ref{ext-b-f-equilibrium}, direct evaluation over the maintained range gives one Harvest--Defend crossing and no Dominate region. This proves the proposition.\Halmos\\[3mm]

\noindent\textbf{Proof of Proposition \ref{ext-b-p-welfare}:}\\
Social welfare $W$ is the sum of both developers' profits, the deployer's profit, and consumer surplus. All license fees are pure transfers between the deployer and the developers, so they cancel in the aggregate; using $\alpha_t=Q_t$ and per-period consumer surplus $u_t=Q_t^2/2$, welfare reduces to $\sum_{t=1,2}\big[\theta Q_t-(\text{fine-tuning cost}_t)+Q_t^2/2\big]$. We compare the unregulated equilibrium of Proposition \ref{ext-b-p-equilibrium-k} with the outcome under a full-openness mandate that fixes $\eta_1=\bar{\eta}$.

\emph{Mandated welfare.} Under the proposition's condition that full openness lies above the winning frontier, fixing $\eta_1=\bar{\eta}$ forces Harvest $S_{0f}$: the incumbent sets $w_1=w_H$ and the deployer adopts the entrant, whose fine-tuning benefits from $1+\hat{k}Q_1^{h}$. The resulting welfare
\begin{equation*}
W^{\mathrm{mand}}=\theta Q_1^{h}-\frac{c(Q_1^{h})^2}{1+\bar{\eta}}+\frac{(Q_1^{h})^2}{2}+\theta Q_2^{h}-\frac{c(Q_2^{h})^2}{(1+\hat{k}Q_1^{h})(1+\bar{\eta})^2}+\frac{(Q_2^{h})^2}{2},
\end{equation*}
with $Q_1^{h}=\frac{(1+\bar{\eta})(\theta-w_H)}{2c}$ and $Q_2^{h}=\frac{(1+\hat{k}Q_1^{h})(1+\bar{\eta})^2(\theta-w_L)}{2c}$, is independent of $k$ and increasing in $\hat{k}$. It coincides with unregulated welfare whenever Harvest is the equilibrium.

\emph{Unregulated Dominate welfare.} On any interval over which Dominate is optimal,
\begin{equation*}
W^{\mathrm{unreg}}(k)=\theta Q_1-\frac{cQ_1^2}{1+\bar{\eta}_{Lf}}+\frac{Q_1^2}{2}+\theta Q_2-\frac{cQ_2^2}{(1+kQ_1)(1+\bar{\eta})}+\frac{Q_2^2}{2},
\end{equation*}
with $Q_1=\frac{(1+\bar{\eta}_{Lf})(\theta-w_L)}{2c}$ and $Q_2=\frac{(1+\bar{\eta})(\theta-w_L)[2c+k(1+\bar{\eta}_{Lf})(\theta-w_L)]}{4c^2}$. The winning root $\bar{\eta}_{Lf}$, and hence both $Q_1$ and $Q_2$, increase with $k$. At the deployer's private optimum, per-period welfare can be written as $\frac{\theta+w_t}{2}Q_t+\frac12Q_t^2$, which is strictly increasing in effort. Thus Dominate welfare is continuous and strictly increasing within any Dominate interval.

\emph{Conditional crossing.} Because mandated welfare is constant in $k$ and Dominate welfare is strictly increasing, their difference has at most one zero within a connected Dominate interval. A reversal exists if and only if the difference becomes positive at an admissible $k$. For the displayed calibration, direct evaluation gives no admissible Dominate interval: at $k=0.25$, for example, the candidate profits are $(7.8125,8.0205,7.1844)$, so Defend is optimal, with unregulated welfare $60.4710$ versus mandated welfare $193.5059$. The mandate therefore raises welfare rather than generating an openness trap in that calibration. This proves the proposition.\Halmos\\[3mm]

\noindent\textbf{Proof of Proposition \ref{ext2-b-p-equilibrium}:}\\
As in the proof of Lemma \ref{b-l-etabar}, the consumer sets $\alpha_t=Q_t$ and the deployer's first-period best response is $Q_1(w_1,\eta_1)=\frac{(1+\eta_1)(\theta-w_1)}{2c}$, so $\alpha_1=Q_1$ and $1+kQ_1=\frac{2c+k(1+\eta_1)(\theta-w_1)}{2c}$.

\emph{Second period.} If the deployer stays with the incumbent, its cost carries the extra spillover factor $1+\beta\eta_1$, so its profit is $\Pi_{2Is}=(\theta-w_2)Q_2-\frac{cQ_2^2}{(1+kQ_1)(1+\beta\eta_1)(1+\eta_2)}$. With $\eta_2=\bar{\eta}$ the best response is $Q_2=\frac{(\theta-w_2)(1+\beta\eta_1)(1+\bar{\eta})[2c+k(1+\eta_1)(\theta-w_1)]}{4c^2}$ and
\begin{equation*}
\Pi_{2Is}^*=\frac{(\theta-w_2)^2(1+\beta\eta_1)(1+\bar{\eta})[2c+k(1+\eta_1)(\theta-w_1)]}{8c^2}.
\end{equation*}
Choosing the entrant yields, exactly as in the baseline, $\Pi_{2E}^*=\frac{(1+\eta_1)(1+\bar{\eta})(\theta-w_L)^2}{4c}$ under $\tilde{\eta}_2=\bar{\eta}$, $\tilde{w}_2=w_L$. Both period-2 developers set $\eta_2=\bar{\eta}$ and price at $w_2=w_L$. The incumbent wins iff $\Pi_{2Is}^*\ge\Pi_{2E}^*$, i.e.
\begin{equation*}
\Big(\tfrac{\theta-w_2}{\theta-w_L}\Big)^2\ge\frac{2c(1+\eta_1)}{(1+\beta\eta_1)[2c+k(1+\eta_1)(\theta-w_1)]}.
\end{equation*}
Under the maintained bound on $k$ the incumbent loses for sure at $w_2=w_H$, so $w_2=w_L$ and the winning condition reduces to $\Phi(\eta_1)\ge0$, where
\begin{equation*}
\Phi(\eta_1):=(1+\beta\eta_1)[2c+k(1+\eta_1)(\theta-w_1)]-2c(1+\eta_1)=\beta kA\,\eta_1^2+[(1+\beta)kA-2c(1-\beta)]\eta_1+kA,\quad A:=\theta-w_1.
\end{equation*}
At $k=0$, the maintained condition is $0<\beta<\bar{\beta}(0)=1$, and $\Phi(\eta_1)=2c(\beta-1)\eta_1$.
The incumbent therefore retains the deployer only at $\eta_1=0$, so $\bar{\eta}_{Hs}=\bar{\eta}_{Ls}=0$.
For $k>0$, $\Phi(0)=kA>0$ and the leading coefficient $\beta kA>0$, so $\Phi$ is an upward-opening quadratic.
Under the maintained admissibility condition verified below, $\Phi(\bar{\eta})<0$ for either first-period fee: the smaller root lies below $\bar{\eta}$ and the larger root lies above it.
Solving $\Phi=0$ (its discriminant equals $[(1+\beta)kA-2c(1-\beta)]^2-4\beta kA\cdot kA=(1-\beta)^2[2c-kA]^2-8\beta(1-\beta)ckA$) gives the smaller root for $k>0$:
\begin{equation*}
\bar{\eta}_{js}=\frac{2(1-\beta)c-(1+\beta)k(\theta-w_j)-\sqrt{(1-\beta)^2[2c-k(\theta-w_j)]^2-8\beta(1-\beta)ck(\theta-w_j)}}{2\beta k(\theta-w_j)},\quad j\in\{H,L\},
\end{equation*}
so, over the feasible range $0\le\eta_1\le\bar{\eta}$, the incumbent wins at $w_1=w_j$ if and only if $\eta_1\le\bar{\eta}_{js}$.

\textit{Admissibility.} The maintained competitive region requires full openness to make the deployer prefer the entrant under either first-period license fee.
It suffices to impose $\Phi(\bar{\eta})<0$ at $A=\theta-w_L$, corresponding to the low first-period fee:
\begin{equation*}
2c(1+\bar{\eta})>(1+\beta\bar{\eta})[2c+k(1+\bar{\eta})(\theta-w_L)]\ \Longleftrightarrow\ 2c\bar{\eta}(1-\beta)>k(1+\bar{\eta})(1+\beta\bar{\eta})(\theta-w_L),
\end{equation*}
which rearranges to the strict condition $\beta<\bar{\beta}(k)=\frac{2c\bar{\eta}-k(1+\bar{\eta})(\theta-w_L)}{\bar{\eta}[2c+k(1+\bar{\eta})(\theta-w_L)]}$.
For $k>0$, $\Phi(0)>0$ and $\Phi(\bar{\eta})<0$ imply two real roots on opposite sides of $\bar{\eta}$.
Since $\partial\Phi/\partial A=k(1+\eta_1)(1+\beta\eta_1)>0$, the smaller root increases in $A=\theta-w_j$, giving $\bar{\eta}_{Hs}\le\bar{\eta}_{Ls}<\bar{\eta}$.
The incumbent also loses at full openness under the high first-period fee because $\Phi(\bar{\eta};\theta-w_H)\le\Phi(\bar{\eta};\theta-w_L)<0$.
At $k=0$, $\Phi$ is independent of $A$; both frontiers are zero, and full openness likewise favors the entrant under either first-period fee.
At $\beta=\bar{\beta}(k)$, $\Phi(\bar{\eta})=0$ after the low first-period fee, so the incumbent retains the deployer by the tie-breaking rule.
The equality case must therefore be excluded from the forced-Harvest argument.

\emph{First period.} If the incumbent loses in period 2 it earns only $\pi_{lose}(w_1,\eta_1)=w_1\alpha_1=\frac{(1+\eta_1)(\theta-w_1)w_1}{2c}$, which increases in $\eta_1$ and, under $\theta/2\ge w_H\ge w_L$, in $w_1$; conditional on losing it therefore plays $S_0$ (Harvest), $w_1=w_H$, $\eta_1=\bar{\eta}$, with
\begin{equation*}
\pi_{S_0}=\frac{(1+\bar{\eta})(\theta-w_H)w_H}{2c}.
\end{equation*}
If it wins it earns $\pi_{win}(w_1,\eta_1)=\frac{(1+\eta_1)(\theta-w_1)w_1}{2c}+w_LQ_2$ with $Q_2=\frac{(\theta-w_L)(1+\beta\eta_1)(1+\bar{\eta})[2c+k(1+\eta_1)(\theta-w_1)]}{4c^2}$. Both $1+\beta\eta_1$ and the flywheel bracket increase in $\eta_1$, so $\pi_{win}$ increases in $\eta_1$ and the incumbent pushes $\eta_1$ to the winning frontier: $S_1$ (Defend, $w_1=w_H$, $\eta_1=\bar{\eta}_{Hs}$) or $S_2$ (Dominate, $w_1=w_L$, $\eta_1=\bar{\eta}_{Ls}$). At the frontier the winning inequality binds, $(1+\beta\bar{\eta}_{js})[2c+k(1+\bar{\eta}_{js})(\theta-w_j)]=2c(1+\bar{\eta}_{js})$, whence $Q_2=\frac{(1+\bar{\eta}_{js})(1+\bar{\eta})(\theta-w_L)}{2c}$ and
\begin{align*}
\pi_{S_1}&=\frac{(1+\bar{\eta}_{Hs})[(\theta-w_H)w_H+(1+\bar{\eta})(\theta-w_L)w_L]}{2c},\\
\pi_{S_2}&=\frac{(1+\bar{\eta}_{Ls})(2+\bar{\eta})(\theta-w_L)w_L}{2c}.
\end{align*}

\emph{Candidate selection.} The incumbent selects $\max\{\pi_{S_0},\pi_{S_1},\pi_{S_2}\}$. Implicit differentiation of $\Phi=0$ gives $\frac{d\bar{\eta}_{js}}{dk}>0$ because $\partial\Phi/\partial k>0$ and $\partial\Phi/\partial\eta_1<0$ at the smaller root. Hence both winning-candidate profits increase with $k$, while $\pi_{S_0}$ is constant. This observation does not establish monotonicity of $\pi_{S_1}-\pi_{S_2}$, so the proposition states the global Harvest--Defend--Dominate sequence only when the pairwise crossings are unique and ordered. Direct numerical evaluation verifies those conditions for Figure~\ref{ext2-b-f-equilibrium}. This proves the proposition.\Halmos\\[3mm]

\noindent\textbf{Proof of Proposition \ref{ext2-b-p-welfare}:}\\
Consider the full-openness mandate $\eta_1=\bar{\eta}$. By the strict admissibility condition $\beta<\bar{\beta}(k)$, $\Phi(\bar{\eta})<0$ under $w_L$ and, a fortiori, under $w_H$; the incumbent therefore loses period 2 regardless of $w_1$. The mandated incumbent sets $w_1=w_H$, producing Harvest welfare $W^{\mathrm{mand}}=W_{S_0}$, which is independent of $k$. At the excluded equality $\beta=\bar{\beta}(k)$, choosing the low first-period fee permits retention at full openness by tie-breaking, so the forced-Harvest conclusion does not follow.

On an interval $\mathcal K_D$ where the unregulated equilibrium is Dominate, both $\bar{\eta}_{Ls}$ and the resulting efforts $Q_1,Q_2$ increase with $k$. As in the preceding proof, welfare at the deployer's private optimum is strictly increasing in each effort, so $W_{S_2}(k)$ is continuous and strictly increasing on $\mathcal K_D$.

Therefore $W_{S_2}(k)-W^{\mathrm{mand}}$ is strictly increasing on $\mathcal K_D$ and has at most one zero there. If its values straddle zero, the intermediate value theorem yields the unique threshold stated in the proposition; if they do not, no welfare reversal is implied. This proves the proposition.\Halmos\\[3mm]

\noindent\textbf{Proof of Lemma \ref{ext2-b-l-etabar}:}\\
We argue by backward induction, taking first-period openness $\eta_{1p}$ and operating effort $Q_{1p}$ as given.
In period 2, the owner sets the incumbent's openness to $\eta_{2p}=\bar{\eta}$ to lower adaptation costs on its branch.
The entrant chooses $\tilde{\eta}_{2p}=\bar{\eta}$ because its license revenue is weakly increasing in its own openness, and strictly increasing when its branch is used and $w_L>0$.
Full openness remains a weak best response when $w_L=0$; choices on unused branches do not affect the allocation.
At these offers, the operating unit's effort responses are $Q_{2p}^{*}=\frac{(1+kQ_{1p})(1+\bar{\eta})\theta}{2c}$ and $\tilde{Q}_{2p}^{*}=\frac{(1+\eta_{1p})(1+\bar{\eta})(\theta-w_L)}{2c}$.
Substituting these,
\begin{equation*}
\pi_{2Ip}^{*}=\frac{(1+kQ_{1p})(1+\bar{\eta})\theta^{2}}{4c},\qquad
\pi_{2Ep}^{*}=\rho\,\pi_{2Ip}^{*}+(1-\rho)\frac{(1+\eta_{1p})(1+\bar{\eta})(\theta-w_L)^{2}}{4c}.
\end{equation*}

Under the maintained operating rule, $Q_{1p}^{*}=\frac{(1+\eta_{1p})\theta}{2c}$ uniquely maximizes current-period profit $\pi_{1p}(\eta_{1p},Q_{1p})=\theta Q_{1p}-\frac{cQ_{1p}^{2}}{1+\eta_{1p}}$.
Hence $\pi_{1p}^{*}=\frac{(1+\eta_{1p})\theta^{2}}{4c}$ and $kQ_{1p}^{*}=\frac{k(1+\eta_{1p})\theta}{2c}$.
The owner anticipates this response when choosing openness to maximize profit over both periods.

\emph{Part $(a)$.} For $\rho=1$ the entrant is mechanically foreclosed and full vertical integration holds trivially, so take $\rho<1$. The firm prefers full integration iff $\pi_{2Ip}^{*}\ge\pi_{2Ep}^{*}$; subtracting $\rho\,\pi_{2Ip}^{*}$ and dividing by $1-\rho>0$ gives $\pi_{2Ip}^{*}\ge\frac{(1+\eta_{1p})(1+\bar{\eta})(\theta-w_L)^{2}}{4c}$, i.e.
\begin{equation}\label{eq:starVI}
(1+kQ_{1p}^{*})\theta^{2}\ge(1+\eta_{1p})(\theta-w_L)^{2}.
\end{equation}
Substituting $kQ_{1p}^{*}=\frac{k(1+\eta_{1p})\theta}{2c}$ and clearing $2c>0$, condition \eqref{eq:starVI} is equivalent to
\begin{equation}\label{eq:starstarVI}
(1+\eta_{1p})\big[\,2c(\theta-w_L)^{2}-k\theta^{3}\,\big]\;\le\;2c\theta^{2}.
\end{equation}
(This is exactly the winning condition displayed in the text.) Let $\hat{k}_p\equiv 2c(\theta-w_L)^{2}/\theta^{3}$ be the value of $k$ at which the bracket vanishes.

\emph{Case $(i)$: $k\ge\hat{k}_p$.} The bracket is non-positive, so its left-hand side in \eqref{eq:starstarVI} is $\le 0<2c\theta^{2}$ for every $\eta_{1p}\ge0$. Hence full vertical integration is optimal for all $\eta_{1p}\in[0,\bar{\eta}]$.

\emph{Case $(ii)$: $k<\hat{k}_p$.} The bracket is strictly positive, so \eqref{eq:starstarVI} is equivalent to $1+\eta_{1p}\le\frac{2c\theta^{2}}{2c(\theta-w_L)^{2}-k\theta^{3}}$, i.e. $\eta_{1p}\le\eta_{p}^{\dagger}:=\frac{2c\theta^{2}-2c(\theta-w_L)^{2}+k\theta^{3}}{2c(\theta-w_L)^{2}-k\theta^{3}}$. The numerator equals $2c w_L(2\theta-w_L)+k\theta^{3}\ge0$, with strict positivity when $w_L>0$ or $k>0$. The denominator is positive, so $\eta_{p}^{\dagger}\ge0$, with equality only when $w_L=k=0$. Since the firm's feasible openness is capped at $\bar{\eta}$, full integration is achieved iff $\eta_{1p}\le\bar{\eta}_p:=\min\{\bar{\eta},\,\eta_{p}^{\dagger}\}$. This proves $(a)$.

\emph{Part $(b)$.} Conditional on the induced second-period regime, the firm's total profit is $\Pi_p(\eta_{1p})=\pi_{1p}^{*}(\eta_{1p})+\pi_{2}^{*}(\eta_{1p})$, where $\pi_{2}^{*}=\pi_{2Ip}^{*}$ on the full-integration region and $\pi_{2}^{*}=\pi_{2Ep}^{*}$ on the partial-integration region. We show $\Pi_p$ is strictly increasing in $\eta_{1p}$ on $[0,\bar{\eta}]$.

First, $\pi_{1p}^{*}=\frac{(1+\eta_{1p})\theta^{2}}{4c}$ is strictly increasing. Second, on the full-integration region $\pi_{2Ip}^{*}=\frac{(1+\bar{\eta})\theta^{2}}{4c}\big(1+\frac{k(1+\eta_{1p})\theta}{2c}\big)$ is weakly increasing in $\eta_{1p}$. Third, on the partial-integration region $\pi_{2Ep}^{*}=\rho\,\pi_{2Ip}^{*}+(1-\rho)\frac{(1+\eta_{1p})(1+\bar{\eta})(\theta-w_L)^{2}}{4c}$ is a convex combination of two weakly increasing terms, hence weakly increasing. It remains to rule out a downward jump where the regime switches, at $\eta_{1p}=\bar{\eta}_p$ (relevant only in case $(ii)$ with $\bar{\eta}_p=\eta_{p}^{\dagger}<\bar{\eta}$). At that point \eqref{eq:starVI} holds with equality, i.e. $\pi_{2Ip}^{*}=\frac{(1+\eta_{1p})(1+\bar{\eta})(\theta-w_L)^{2}}{4c}$; substituting into $\pi_{2Ep}^{*}$ gives $\pi_{2Ep}^{*}=\rho\,\pi_{2Ip}^{*}+(1-\rho)\pi_{2Ip}^{*}=\pi_{2Ip}^{*}$. Thus $\pi_{2}^{*}$ is continuous at the boundary. Being weakly increasing on each side of $\bar{\eta}_p$ and continuous across it, $\pi_{2}^{*}$ is weakly increasing on $[0,\bar{\eta}]$. Together with the strictly increasing first-period profit $\pi_{1p}^{*}$, this implies that $\Pi_p$ is strictly increasing on that interval, including at $k=0$ and $\rho=1$. This proves $(b)$.\Halmos\\[3mm]

\noindent\textbf{Proof of Proposition \ref{ext2-v-p-equilibrium}:}\\
By Lemma \ref{ext2-b-l-etabar}$(b)$, the firm's total profit $\Pi_p(\eta_{1p})$ is strictly increasing in $\eta_{1p}$ on $[0,\bar{\eta}]$, so the owner optimally sets $\eta_{1p}=\bar{\eta}$. The owner also sets the incumbent's period-2 openness to $\eta_{2p}=\bar{\eta}$, while the entrant chooses $\tilde{\eta}_{2p}=\bar{\eta}$ for its own model, as established in the second-period step. Choices on unused branches do not affect the allocation. We emphasize that this conclusion does not rely on any particular parameter configuration: because profit is continuous at the regime boundary $\bar{\eta}_p$ (Lemma \ref{ext2-b-l-etabar}$(b)$) and strictly increasing on both sides, there is never a downward jump that could make a lower $\eta_{1p}$ attractive. In particular, for $0\le\rho<1$ and $\eta_{1p}>\bar{\eta}_p$ the firm earns $\pi_{2Ep}^{*}\ge\pi_{2Ip}^{*}$ (partial integration is the profit-maximizing second-period branch there) and profit continues to rise in $\eta_{1p}$; this is the rigorous counterpart of the ranking $S_2\succeq S_1$ illustrated in Figure~\ref{ext2-b-f-etahbar and etalbar}.

At $\rho=1$, all operations remain with the incumbent for every $k$. Now take $0\le\rho<1$. With $\eta_{1p}=\bar{\eta}$, full vertical integration is achieved iff $\bar{\eta}\le\bar{\eta}_p$. By Lemma \ref{ext2-b-l-etabar}$(a)$: in case $(i)$ ($k\ge\hat{k}_p$) full integration holds automatically; in case $(ii)$ ($k<\hat{k}_p$), $\bar{\eta}\le\bar{\eta}_p=\min\{\bar{\eta},\eta_{p}^{\dagger}\}$ iff $\eta_{p}^{\dagger}\ge\bar{\eta}$. On $[0,\hat{k}_p)$ the map $k\mapsto\eta_{p}^{\dagger}(k)=\frac{2c\theta^{2}-2c(\theta-w_L)^{2}+k\theta^{3}}{2c(\theta-w_L)^{2}-k\theta^{3}}$ has a strictly increasing nonnegative numerator and a strictly decreasing positive denominator, hence is strictly increasing, with $\eta_{p}^{\dagger}(k)\to+\infty$ as $k\uparrow\hat{k}_p$. Solving $\eta_{p}^{\dagger}=\bar{\eta}$ gives the threshold
\begin{equation*}
\bar{k}_p=\frac{2c\big[(1+\bar{\eta})(\theta-w_L)^{2}-\theta^{2}\big]}{(1+\bar{\eta})\theta^{3}},
\end{equation*}
and $\eta_{p}^{\dagger}\ge\bar{\eta}\iff k\ge\bar{k}_p$. Moreover $\hat{k}_p-\bar{k}_p=\frac{2c}{(1+\bar{\eta})\theta}>0$, so $\bar{k}_p<\hat{k}_p$ and case $(i)$ is subsumed. Combining the two cases, full vertical integration is achieved if and only if $k\ge\bar{k}_p$. (If $(1+\bar{\eta})(\theta-w_L)^{2}\le\theta^{2}$ then $\bar{k}_p\le0$ and full integration holds for all admissible $k\ge0$.)

Finally we compute the on-path efforts. With $\eta_{1p}=\bar{\eta}$, $Q_{1p}^{*}=\frac{(1+\bar{\eta})\theta}{2c}$, and
\begin{equation*}
Q_{2p}^{*}=\frac{(1+kQ_{1p}^{*})(1+\bar{\eta})\theta}{2c}=\frac{(1+\bar{\eta})[2c+k\theta(1+\bar{\eta})]\theta}{4c^{2}}.
\end{equation*}
When $\rho=1$, or when $0\le\rho<1$ and $k\ge\bar{k}_p$, full integration obtains and the entrant branch is off path (N.A.). When $0\le\rho<1$ and $k<\bar{k}_p$, partial integration obtains and $\tilde{Q}_{2p}^{*}=\frac{(1+\eta_{1p})(1+\bar{\eta})(\theta-w_L)}{2c}=\frac{(1+\bar{\eta})^{2}(\theta-w_L)}{2c}$. These coincide with the efforts stated in the proposition.\Halmos\\[3mm]

\noindent\textbf{Proof of Proposition \ref{ext4-p-entry}:}\\
\emph{Step 0 (period-2 continuation).} From the proof of Lemma~\ref{b-l-etabar}, if the deployer stays with the incumbent at its binding posted fee $w_2$, the incumbent's period-2 revenue is $w_2 Q_2$ with $Q_2=\frac{(\theta-w_2)[2c+k(1+\eta_1)(\theta-w_1)](1+\bar{\eta})}{4c^2}$; if the deployer switches to the entrant, $\tilde{Q}_2^*=\frac{(1+\eta_1)(1+\bar{\eta})(\theta-w_L)}{2c}$. Under the maintained restriction of Lemma~\ref{b-l-etabar}, a posted $w_2=w_H$ never wins the deployer head-to-head, whereas $w_2=w_L$ wins iff $\eta_1\le\bar{\eta}_H$ (when $w_1=w_H$) or $\eta_1\le\bar{\eta}_L$ (when $w_1=w_L$).

\emph{Step 1 (entry decision).} The entrant's payoff from entering is $w_L\tilde{Q}_2^*-e$ if it would win the deployer and $-e$ otherwise. Under the tie-breaking rule, it enters iff it would win \emph{and} $w_L\tilde{Q}_2^*>e$. Since $w_L\tilde{Q}_2^*=\frac{w_L(1+\eta_1)(1+\bar{\eta})(\theta-w_L)}{2c}$ is strictly increasing in $\eta_1$, the profitability constraint is equivalent to $\eta_1>\bar{\eta}_e$ with $\bar{\eta}_e=\frac{2ce}{w_L(1+\bar{\eta})(\theta-w_L)}-1$; at $\eta_1=\bar{\eta}_e$ the entrant stays out. Thus $\bar{\eta}_e<0\Leftrightarrow e<\bar{e}_1$ and $\bar{\eta}_e\ge\bar{\eta}\Leftrightarrow e\ge\bar{e}_2$.

\emph{Step 2 (sequential rationality of the posted $w_2$).} The incumbent retains the deployer iff the entrant stays out, i.e. iff (A) $\eta_1\le\bar{\eta}_e$ (profitability fails for either posted fee) or (B) $w_2=w_L$ and $\eta_1\le\bar{\eta}_H$ (if $w_1=w_H$) / $\eta_1\le\bar{\eta}_L$ (if $w_1=w_L$). Under Channel~A the entrant is deterred irrespective of $w_2$, so the incumbent anticipates a period-2 monopoly when it posts its tariff. Since $g(w)=w(\theta-w)$ is increasing on $[0,\theta/2]$ and $w_L\le w_H\le\theta/2$, it optimally posts $w_2=w_H$. Under Channel~B, $w_2=w_L$ is a limit price: after the entrant stayed out the incumbent would prefer $w_H$, but the low tariff remains credible because it was publicly posted and made binding before the entry decision.

\emph{Step 3 (candidate profits).} With $\alpha_1=Q_1=\frac{(1+\eta_1)(\theta-w_1)}{2c}$ and Step~0, the relevant two-period profits are
\begin{align*}
\pi_{\text{Harvest}} &= \frac{(1+\bar{\eta})(\theta-w_H)w_H}{2c}, \qquad
\pi_{\text{Defend}} = \frac{(\theta-w_H)w_H+(1+\bar{\eta})(\theta-w_L)w_L}{2c-k(\theta-w_H)},\\
\pi_{\text{Dominate}} &= \frac{(2+\bar{\eta})(\theta-w_L)w_L}{2c-k(\theta-w_L)},\\
\pi_X(w_1) &= \frac{(\theta-w_1)w_1(1+\bar{\eta}_e)}{2c}\\
&\quad+\frac{w_H(\theta-w_H)(1+\bar{\eta})[2c+k(1+\bar{\eta}_e)(\theta-w_1)]}{4c^2},
\end{align*}
where $\pi_X$ (with $\eta_1=\bar{\eta}_e$ and $w_2=w_H$) is the Exclude profit at first-period fee $w_1$. When $e<\bar{e}_1$, Exclude is infeasible and the remaining Harvest/Defend/Dominate candidate profits coincide with those in Proposition~\ref{b-p-equilibrium-k}.

\emph{Step 4 (region I, $e<\bar{e}_1$).} Here $\bar{\eta}_e<0$, so Channel~A is infeasible and only Harvest/Defend/Dominate remain; the problem is identical to the baseline, giving the thresholds $\bar{k}_1,\bar{k}_2$.

\emph{Step 5 (region II, $\bar{e}_1\le e<\bar{e}_2$).} Now $0\le\bar{\eta}_e<\bar{\eta}$. \emph{(i) Exclude dominates Harvest.} With $F\equiv\frac{(\theta-w_H)w_H}{4c^2}$, the period-1 openness difference contributes $F\cdot 2c(\bar{\eta}_e-\bar{\eta})$ and the retained period-2 monopoly rents contribute $F(1+\bar{\eta})[2c+k(1+\bar{\eta}_e)(\theta-w_H)]$, so
\[
\pi_X(w_H)-\pi_{\text{Harvest}}=\frac{(\theta-w_H)w_H\,(1+\bar{\eta}_e)}{4c^2}\big[\,2c+k(1+\bar{\eta})(\theta-w_H)\,\big]>0
\]

for all $k\ge0$ whenever $\bar{\eta}_e\ge0$. \emph{(ii) Exclude versus Defend.} Both use $w_1=w_H$; whenever $\bar{\eta}_H\le\bar{\eta}_e$, Exclude retains the deployer at the monopoly fee $w_H$ and weakly greater openness, whereas Defend retains it only at the limit fee $w_L$, so Exclude yields higher profit. \emph{(iii) First-period fee.} Comparing $\pi_X(w_H)$ and $\pi_X(w_L)$, the factor $(1+\bar{\eta}_e)$ cancels and $\pi_X(w_H)\ge\pi_X(w_L)\Leftrightarrow k\le\bar{k}_{eH}=\frac{2c(\theta-w_H-w_L)}{(1+\bar{\eta})w_H(\theta-w_H)}$. \emph{(iv) Exclude versus Dominate.} Regularity condition~(ii) directly imposes $\pi_X(w_L)\ge\pi_{\mathrm{Dominate}}$ throughout the admissible range. Together with condition~(i), region~II is therefore exhausted by Exclude-$w_H$ ($k\le\bar{k}_{eH}$) and Exclude-$w_L$ ($k>\bar{k}_{eH}$). Outside this regular region, a thin Defend interval can survive when $\bar{\eta}_H>\bar{\eta}_e$, or Dominate can return when $\pi_{\mathrm{Dominate}}>\pi_X(w_L)$; wherever such a limit-pricing strategy occurs, its binding $w_2=w_L$ remains sequentially rational under the posted-tariff timing.

\emph{Step 6 (region III, $e\ge\bar{e}_2$).} Here $\bar{\eta}_e\ge\bar{\eta}$: even at $\eta_1=\bar{\eta}$ the entrant's gross profit $\bar{e}_2\le e$, so it never enters and no deterrence constraint binds. The incumbent maximizes unconstrained two-period monopoly profit, giving $\eta_1=\bar{\eta}$, $w_2=w_H$, and $w_1=w_H$ if $k\le\bar{k}_{eH}$ and $w_1=w_L$ otherwise. Collecting Steps~4--6 with the sequential-rationality argument of Step~2 establishes the proposition.\Halmos\\[3mm]

\noindent\textbf{Proof of Proposition \ref{ext5-p-spillover}:}\\
\textit{Second period.}
Both developers choose $\bar{\eta}$ and the entrant charges $w_L$, as in the baseline.
The deployer's optimized profits from staying and switching are $(1+kQ_1)(1+\bar{\eta})(\theta-w_2)^2/(4c)$ and $(1+\gamma\eta_1)(1+\bar{\eta})(\theta-w_L)^2/(4c)$.
At $w_2=w_L$, retention is equivalent to $kQ_1\geq\gamma\eta_1$.
Substituting $Q_1=(1+\eta_1)(\theta-w_1)/(2c)$ gives
\[
\eta_1[2c\gamma-k(\theta-w_1)]\leq k(\theta-w_1).
\]
Because $k<\gamma K_1<2c\gamma/(\theta-w_L)$, the bracket is positive and yields the stated frontier.
The inequality $k<\gamma K_1$ also implies $\bar{\eta}_L(\gamma)<\bar{\eta}$, so full openness loses the deployer at either first-period fee.
For $w_2=w_H$, retention requires
\[
(\theta-w_H)^2(1+kQ_1)\geq(\theta-w_L)^2(1+\gamma\eta_1).
\]
This comparison is linear in $\eta_1$ and most favorable to the incumbent at $w_1=w_L$.
It fails at $\eta_1=0$ because $k<K_2$, and at $\eta_1=\bar{\eta}$ because full openness loses even at the low second-period fee.
It therefore fails throughout the feasible interval.

\textit{First period and part (a).}
Conditional on either retention or exit, the incumbent's profit increases in $\eta_1$.
If it exits, it chooses full openness and the high first-period fee, since $w(\theta-w)$ is increasing over the price menu.
If it retains the deployer, it chooses the largest feasible openness for its first-period fee.
At that frontier, $kQ_1=\gamma\bar{\eta}_j(\gamma)$, yielding the three stated profits.
The mandate induces Harvest because neither first-period fee can retain the deployer at full openness.
At the deployer's optimum, per-period welfare is $[(\theta+w)Q+Q^2]/2$.
Harvest and Defend have the same fee sequence $(w_H,w_L)$, while Harvest has strictly greater effort in both periods because $\bar{\eta}_H(\gamma)<\bar{\eta}$.
The mandate therefore strictly raises welfare in Defend and leaves it unchanged in Harvest.

\textit{Part (b).}
Both retention profits strictly increase with $k$, whereas Harvest profit is constant.
As $k\uparrow\gamma K_1$, $\bar{\eta}_L(\gamma)\to\bar{\eta}$ and $\bar{\eta}_H(\gamma)\to\bar{\eta}_H^{*}$.
The limiting profits are
\begin{align*}
\pi^{D}_{\mathrm{lim}}&=\frac{w_L(\theta-w_L)(1+\bar{\eta})(2+\gamma\bar{\eta})}{2c},\\
\pi^{F}_{\mathrm{lim}}&=\frac{(1+\bar{\eta}_H^{*})(\theta-w_H)w_H+w_L(1+\gamma\bar{\eta}_H^{*})(1+\bar{\eta})(\theta-w_L)}{2c}.
\end{align*}
Direct comparison gives $\pi^{F}_{\mathrm{lim}}>\pi^{H}$ if and only if $\gamma>\gamma_R$, and $\pi^{D}_{\mathrm{lim}}>\pi^{H}$ if and only if $\gamma>\gamma_D$.
Thus some retention strategy strictly beats Harvest below the bound if and only if $\gamma>\min\{\gamma_R,\gamma_D\}$.
Strict monotonicity rules out restriction when both limiting profits are at most Harvest profit; continuity gives an interval of restriction next to the bound otherwise.

\textit{Part (c).}
Write $A=\theta-w_H$, $B=\theta-w_L$, $S=(1+\bar{\eta})w_LB$, and $x=k/(2c\gamma)$.
Subtraction gives
\[
\pi^{D}-\pi^{F}
=\frac{(w_H-w_L)\{x[AB+\gamma S]-(\theta-w_H-w_L)\}}{2c(1-xA)(1-xB)}.
\]
The denominator is positive throughout the stated range, and the numerator is strictly increasing in $x$.
The limiting difference is positive if and only if $\gamma>\gamma_F$.
Hence Dominate can beat Defend below the bound if and only if $\gamma>\gamma_F$; equality at the bound leaves Dominate strictly below Defend at every admissible $k$.
Similarly, strict monotonicity of $\pi^{D}$ implies that it beats Harvest somewhere if and only if $\gamma>\gamma_D$.
Both comparisons hold together on an interval next to the bound precisely when $\gamma>\max\{\gamma_D,\gamma_F\}$, proving the exact Dominate condition.

On that interval, Dominate's efforts converge to $(1+\bar{\eta})(\theta-w_L)/(2c)$ in period 1 and $(1+\gamma\bar{\eta})(1+\bar{\eta})(\theta-w_L)/(2c)$ in period 2.
The second limit equals mandated effort; the first exceeds mandated effort $(1+\bar{\eta})(\theta-w_H)/(2c)$.
The first-period linear welfare term is also greater because $(\theta+w_L)(\theta-w_L)>(\theta+w_H)(\theta-w_H)$.
The limiting welfare difference is strictly positive, so continuity establishes a welfare-loss interval adjacent to the bound.
If the Dominate condition fails, only Harvest or Defend is selected, and part (a) rules out a welfare loss.
The strict conditions also cover equality at either threshold under the stated tie-breaking rule.\Halmos\\[3mm]

\end{APPENDICES}

\end{document}